\documentclass{article}

\usepackage{PRIMEarxiv}

\usepackage[utf8]{inputenc} % allow utf-8 input
\usepackage[T1]{fontenc}    % use 8-bit T1 fonts
\usepackage{algorithm,algorithmic}
\usepackage{enumitem}

\usepackage{amsmath}
\usepackage{tikz}
\usetikzlibrary{positioning, arrows.meta}
\usepackage{listofitems} % for \readlist to create arrays
\usetikzlibrary{arrows.meta} % for arrow size
\usepackage[outline]{contour} % glow around text
\contourlength{1.4pt}

\tikzset{>=latex} % for LaTeX arrow head
\usepackage{xcolor}
\colorlet{myred}{red!80!black}
\colorlet{myblue}{blue!80!black}
\colorlet{mygreen}{green!60!black}
\colorlet{myorange}{orange!70!red!60!black}
\colorlet{mydarkred}{red!30!black}
\colorlet{mydarkblue}{blue!40!black}
\colorlet{mydarkgreen}{green!30!black}
\tikzstyle{node}=[thick,circle,draw=myblue,minimum size=22,inner sep=0.5,outer sep=0.6]
\tikzstyle{node in}=[node,green!20!black,draw=mygreen!30!black,fill=mygreen!0]
\tikzstyle{node hidden}=[node,blue!20!black,draw=myblue!30!black,fill=myblue!0]
\tikzstyle{node convol}=[node,orange!20!black,draw=myorange!30!black,fill=myorange!20]
\tikzstyle{node out}=[node,red!20!black,draw=myred!30!black,fill=myred!0]
\tikzstyle{connect}=[thick,mydarkblue] %,line cap=round
\tikzstyle{connect arrow}=[-{Latex[length=4,width=3.5]},thick,mydarkblue,shorten <=0.5,shorten >=1]
\tikzset{ % node styles, numbered for easy mapping with \nstyle
  node 1/.style={node in},
  node 2/.style={node hidden},
  node 3/.style={node out},
}
\usetikzlibrary{trees}
\usepackage{qtree}
\usetikzlibrary{graphs,shapes,arrows.meta}
\usetikzlibrary{shapes.geometric}
\usepackage{subcaption}
\usepackage{hyperref}       % hyperlinks
\usepackage{url}            % simple URL typesetting
\usepackage{booktabs}       % professional-quality tables
\usepackage{amsfonts}       % blackboard math symbols
\usepackage{nicefrac}       % compact symbols for 1/2, etc.
\usepackage{microtype}      % microtypography
\usepackage{lipsum}
\usepackage{fancyhdr}       % header
\usepackage{graphicx}       % graphics
\graphicspath{{media/}}     % organize your images and other figures under media/ folder
\usepackage{amsmath}
\def\F{{\mathbf F}}
\def\P{{\mathbf P}}

\def\C{{\mathbf C}}
\def\I{{\mathbf I}}
\def\u{{\mathbf u}}
\def\curlyL{{\mathcal L}}
\def\curlyG{{\mathcal G}}

\def\WF{{W(\mathbf{F})}}
\newlength\myindent
\newcommand{\bindent}{%
  \begingroup
  \setlength{\itemindent}{\myindent}
  \addtolength{\algorithmicindent}{\myindent}
}
\newcommand{\eindent}{\endgroup}

\newlist{MyIndentedList}{itemize}{4}
\setlist[MyIndentedList,1]{%
    label={},
    noitemsep,
    leftmargin=0.3cm,
    }
\setlist[MyIndentedList]{%
    label={},
    noitemsep,
    }

\usepackage{environ} 
 
\NewEnviron{NORMAL}{% 
    \scalebox{0.75}{$\BODY$} 
}
\NewEnviron{NORMALONE}{% 
    \scalebox{0.65}{$\BODY$} 
}

\title{The Language of Hyperelastic Materials
}

\author{
  Georgios Kissas \\
  AI Center \\
  ETH Zurich \\
  \texttt{gkissas@ai.ethz.ch} \\
   \And
  Siddhartha Mishra \\
  Seminar for Applied Mathematics \\
  ETH Zurich \\
  \texttt{siddhartha.mishra@sam.math.ethz.ch} \\
    \And
    Eleni Chatzi \\
  Department of Civil, Environmental and Geomatic Engineering  \\
  ETH Zurich \\
  \texttt{chatzi@ibk.baug.ethz.ch} \\
    \And
    Laura De Lorenzis \\
  Department of Mechanical and Process Engineering  \\
  ETH Zurich \\
  \texttt{ldelorenzis@ethz.ch} \\
}
\usepackage{kantlipsum}
\allowdisplaybreaks

\begin{document}
\maketitle

\begin{abstract} 
The automated discovery of constitutive laws forms an emerging research area, that focuses on automatically obtaining symbolic expressions describing the constitutive behavior of solid materials from experimental data. Existing symbolic/sparse regression methods rely on the availability of libraries of material models, which are typically hand-designed by a human expert using known models as reference, or deploy generative algorithms with exponential complexity which are only practicable for very simple expressions. In this paper, we propose a novel approach to constitutive law discovery relying on formal grammars as an automated and systematic tool to generate constitutive law expressions. Compliance with physics constraints is partly enforced a priori and partly empirically checked a posteriori. We deploy the approach for two tasks: i) Automatically generating a library of valid constitutive laws for hyperelastic isotropic materials; ii) Performing data-driven discovery of hyperelastic material models from displacement data affected by different noise levels. For the task of automatic library generation, we demonstrate the flexibility and efficiency of the proposed methodology in avoiding hand-crafted features and human intervention. For the data-driven discovery task, we demonstrate the accuracy, robustness and significant generalizability of the proposed methodology.

% Automatic library generation properties: Efficient, easy to extend to other types of materials, and flexible to applying constraints 
% Data-driven discovery properties: Accurate, efficient, robust to noise, discovers expressions that generalize to different tasks, meaning different loading conditions.
\end{abstract}

% keywords can be removed
Automated Model Discovery ; Data-Driven Constitutive Models ;  Formal Grammars ; Symbolic Regression ; Deep Generative Models

\section{Introduction}
\label{sec:related work}
From the mechanics of a beating heart to the haptics of a robotic arm, from the rupture of an aneurysm sack to the aeroelasticity of a spaceship, mechanical phenomena controlled by the behavior and properties of different types of materials are ubiquitous in nature and in engineering applications alike. Thus, understanding, modeling and characterizing the mechanical response of materials has been an important research focus for centuries. Accordingly, extracting so-called material models (or constitutive laws), i.e. relations between stresses, strains and often additional variables, from experimental data has been and still is a central task in solid mechanics  \cite{mahnken2004identification}. The conventional approach to solve this underlying inverse problem requires a large number of experiments and a tedious trial-and-error iterative procedure, involving at each iteration the choice of a model from the available literature (largely based on experience) and the subsequent calibration of its unknown parameters \cite{flaschel2021unsupervised}, often referred to as model updating, which can be achieved under both deterministic and stochastic schemes \cite{oliveira2021numerical,mototake2020universal}. However, the recent advances in imaging techniques are increasingly motivating a paradigm shift from point-wise measurements to full-field data acquisition \cite{pierron2021towards}. At the same time, the stunning recent successes of machine learning, and more generally data-driven techniques, have spawned a dramatic surge in the adoption of such algorithms also in the field of material modeling \cite{li2023database}; see \cite{bock2019review} for a recent review. In this work, we leverage machine learning techniques to automatize model selection and calibration by condensing these into the single task of model discovery, see the difference in model identification and discovery in \cite{flaschel2021unsupervised}. Moreover, we are interested in approaches which deliver as output an interpretable material model, i.e. one that can be represented by mathematical expressions satisfying appropriate constraints, thus moving away from the model-free paradigm \cite{kirchdoerfer2016data}. 

\begin{figure}[!ht]
    \centering
        \includegraphics[width=\textwidth]{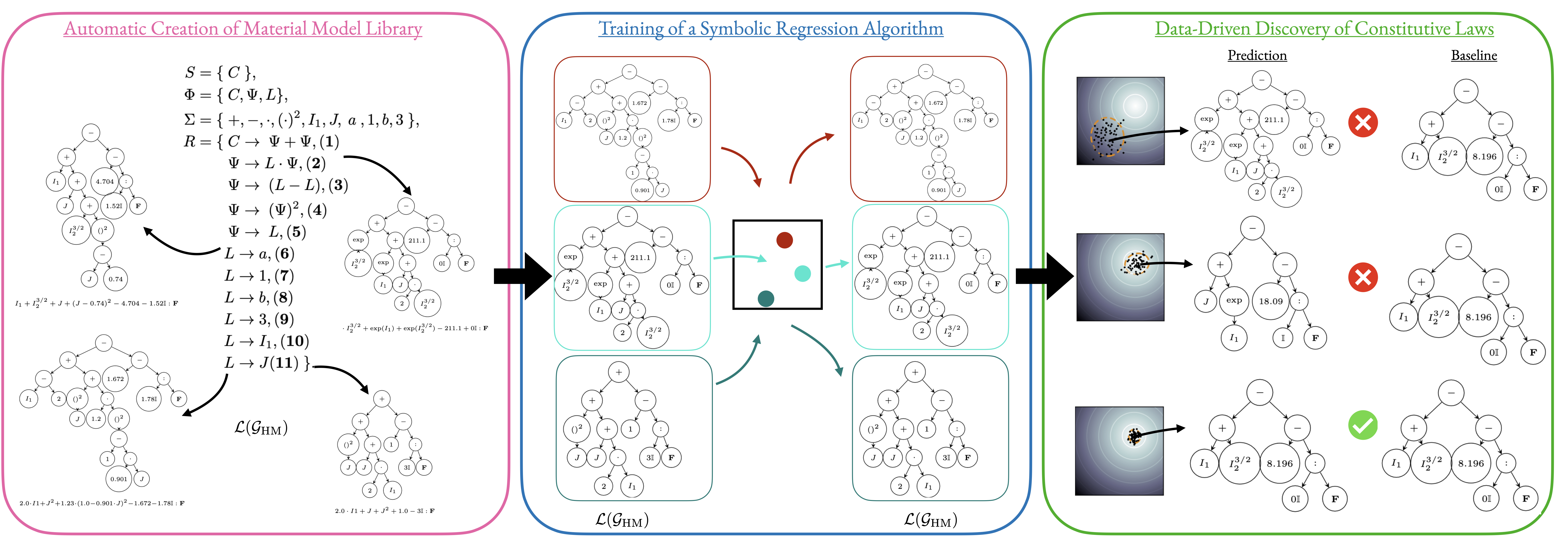}
\caption{A graphical representation of the process proposed in this manuscript. In a first step, a virtual material library is assembled offline using a formal grammar. In a second step, a symbolic regression algorithm is trained offline on the generated virtual material library. In a third step, constitutive laws are discovered from experimental data using gradient-free optimization, and  the trained model from the second step, for solving the resulting inverse problem.}
\label{fig:master_figure_lhm}
\end{figure}

% -----------\\

%\section{Symbolic Regression for Model Discovery: Review and Objectives} 

% \subsection{Literature Review on Symbolic Regression for Model Discovery} 

% In this section, w
In general, approaches that derive mathematical expressions from targeted experimental data are commonly referred to as symbolic regression methods. In what follows, 
%To decide how symbolic regression methods can be employed in constitutive law discovery, 
we review different symbolic regression methods proposed in the literature for general (i.e. not necessarily material) model  discovery and comment on their advantages and shortcomings in relation to the discovery of constitutive laws. For the sake of a unified presentation, we choose to view symbolic regression through the lens of operations on graphs. We further refer readers not familiar with the graph representation of mathematical expressions to \ref{section: appendix representation of expressions}, which reviews some basic notions that facilitate reading. From this perspective, symbolic regression algorithms can be viewed as structured ways to add and discard edges of a graph between nodes, with the nodes containing so-called primitives (i.e. constants, variables, mathematical operations, or small expressions). Thus, the construction of a symbolic regression method entails three main choices -- i) how the primitives are chosen; ii) what is the underlying graph topology; iii) how to ensure the derivation of meaningful expressions for the chosen graph topology. We hereby discuss existing classes of such algorithms in terms of these defining choices.

\begin{figure}[!ht]
    \centering
        \includegraphics[width=0.8\textwidth]{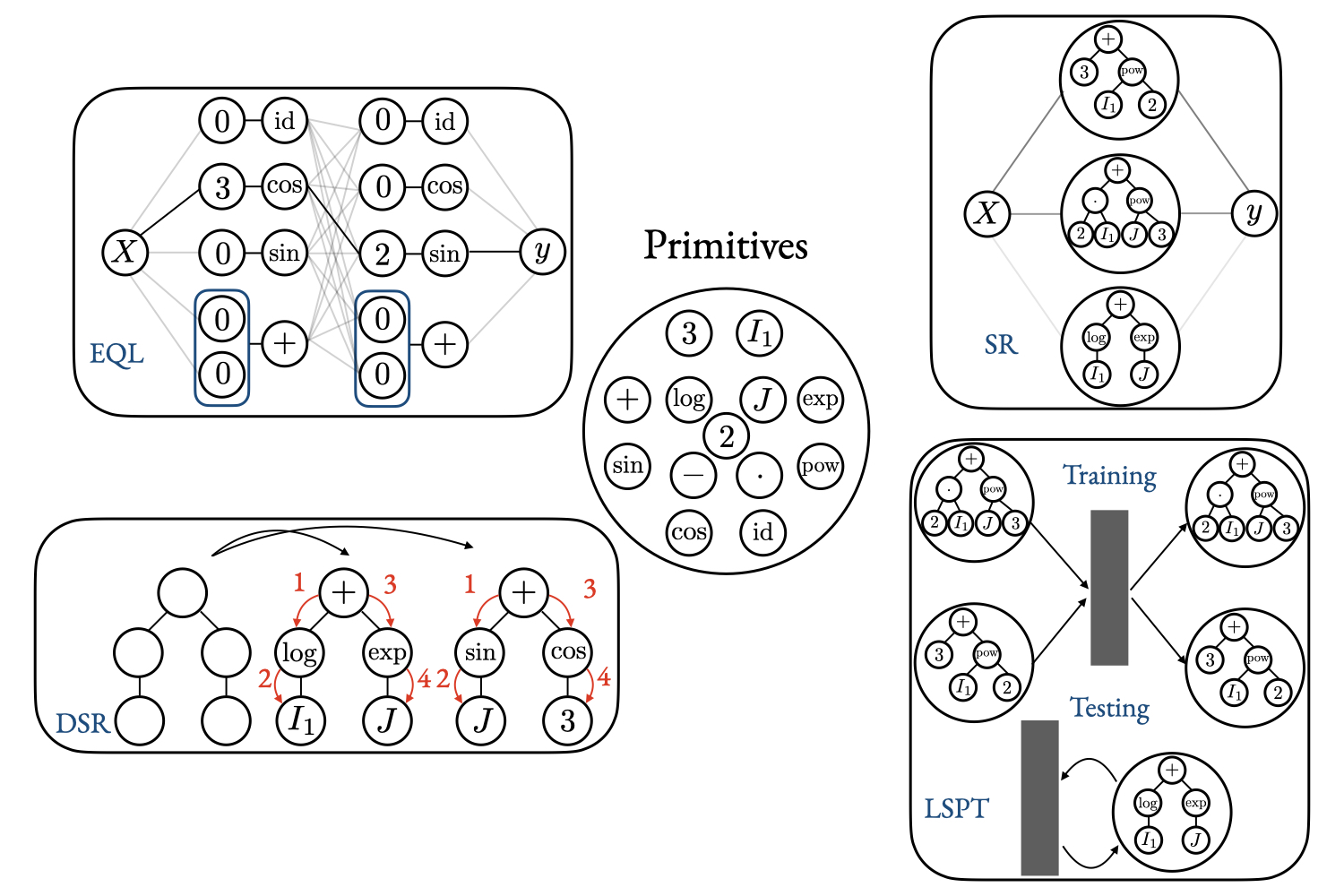}
\caption{Schematic representation of abstractions of different symbolic regression methods. For EQL and SR, we are performing supervised learning to learn a map from $X$ to $y$ parameterized by a network, while also imposing sparsity to the network. After the training is completed, we track the remaining edges to discover the form of the equation. The DSR and LSPT methods are constructed by randomly generating and evaluating trees based on a set of primitives, so they do not directly involve supervised learning.} 
\label{fig:symbolic regression methods}
\end{figure}

\emph{Equation Learner.} A simple and interpretable framework for symbolic regression is the Equation Learner (EQL) \cite{martius2016extrapolation}. To construct the EQL algorithm, the primitives include mathematical operations, such as the $ \sin$, $ \cos$, identity and $\cdot$ operations, and the input variable $X$, whereas the underlying topology is a weighted directed graph topology that represents all the functions that provide an output $y$ given an input $X$, see Figure \ref{fig:symbolic regression methods}. This is equivalent to considering a fully connected neural network whose weights are the weights of the graph and whose activation functions are the primitive operations. Multiple layers of the network are considered for approximating complex functions and different paths of the graph represent different mathematical expressions. The method is trained such that, given an input, it determines the edges of the graph that constitute the simplest path whose resulting expression evaluation is the closest to the measurements. Even though this method has been successfully employed for different tasks, it comes with some important drawbacks. First, there is no systematic way to constrain the method to provide meaningful mathematical expressions. Moreover, EQL is difficult to train for primitives such as division, exponentiation and logarithms even though ways to alleviate this issue have been proposed \cite{sahoo2018learning, costa2020fast}. Since these operations are present in many material models,  the application of this method to constitutive law discovery can be problematic. An EQL-based approach for discovery of hyperelastic constitutive laws is presented in \cite{linka2023new, tacc2023benchmarking}; the authors propose methods that consider the invariants of the Cauchy-Green tensor as inputs and activation functions consisting of unary operations. These authors do not consider constitutive laws containing operations such as division, exponentiation or logarithms. 

% DO THESE AUTHORS ENCOUTER ISSUES WITH DIVISION, EXP AND LOG?

\emph{Sparse Regression.} An alternate family of algorithms is based on Sparse Regression (SR) \cite{tibshirani1996regression, tibshirani2013lasso, zou2007degrees}. For constructing a SR algorithm, primitives are chosen to be  expressions, rather than operations, and the underlying topology is that of a weighted directed acyclic graph connecting the root node (i.e. the input $X$) with the primitives, see Figure \ref{fig:symbolic regression methods}. This is equivalent to considering a fully connected network without activation functions, trained to provide the weighted combination of edges that best fits the given measurements, while also imposing sparsity \cite{landajuela2022unified} with the purpose of obtaining a parsimonious, i.e. a simple mathematical expression. In SR, whether a generated expression is meaningful is (at least partially) determined by the primitives. For example, considering simple constitutive laws already as primitives is likely to result in a valid final constitutive law. This approach has been successfully applied in different areas of scientific discovery \cite{brunton2016discovering} and also to the discovery of constitutive laws with applications to hyperelasticity \cite{flaschel2021unsupervised,JOSHI2022115225,FLASCHEL2023105404,BODDAPATI2023105471,pierre2023principal, pierre2023discovering}, viscoelasticity \cite{marino2023automated}, plasticity \cite{flaschel2022discovering,bahmani2023discovering} and generalized standard material models \cite{flaschel2023automated}.
The main drawback of SR is that a human expert needs to hand-design the primitives; this inevitably introduces a bias in the process of discovery and restricts the search space of possible candidate expressions to those included in the starting library. 
 
\emph{Genetic Programming.} Instead of imposing sparsity to a directed acyclic graph, other methods achieve sparsity by operating directly on trees rather than graphs; in other words, sparsity naturally follows from the tree structure.
A standard approach for performing symbolic regression on trees is Genetic Programming (GP) \cite{koza1994genetic, mundhenk2021symbolic}. Here, an initial population of S-expressions is constructed randomly by combining different primitives, i.e. constants as well as unary or binary operators  \cite{schmidt2009distilling}. For each random generation, actions are performed that alter step-by-step the structure of the population of the tree, by either mutating the primitives or removing/adding sub-expressions to the tree. This approach is not equipped with any structured way to impose constraints to each step of the generation of mathematical expressions. Moreover, the complexity of the method grows very fast with the depth of the tree \cite{virgolin2022symbolic}, making it practically applicable only to small mathematical expressions. The prohibitive computational cost for potentially complex expressions makes GP not suitable for constitutive law discovery. Nevertheless, attempts in this direction have been made \cite{pal1996calibration, birky2023generalizing, bahmani2023physics, bahmani2023discovering}. In these cases, initial expression trees are evolved using mutations and crossover operations either as a standalone procedure or as a part of a more general pipeline.% DO THESE PAPERS MENTION THE COMPUTATIONAL COST ISSUE?

\emph{Deep Symbolic Regression.}  Deep Symbolic Regression (DSR) \cite{petersen2019deep} exploits Recurrent Neural Networks to predict the probability of children nodes given the parent node. To construct the DSR algorithm one needs to choose a set of primitives, namely constants as well as unary and binary operations, and a specific sequence of operations that corresponds to a top-to-bottom, left-to-right ordering of tree nodes, see Figure \ref{fig:symbolic regression methods}. The algorithm randomly chooses a root node; then, using auto-regressive sampling, it computes the probability of the next primitive on the tree conditioned on the previous one. The model is trained using a risk-seeking policy gradient to generate best-fitting expressions with high probability. In-situ constraints to zero out the probability of particular children given the parents can be considered to shrink the search space \cite{petersen2019deep, landajuela2022unified, petersen2021incorporating}. The difficulty in applying this methodology to a constitutive law discovery scenario (which, to the best of our knowledge, has not yet been attempted) is that the constraints are imposed at each step of the generation but not to the whole expression. Moreover, the generation process requires a second optimization step to calibrate constants, which may end up violating constitutive law constraints.

\emph{Large-Scale Pre-Trained Models.} The approaches we introduced so far, learn a model that 
%works for a single data point and 
needs to be re-trained as new data points are added. To address this issue in symbolic regression, Large-Scale Pre-Trained (LSPT) models have been proposed \cite{kamienny2022endtoend, biggio2021neural,vastl2022symformer}. LSPT models are trained once on a library of \emph{valid} symbolic expressions and then provide the probability of an expression given a new data point. Constructing a LSPT model entails choosing primitives as a set of constants and mathematical operations, and a tree topology with a sequence of operations that corresponds to a top-to-bottom, left-to-right ordering of the tree nodes. By sampling \emph{valid} sequences of primitives, a library of potential expressions is constructed and then evaluated for a range of the input variable $X$ and constant values. The pair consisting of the input variable $X$ and the evaluation of the expression $y$ is encoded using a transformer architecture and the encoding is then used by a different transformer architecture to provide a distribution of expressions. The whole process is trained end-to-end by supervised learning via matching the function evaluation and the tree labels. After the model is trained, a beam search is performed for pairs of $(X,y)$ to discover expressions that provide $y$ with high likelihood, see Figure \ref{fig:symbolic regression methods}. This method could potentially translate to constitutive law discovery, but it would need to be pre-trained on a large library of expressions. At present, there exists no systematic way to construct such library, nor to impose constraints during the library creation or the expression generation process. 

\emph{Grammars and Variational Autoencoders.} Formal grammars have been explored for the generation of molecules as well as of mathematical expressions using sequence-to-sequence  \cite{kusner2017grammar, dai2018syntaxdirected} and recursive  \cite{paassen2022recursive} learning. Discovery processes of both new molecules and mathematical expressions can be succeeded via use of a deep generative model, more specifically using a variational autoencoder (VAE). In Kusner et. al. \cite{kusner2017grammar}, the authors propose a method that first extracts from a tree the sequence of grammar rules that generates it and then encodes the sequence using a low-dimensional vector. In \cite{paassen2022recursive}, the authors directly encode a tree using a recursive procedure that encodes the rule that generated the parent node given the children. In both cases, a decoding algorithm based on a grammar is considered to guarantee syntactic validity of the expressions. After the VAE model is trained, a gradient-free optimization procedure, such as Bayesian optimization or an evolutionary strategy, can be employed for exploring the low-dimensional space of the VAE to find the expression that best fits a new set of measurements. Methods based on grammar possess low complexity,  constrain the tree generation using rules, and generalize. To the best of our knowledge, they have never been used for constitutive law discovery.

The above survey on available tools for symbolic regression clearly suggests the following characteristics that would be desirable for symbolic regression algorithms applied to data-driven constitutive law discovery:

\begin{itemize}
    \item Accommodation of primitive operations of significant complexity (e.g. including $\exp$, $\log$ or $/$), without exhibiting training instabilities.
    \item Alleviation of human bias via flexible and automated generation of new constitutive expressions.
    \item Low complexity, meaning that the complexity of the discovery process should not increase exponentially with the length of the expression.
    \item Possibility to incorporate constraints on both the final expression and the steps of the expression generating process.
    \item Generalization of the symbolic regression algorithm, i.e. potential to provide expressions that are valid for new and unseen data points without re-training.
\end{itemize}

In this paper, we propose a symbolic regression pipeline for constitutive law discovery that satisfies all the above-listed requirements. Due to the advantages of formal grammars mentioned earlier (low complexity, ability to systematically embed constraints stemming from domain knowledge, and generalization capability) we explore grammar-based symbolic regression and develop formal grammars which are specifically designed for constitutive laws. In this first investigation, we focus on hyperelasticity, whereby the material behavior is fully encoded by the elastic strain energy density function for which we seek a parsimonious mathematical expression.

The proposed pipeline is composed by three steps: a library construction step, a pre-training step, and a data-driven discovery step, as illustrated in Figure \ref{fig:master_figure_lhm}. In the first step, we define a formal grammar that generates mathematical expressions in the form of trees. We ensure that the expressions derived by the grammar are valid constitutive laws (i.e., in our case, valid elastic strain energy density functions) partly by applying constraints during the generation process and partly by performing empirical acceptance checks on the final expressions. The grammar is then used to perform an off-line library generation of constitutive laws. Note that this library is a useful result per se, as it may be deployed in place of a hand-crafted library in the context of constitutive model discovery based on sparse regression \cite{flaschel2021unsupervised}. In the second step, this library is used to train off-line a symbolic regression algorithm that takes advantage of the tree structure of the mathematical expressions and of the grammar of the constitutive laws. The pre-trained model is trained to encode a tree to a low-dimensional latent vector representations and decode this low-dimensional vector representation to a tree, creating a low-dimensional manifold of tree representations. In the third step, we perform a gradient-free optimization approach \cite{hansen2003reducing, hansen2001completely} to search the low-dimensional manifold for the vectorial representation of the model that best fits the given measurements. Therefore, the expensive step of training the deep learning model needs to be executed once and then we perform the discovery without re-training.

The remainder of the manuscript is structured as follows. In Section \ref{sec:formal grammars} we focus on the notion of formal grammars and describe the Context-Free and Regular Tree Grammar classes, their properties and semantics.
Section \ref{sec:the language of hyperelastic materials} is devoted to the construction of a formal language, the Language of Hyperelastic Materials, where the constitutive law constraints are either included in the grammar or enforced on the derived expressions. 
%In Section \ref{sec:semantics for the language of materials}, we introduce the requirements that the mathematical expressions derived from the grammar need to satisfy such that they are valid elastic strain energy density functions. In Section \ref{sec:enforcement of constraints}, we discuss how these requirements are enforced to the expressions derived from the grammar. In Section \ref{sec:generating the language of hyperelastic materials}, we discuss the construction of the Language of the Hyperelastic materials and in Section \ref{sec:deriving parsimonious expressions} constraining the grammar to derive parsimonious expressions. 
In this section we also discuss how the Language of Hyperelastic Materials can deployed for the automatic construction of a library of hyperelastic constitutive laws, and present examples of generation of such a library. In Section \ref{sec:data driven constitutive law discovery}, we introduce a symbolic regression method combining VAEs and a Regular Tree Grammar defined for constitutive laws, that fulfils all the requirements of the earlier list. Finally, we propose a pipeline for discovering constitutive laws from available data and illustrate an example using artificially generated full-field displacement data. Conclusions and an outlook close the paper in Section \ref{sec:discussion}. 
%we present a summary of the manuscript together with a discussion on the limitations of the proposed approach and the directions for future research.

 \label{sec:formal grammars}
%For constitutive models, we aim at interpretable expressions, which motivates our focus on grammar structures throughout this manuscript. 
In this section, we present two different types of formal grammars, namely Context-Free Grammars and Regular Tree Grammars, along with a simple example to showcase their use. To enable a high-level intuition of how grammar works, we provide a simplified schematic representation of a tree created using the proposed grammar in Figure \ref{fig:tree vs grammar tree}. The node labels are variables, e.g. $L, \Psi, S$, from which one, in this case $S$, is assigned as the tree root. During tree generation, these variables are substituted with different primitives (i.e. constants, e.g. $2$, $3$, variables, e.g. $I_1$, $J$, or operators, e.g. $+, -, \cdot$), using predefined substitution rules, e.g. $r_1, r_2$. The number of arguments that the primitives take (e.g. $+$ takes two arguments while $J$ takes no arguments) determine the number of their children and the tree connectivity. 

\subsection{Context-Free Grammars} 
\label{CFG} 
Roughly described, a Context-Free Grammar (CFG) is a systematic way of generating tree structures that represent syntactically and semantically meaningful expressions using a set of rules. 
CFGs are defined as a tuple $\curlyG = \{ \Phi, \Sigma, R, S \}$, where $\Phi$ is a set of non-terminal symbols, $\Sigma$ is an alphabet of terminal symbols, $R$ is a set of production rules, and $S$ denotes a special non-terminal symbol called the starting symbol. Let us explain the meaning of these terms.

\begin{figure}[!ht]
    \centering
        \includegraphics[width=0.6\textwidth]{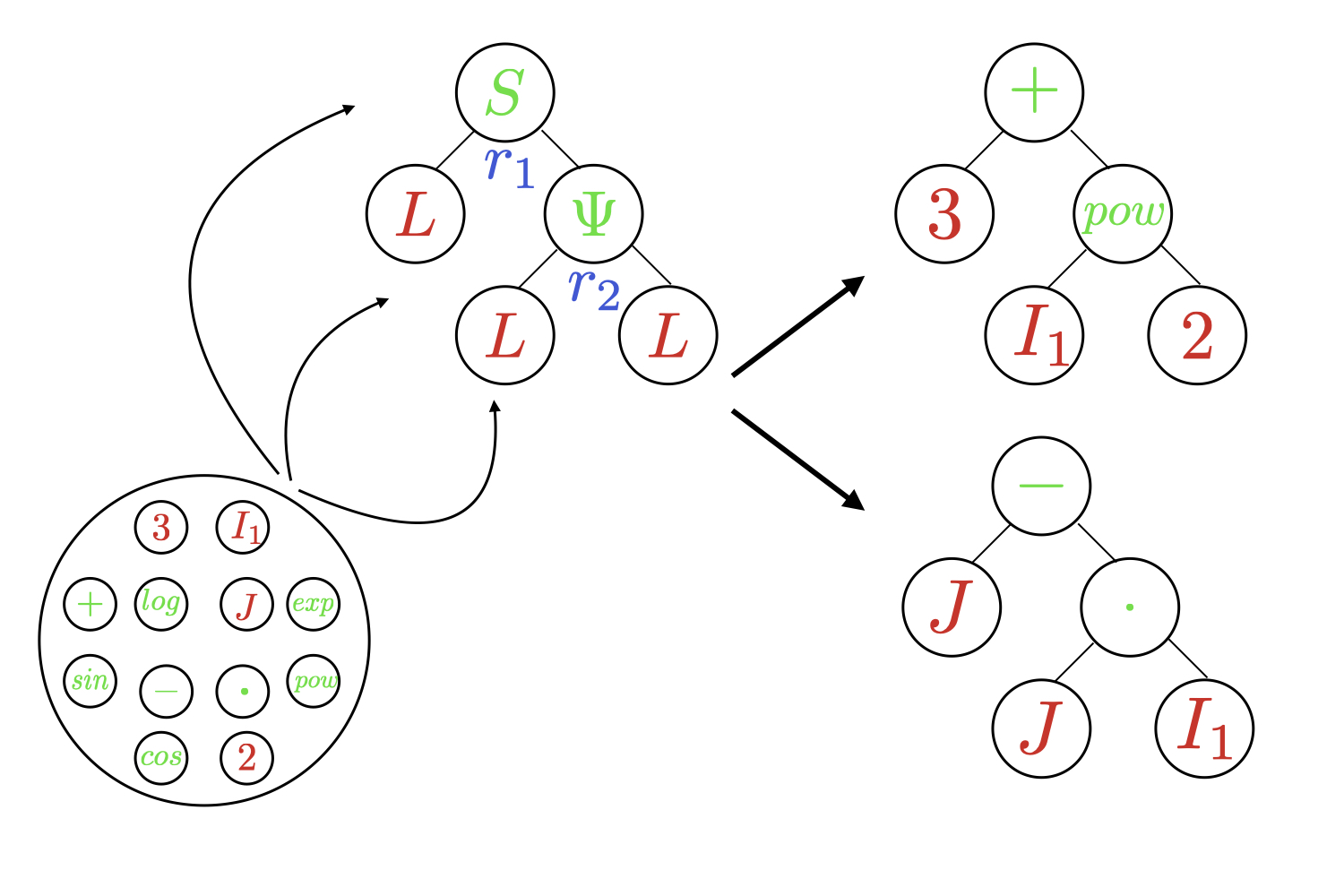}
\caption{Schematic explanation of a tree derivation using grammar. The non-terminals are color coded based on the grammar rules that substitute them. For example, $L$ is substituted by $3$, $I_1$, or $J$ and is re-written by the grammar rules  $J \rightarrow 3 | I_1 | J$.  On the other hand, $\Psi$ and $S$ are re-written by $+$, $-$, $\cdot$, or $pow$,  and correspond to  $S \rightarrow +(L,\Psi)| -(L,\Psi)$, and $\Psi \rightarrow pow(L, L) | \cdot(L, L)$, respectively. The black arrows show two  possible tree derivation generated by these substitutions.}
\label{fig:tree vs grammar tree}
\end{figure}

\begin{itemize}
\item \emph{Non-terminal} symbols are the variable node labels (i.e. $L, \Psi, S$ in Figure \ref{fig:tree vs grammar tree}). They are syntactic variables that cannot appear in an expression as standalone entities; to generate a sentence, or, in our context, a mathematical expression, they are substituted by terminal symbols through production rules. 
\item \emph{Terminal symbols} are the primitives. They are the building blocks that compose sentences or, in our context, mathematical expressions. In our case, the terminal symbols can be  defined as constants, variables, operations (such as in Figure \ref{fig:tree vs grammar tree}), or sub-expressions.
\item \emph{Production rules} are user-defined rules to substitute non-terminals with other non-terminal or terminal symbols. Each substitution rule consists of a left-hand side that contains a non-terminal symbol, and a right-hand side that contains a mixture of terminals and non-terminals to be substituted to the left-hand side. E.g., the rule $\Psi \rightarrow sL$, with $\Psi,L \in \Phi$ and $s \in \Sigma$, substitutes $\Psi$ with $sL$ in a sentence. If a non-terminal symbol appears on the left-hand side of more than one production rule, it can be replaced by any of the right-hand sides of these rules. By recursive substitution of the non-terminal symbols using the production rules, initiating from the starting symbol $S$, sentences that contain only terminal symbols are finally derived.
\end{itemize}

%\paragraph{Context-Free Languages.} 

Note that because the production rules of CFGs contain only one non-terminal on the left-hand side, the sentences of CFGs can be expressed as trees, called derivation trees, as illustrated in Figure \ref{fig:tree vs grammar tree}. Each derivation tree corresponds to a terminal sentence, or, in our context, to a mathematical expression. A language $\curlyL(\curlyG)$ is defined as the set of all possible terminal sentences that can be derived by applying the production rules of the grammar starting from $S$, or all possible ways that the nodes of a derivation tree can be connected starting from $S$.
 
%\paragraph{A Simple Example:}

We now illustrate these concepts with a simple example. Consider the grammar  $\curlyG_{\text{NH}} = \{ \Phi, \Sigma, R, S \}$, where:
\begin{equation*}
\NORMAL{
    \begin{split}
    \allowdisplaybreaks
        S = \{& \ C \  \}, \\
        \Phi  = \{& \ C,  \Psi, L \}, \\
        \Sigma = \{& \ +, - ,  \cdot, (\cdot)^2,  I_1, J, \  a \ , 1, b,  3 \  \}, \\
        R  = \{ &  \ C \rightarrow  \ \Psi + \Psi, \mathbf{(1)} \\
               &  \ \Psi \rightarrow L \cdot \Psi, \mathbf{(2)} \\
               &  \ \Psi \rightarrow \  (L - L), \mathbf{(3)} \\
               &  \ \Psi \rightarrow  \ (\Psi)^2, \mathbf{(4)} \\
               &  \ \Psi \rightarrow  \ L, \mathbf{(5)} \\
                 &  L \rightarrow a, \mathbf{(6)} \\ 
                 & L \rightarrow  1,  \mathbf{(7)}\\ 
                 & L \rightarrow b, \mathbf{(8)} \\
                 & L \rightarrow 3, \mathbf{(9)} \\
                 & L \rightarrow I_1, \mathbf{(10)} \\
                 & L \rightarrow J \mathbf{(11)} \ \}.
    \end{split}
    }
\end{equation*}
 Here, $C, \Psi$, and  $L$ denote the non-terminals, with $C$ as the starting symbol, and $\Psi$ and $L$ are used for recursive substitution of operations and literals (i.e. constants and variables), respectively. More specifically, $\Psi$ describes both unary and binary operations between terminals and non-terminals, whereas $L$ re-writes the integers $1,3$, the real constants $a,c$ and the variables $I_1, J$. The numbers next to the production rules are used for their annotation. Note that rule $\mathbf{(3)}$ is written as $(L-L)$, but this does not mean that this rule always provides zero values. Each non-terminal $L$ is treated separately, which means that the two $L$'s in rule $\mathbf{(3)}$ do not always have the same value. 

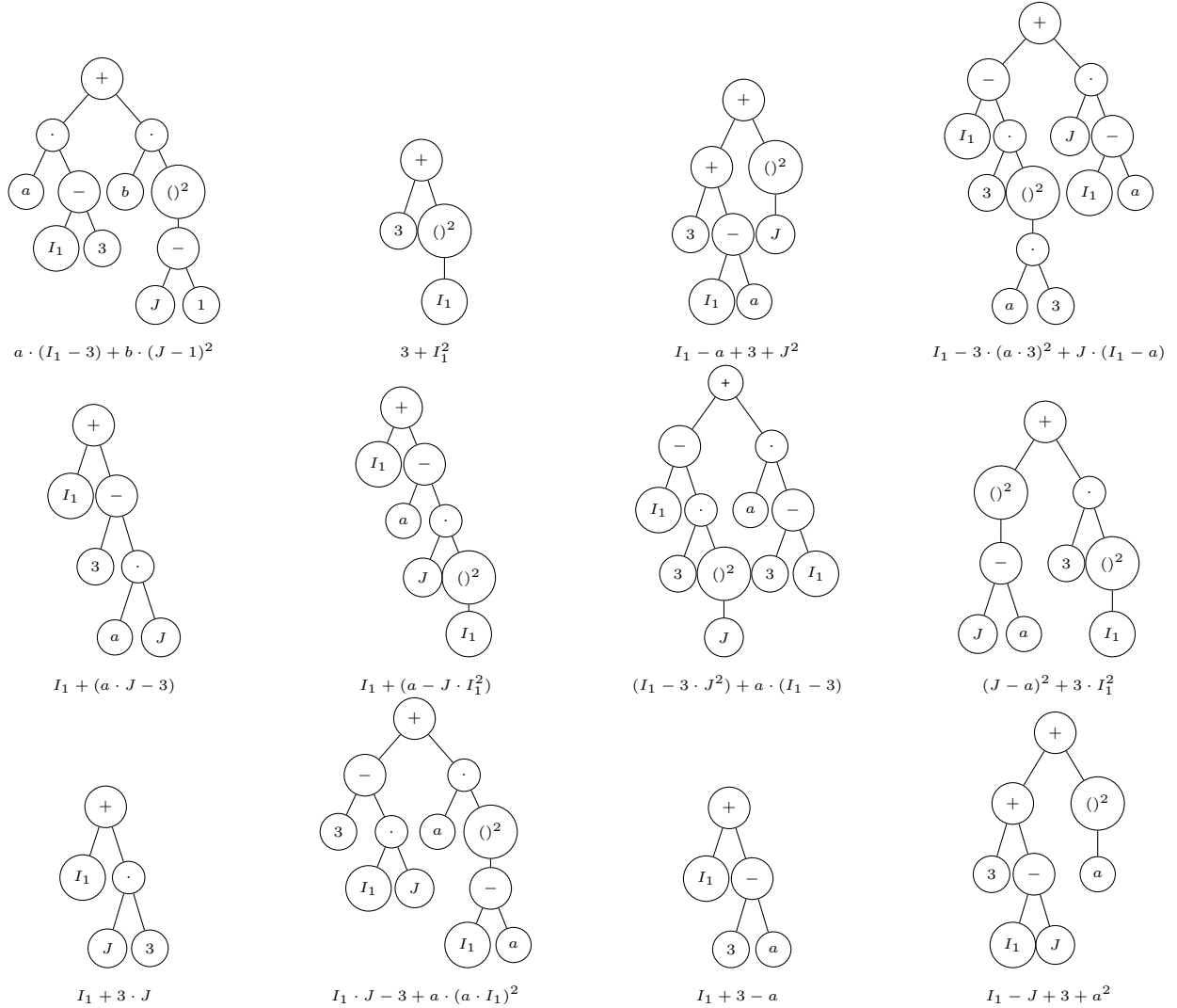
\begin{figure}
  \begin{subfigure}{0.2\textwidth}
    \centering
    \begin{tikzpicture}[
      level distance=0.8cm,
      level 1/.style={sibling distance=1.4cm},
      level 2/.style={sibling distance=0.75cm},
      level 3/.style={sibling distance=0.65cm},
      every node/.style={draw, circle, minimum size=0.05em, font=\tiny},
      edge from parent/.style={draw, -, >=stealth},
    ]
    \node {$+$}
      child { node {$\cdot$}
        child { node {$a$} }
        child { node {$-$}
          child { node {$I_1$} }
          child { node {$3$} }
        }
      }
      child { node {$\cdot$}
        child { node {$b$} }
        child { node {$()^2$}
          child { node {$-$}
            child { node {$J$} }
            child { node {$1$} }
          }
        }      };
    \end{tikzpicture}
    \caption*{ \tiny $a \cdot (I_1 -3 ) + b \cdot (J -1)^2$}
  \end{subfigure}
  \hfill
  \begin{subfigure}{0.2\textwidth}
    \centering
    \begin{tikzpicture}[
      level distance=1cm,
      level 1/.style={sibling distance=0.65cm},
      level 2/.style={sibling distance=0.6cm},
      level 3/.style={sibling distance=0.3cm},
      every node/.style={draw, circle, minimum size=0.05em, font=\tiny},
      edge from parent/.style={draw, -, >=stealth},
    ]
      \node {$+$}
        child {node {$3$}}
        child {node {$()^2$}
          child {node {$I_1$}}
      };
    \end{tikzpicture}
    \caption*{ \tiny $3 + I_1^2$}
  \end{subfigure}
  \hfill
  \begin{subfigure}{0.2\textwidth}
    \centering
    \begin{tikzpicture}[
      level distance=0.95cm,
      level 1/.style={sibling distance=0.9cm},
      level 2/.style={sibling distance=0.6cm},
      level 3/.style={sibling distance=0.6cm},
      every node/.style={draw, circle, minimum size=0.05em, font=\tiny},
      edge from parent/.style={draw, -, >=stealth},
    ]
    \node {$+$}
      child { node {$+$}
        child { node {$3$} }
        child { node {$-$}
          child { node {$I_1$} }
          child { node {$a$} }
        }
      }
      child { node {$()^2$}
        child { node {$J$} }
      };
    \end{tikzpicture}
    \caption*{\tiny $I_1 - a + 3 + J^2$}
  \end{subfigure}
  \hfill
  \begin{subfigure}{0.2\textwidth}
    \centering
    \begin{tikzpicture}[
      level distance=0.8cm,
      level 1/.style={sibling distance=1.45cm},
      level 2/.style={sibling distance=0.6cm},
      level 3/.style={sibling distance=0.65cm},
      every node/.style={draw, circle, minimum size=0.05em, font=\tiny},
      edge from parent/.style={draw, -, >=stealth},
    ]
    \node {$+$}
      child { node {$-$}
        child { node {$I_1$} }
        child { node {$\cdot$}
          child { node {$3$} }
          child { node {$()^2$}
            child { node {$ \cdot $}
              child { node {$a$} }
              child { node {$3$} }
            }
          }
        }
      }
      child { node {$\cdot$}
        child { node {$J$} }
        child { node {$-$}
          child { node {$I_1$} }
          child { node {$a$} }
        }
      };
    \end{tikzpicture}
    \caption*{\tiny $I_1 - 3 \cdot (a \cdot 3)^2 + J \cdot (I_1 - a)$}
  \end{subfigure}
  \hfill
  \begin{subfigure}{0.2\textwidth}
    \centering
    \begin{tikzpicture}[
      level distance=1cm,
      level 1/.style={sibling distance=0.65cm},
      level 2/.style={sibling distance=0.6cm},
      level 3/.style={sibling distance=0.65cm},
      every node/.style={draw, circle, minimum size=0.05em, font=\tiny},
      edge from parent/.style={draw, -, >=stealth},
    ]
      \node {$+$}
        child {node {$I_1$}}
        child {node {$-$}
          child {node {$3$}}
          child {node {$\cdot$}
            child {node {$a$}}
            child {node {$J$}}
      }};
    \end{tikzpicture}
    \caption*{\tiny $I_1 + (a \cdot J - 3)$}
  \end{subfigure}
  \hfill
  \begin{subfigure}{0.2\textwidth}
    \centering
    \begin{tikzpicture}[
      level distance=0.8cm,
      level 1/.style={sibling distance=0.65cm},
      level 2/.style={sibling distance=0.6cm},
      level 3/.style={sibling distance=0.65cm},
      every node/.style={draw, circle, minimum size=0.05em, font=\tiny},
      edge from parent/.style={draw, -, >=stealth},
    ]
      \node {$+$}
        child {node {$I_1$}}
        child {node {$-$}
          child {node {$a$}}
          child {node {$\cdot$}
            child {node {$J$}}
            child {node {$()^2$}
              child {node {$I_1$}}
      }}};
    \end{tikzpicture}
    \caption*{\tiny $I_1 + (a - J \cdot I_1^2)$}
  \end{subfigure}
  \hfill
  \begin{subfigure}{0.2\textwidth}
    \centering
    \begin{tikzpicture}[
      level distance=0.9cm,
      level 1/.style={sibling distance=1.3cm},
      level 2/.style={sibling distance=0.6cm},
      level 3/.style={sibling distance=0.65cm},
      every node/.style={draw, circle, minimum size=0.05em, font=\tiny},
      edge from parent/.style={draw, -, >=stealth},
    ]
    \node {+}
      child { node {$-$}
        child { node {$I_1$} }
        child { node {$\cdot$}
          child { node {$3$} }
          child { node {$()^2$}
            child { node {$J$} }
          }
        }
      }
      child { node {$\cdot$}
        child { node {$a$} }
        child { node {$-$}
          child { node {$3$} }
          child { node {$I_1$} }
        }
      };
    \end{tikzpicture}
    \caption*{\tiny $(I_1 - 3 \cdot J^2) + a \cdot (I_1 - 3)$}
  \end{subfigure}
  \hfill
  \begin{subfigure}{0.2\textwidth}
    \centering
    \begin{tikzpicture}[
      level distance=1cm,
      level 1/.style={sibling distance=1.25cm},
      level 2/.style={sibling distance=0.65cm},
      level 3/.style={sibling distance=0.65cm},
      every node/.style={draw, circle, minimum size=0.05em, font=\tiny},
      edge from parent/.style={draw, -, >=stealth},
    ]
      \node {$+$}
              child { node {$()^2$}
        child { node {$-$}
          child { node {$J$} }
          child { node {$a$} }
        }}
        child { node {$\cdot$}
          child { node {$3$} }
          child { node {$()^2$}
            child { node {$I_1$} }
          }
      };
    \end{tikzpicture}
    \caption*{\tiny $(J - a)^2 + 3 \cdot I_1^2$}
  \end{subfigure}
  \hfill
  \begin{subfigure}{0.2\textwidth}
    \centering
    \begin{tikzpicture}[
      level distance=1cm,
      level 1/.style={sibling distance=0.65cm},
      level 2/.style={sibling distance=0.6cm},
      level 3/.style={sibling distance=0.65cm},
      every node/.style={draw, circle, minimum size=0.05em, font=\tiny},
      edge from parent/.style={draw, -, >=stealth},
    ]
      \node {$+$}
        child { node {$I_1$} }
        child { node {$\cdot$}
          child { node {$J$} }
          child { node {$3$} }
      };
    \end{tikzpicture}
    \caption*{\tiny $I_1 + 3 \cdot J$}
  \end{subfigure}
  \hfill
  \begin{subfigure}{0.2\textwidth}
    \centering
    \begin{tikzpicture}[
      level distance=0.8cm,
      level 1/.style={sibling distance=1.4cm},
      level 2/.style={sibling distance=0.75cm},
      level 3/.style={sibling distance=0.65cm},
      every node/.style={draw, circle, minimum size=0.05em, font=\tiny},
      edge from parent/.style={draw, -, >=stealth},
    ]
    \node {$+$}
      child { node {$-$}
        child { node {$3$} }
        child { node {$\cdot$}
          child { node {$I_1$} }
          child { node {$J$} }
        }
      }
      child { node {$\cdot$}
        child { node {$a$} }
        child { node {$()^2$}
          child { node {$-$}
            child { node {$I_1$} }
            child { node {$a$} }
          }
        }
      };
    \end{tikzpicture}
    \caption*{\tiny $I_1 \cdot J - 3 + a \cdot (a \cdot I_1)^2$}
  \end{subfigure}
  \hfill
  \begin{subfigure}{0.2\textwidth}
    \centering
    \begin{tikzpicture}[
      level distance=1cm,
      level 1/.style={sibling distance=0.65cm},
      level 2/.style={sibling distance=0.6cm},
      level 3/.style={sibling distance=0.3cm},
      every node/.style={draw, circle, minimum size=0.05em, font=\tiny},
      edge from parent/.style={draw, -, >=stealth},
    ]
    \node {$+$}
      child {node {$I_1$}}
      child {node {$-$}
        child {node {$3$}}
        child {node {$a$}}
      };
    \end{tikzpicture}
    \caption*{\tiny $I_1 + 3 - a$}
  \end{subfigure}
  \hfill
  \begin{subfigure}{0.2\textwidth}
    \centering
    \begin{tikzpicture}[
      level distance=1cm,
      level 1/.style={sibling distance=1.2cm},
      level 2/.style={sibling distance=0.6cm},
      level 3/.style={sibling distance=0.6cm},
      every node/.style={draw, circle, minimum size=0.05em, font=\tiny},
      edge from parent/.style={draw, -, >=stealth},
    ]
    \node {$+$}
      child { node {$+$}
        child { node {$3$} }
        child { node {$-$}
          child { node {$I_1$} }
          child { node {$J$} }
        }
      }
      child { node {$()^2$}
        child { node {$a$} }
      };
    \end{tikzpicture}
    \caption*{\tiny $I_1 - J + 3 + a^2$}
  \end{subfigure}
  \caption{Binary trees derived from the context-free grammar $\curlyG_{NH}$; they are all part of the language $\curlyL(\curlyG_{NH})$.}
  \label{fig:binary tree examples}
\end{figure}

If we identify $I_1, J$ with the first invariant of the right Cauchy-Green deformation tensor and the third invariant of the deformation gradient in finite deformation kinematics, we can produce the non-dimensional elastic strain energy density function of a simple Neo-Hookean model $W = a   \cdot (I_1 - 3) +  b \cdot (J-1)^2$ by performing recursive substitutions of non-terminal symbols beginning from the starting symbol $C$, as follows:
\begin{equation*}
\NORMAL{
\begin{split}
    C & \xrightarrow{\mathbf{(1)}} \Psi + \Psi, \\
         & \xrightarrow{\mathbf{(2)}} L \cdot \Psi +L \cdot \Psi, \\
         & \xrightarrow{\mathbf{(3)}} L \cdot (L - L) + L \cdot (L - L), \\
         & \xrightarrow{\mathbf{(6)}} a \cdot (L - L) + L \cdot (L - L), \\
         & \xrightarrow{\mathbf{(10)}} a \cdot (I_1 - L) + L \cdot (L - L), \\
         & \xrightarrow{\mathbf{(9)}} a \cdot (I_1 - 3) + L \cdot (L - L), \\
         & \xrightarrow{\mathbf{(8)}} a \cdot (I_1 - 3) + b \cdot (L - L), \\
         & \xrightarrow{\mathbf{(11)}} a \cdot (I_1 - 3) + b \cdot (J - L), \\
         & \xrightarrow{\mathbf{(7)}} a \cdot (I_1 - 3) + b \cdot (J - 1). \\
\end{split}
}
\end{equation*}
The grammar $\curlyG_{\text{NH}}$ can not only produce this specific Neo-Hookean model, but also generate any other expression that can be derived as a combination of the introduced production rules $ \{ \mathbf{(1)}, ... \mathbf{(11)} \}$. Examples derived from $\curlyG_{NH}$ are shown in Figure \ref{fig:binary tree examples}. All the possible terminal strings derived by recursively substituting the non-terminal symbols $C, \Psi, L \in \Phi$ (starting from $C$) using the production rules  constitute the context-free language $\curlyL(\curlyG_{\text{NH}})$.

\subsection{Regular Tree Grammars} 
%\paragraph{Regular Tree Grammars:}
Another type of grammar suitable for our application is a Regular Tree Grammar (RTG), defined as the tuple $\curlyG = \{ \Phi, \tilde{\Sigma}, R, S \}$, where $\tilde{\Sigma}$ is now a ranked alphabet. This is defined as an alphabet augmented by specifying the arity of each primitive, i.e. the number of arguments each primitive takes. To explain the difference between CFGs and RTGs, we now consider a Regular-Tree version of $\curlyG_{\text{NH}}$, which we denote as $\hat{\curlyG}_{\text{NH}}$,  where:
\begin{equation*}
\centering
\NORMAL{
    \begin{split}
        S = \{& \ C \  \}, \\
        \Phi  = \{& \ C,  \Psi, L \}, \\
        \Tilde{\Sigma} = \{& \ + : 2, - : 2 ,  \cdot :2, ()^2:1,  I_1:0, J:0, \  a:0 \ , 1:0, b:0,  3:0 \  \}, \\
        R  = \{ &  \ C \rightarrow  \ + (\Psi, \Psi), \mathbf{(1)} \\
               &  \ \Psi \rightarrow \cdot ( L, \Psi), \mathbf{(2)} \\
               &  \ \Psi \rightarrow \  - (L , L), \mathbf{(3)} \\
               &  \ \Psi \rightarrow  \ ()^2 (\Psi), \mathbf{(4)} \\
               &  \ \Psi \rightarrow  \ L, \mathbf{(5)} \\
                 &  L \rightarrow a(), \mathbf{(6)} \\ 
                 & L \rightarrow  1(),  \mathbf{(7)}\\ 
                 & L \rightarrow b(), \mathbf{(8)} \\
                 & L \rightarrow 3(), \mathbf{(9)} \\
                 & L \rightarrow I_1(), \mathbf{(10)} \\
                 & L \rightarrow J() \mathbf{(11)} \ \}.
    \end{split}
    }
\end{equation*}
% \begin{figure*}
%     \begin{subfigure}[!t]{0.4\textwidth}
%     \vspace{1.0cm}
%     \centering
%         \includegraphics[width=0.8\textwidth]{cfg_NH.png}
%   \end{subfigure}
%     \begin{subfigure}[!t]{0.4\textwidth}
%     \vspace{1.0cm}
%     \centering
%         \includegraphics[width=1.5\textwidth]{rtg_NH.png}
%   \end{subfigure}
% \end{figure*}
There are two immediate differences between  $\hat{\curlyG}_{\text{NH}}$ and $\curlyG_{\text{NH}}$. First, we provide the $\text{arity}$ of each primitive in the alphabet. We consider operations with $\text{arity}=2$, such as $+$, $-$, with $\text{arity}=1$ such as $()^2$, and with $\text{arity}=0$, i.e. constants and variables. Operations such as $()^2$ can also be defined to have $\text{arity}=2$ in the general case where the exponent is also an argument. Second, we write all rules in Polish notation, see Section \ref{section: appendix representation of expressions}, to denote that we are working with trees and not strings. As for CFGs, the sentences (or mathematical expressions) of RTGs are expressed as derivation trees. Also, as for the context-free case, the Regular-Tree Language $\curlyL(\hat{\curlyG})$ is defined as the set of all the trees that can be generated using the RTG $\hat{\curlyG}$. 

\subsection{Characteristics and Operations of Tree Grammars}
\label{sec: Properties of tree grammars}
%\paragraph{From Trees to Data Structures Useful for Symbolic Regression:}
In general cases, trees are fully described by their node labels and connectivity. For the special case of a tree derived from a RTG, the tree can be fully characterized by the list of rules used for its generation, as we state more precisely in the following. % The reason is that for unambiguous grammars each tree is uniquely derived by only one set of rules, see Appendix \ref{section: appendix regular tree grammars}.
In the previous section we have shown that a unique Neo-Hookean model can be derived by a set of rules. For the case of a RTG, the opposite also holds, meaning that all the information required to describe the specific Neo-Hookean model is uniquely represented by a set of rules. More precisely, it can be proven mathematically \cite{paassen2022recursive} that this property holds for unambiguous RTGs, i.e. RTGs in which each rule has a unique right-hand side. One can also prove that any RTG can be made unambiguous and that algorithms handling RTGs possess linear complexity in both computations and memory \cite{paassen2022recursive}, see \ref{section: appendix regular tree grammars} for details on the properties of RTGs. 

In view of the above, there are two meaningful actions to be performed on a tree derived from a RTG: i. given the tree, extract its characterization (i.e. the list of rules of the corresponding grammar used for its generation), which is denoted as \emph{parsing}; and ii. given a characterization, i.e. a list of rules, generate the tree that it corresponds to, which is called \emph{generation}. In \ref{section: appendix regular tree grammars} we provide details on tree parsing and generation algorithms. Note that knowledge on characterization is key for performing tree-based symbolic regression, as characterization (i.e. the list of rules) is the dependent variable that the symbolic regression algorithm learns to predict. 

Importantly, the bijective relation between trees and the sets of grammar rules used for their generation only holds for (unambiguous) RTGs, and not for CFGs \cite{hoogeboom2015undecidable}. On the other hand, CFGs are known to be more expressive than RTGs \cite{linz2022introduction}. Looking ahead to the two main outcomes of this work, namely, automated generation of a material model library and automated model discovery based on data (Sections \ref{sec:the language of hyperelastic materials} and \ref{sec:data driven constitutive law discovery} respectively), it is quite clear that CFGs are more suitable for the first task, in which expressivity is important and the unique definition of the set of grammar rules corresponding to a given model is not essential, whereas RTGs are more suitable for the second task. For this reason, we will make use of CFGs in Section \ref{sec:the language of hyperelastic materials} and of RTGs in Section \ref{sec:data driven constitutive law discovery}.  

%\paragraph{Semantics}
A grammar describes the rules that create sentences and constitute the syntax of natural languages. However, the syntax alone is not enough to define a language that is useful for our purposes; the additional needed ingredient is semantics, i.e. the meaning of the sentences that the language produces.
While it is difficult to define semantics in natural languages, in physics this is more straightforward. For our purposes, \emph{we consider as semantically valid the expressions that satisfy the constraints of the constitutive law relations}. 

\section{The Language of Hyperelastic Materials}
\label{sec:the language of hyperelastic materials}
In this section, our objective is to generate the Language of Hyperelastic Materials. We first review the constraints that need to be satisfied for a mathematical expression to represent a valid elastic strain energy density function. We then discuss ways of enforcing these constraints. This is done partially during the construction of the grammar and partially through checks on the final derived expressions. Subsequently, we propose a general language for hyperelastic materials, i.e. a language whose expressions are valid elastic strain energy density functions (i.e. functions which correspond to valid hyperelastic constitutive laws). Throughout this paper, we assume to have introduced a suitable non-dimensionalization such that we only deal with non-dimensional strain energy density functions.

% We then ...

We consider finite-deformation kinematics and denote with $\u \in \mathbb{R}^3$ the displacement field, with $\F = \mathbf{I} + \rm{Grad} \,\u$ the deformation gradient (where $\rm{Grad}$ is the gradient operator with respect to the reference coordinates), and with $\C = \F^T \F$
%, where $\mathcal{GL}^+(3)$ the set of invertible second order tensors with positive determinant,  of a displacement field . 
the right Cauchy-Green deformation tensor. The principal invariants of $\C$ are defined as:

\begin{equation*}
    J = \rm{det} \,\F = (\rm{det} \,\C)^{1/2}, \quad I_1 = \rm{tr}\,\C, \quad I_2 = \frac{1}{2} (\rm{tr}^2\,\C - tr\,\C^2),
\end{equation*}

where $\rm{tr}$ denotes the trace operator.
For nearly incompressible materials, it is customary to decompose the deformation gradient $\F$ into a volume-preserving (or isochoric) part, $\F^{iso}$, and a volume-altering part, $\F^{vol}$: 

\begin{equation}
    \F = \F^{iso} \F^{vol},
\end{equation} 
where

\begin{equation*}
    \F^{iso} = J^{-1/3} \F \Rightarrow \quad \rm{det}\, \F^{iso} = 1.
\end{equation*}
This decomposition translates to the right Cauchy-Green tensor and its invariants as follows

\begin{equation*}
    \C^{iso} = J^{-2/3} \C, \quad I_1^{iso} = \tilde{I}_1 = J^{-2/3} I_1, \quad I_2^{iso} = \tilde{I}_2 = J^{-4/3} I_2 
\end{equation*}

With the decomposition of the deformation gradient, a frequent choice is to write the strain energy density function $\WF$ as the sum of isochoric, $W^{iso}$, and volumetric, $W^{vol}$ contributions:

\begin{equation}
    \label{eq: addititive poly}
    W(\F) = W^{iso} (\F) + W^{vol} (\F). 
\end{equation}

%with $W^{vol}$ depending only on the volumetric invariant $J$, and $ W^{iso}$ on the isochoric invariants $I_1, I_2$. 

\subsection{Requirements for Hyperelastic Constitutive Laws}
\label{sec:semantics for the language of materials}
% List the constraints for constitutive laws
In the following, we briefly overview the main requirements that hyperelastic constitutive laws have to satisfy according to continuum mechanics theory. For more details, see \cite{bonet1997nonlinear, holzapfel2002nonlinear}. 
\paragraph{Thermodynamic Consistency} A hyperelastic material is one for which we postulate the existence of an elastic strain energy density function

\begin{equation*}
    W: \mathcal{GL}^+ (3) \rightarrow \mathbb{R}, \quad \F \mapsto \WF,
\end{equation*}
where $\mathcal{GL}^+(3)$ is the set of invertible second-order tensors with positive determinant. 
The laws of thermodynamics lead to the Clausius-Duhem inequality

\begin{equation*}
    \P : \dot{\F}^T - \dot{W}  \geq 0,
\end{equation*}
where $:$ denotes the tensor dot product. The Clausius-Duhem inequality is satisfied as an equality for an arbitrary $\dot{\F}$ by the following definition of the first Piola-Kirchhoff stress tensor

\begin{equation}
\label{eq: piola}
    \P = \frac{\partial W}{\partial \F}.
\end{equation}
%By considering the stress tensor as a gradient field, energy conservation and path independence are implied, thus thermodynamic consistency. 

\paragraph{Symmetry of the Stress Tensor} Balance of angular momentum implies the symmetry of the Cauchy stress tensor 
$\boldsymbol{\sigma}=J^{-1}\P \, \F^T$, and results in the fact that the  strain energy density function $\WF$ needs to satisfy

\begin{equation*}
    \P \, \F^T = \F \, \P^T \rightarrow \frac{\partial W}{\partial \F} \, \F^T = \F \, \frac{\partial W }{\partial \F}^T
\end{equation*}
%which results in

%\begin{equation*}
%    \mathbf{T} = \F^{-1} \cdot \frac{W}{\partial \F} = \frac{W}{\partial \F^T} \cdot \F^T = \mathbf{T}^T, 
%\end{equation*}
%the requirement for the symmetry of the stress tensor $\mathbf{\sigma}$ and $\mathbf{T}$. 

\paragraph{Objectivity (Frame Indifference)} Objectivity means independence from the choice of the observer, which in hyperelasticity can be expressed as

\begin{equation*}
     W(\F \, \mathbf{Q}^T) = W(\F) \quad \forall \F \in \mathcal{GL}^+(3), \quad \mathbf{Q} \in \mathcal{SO}(3), 
\end{equation*}
with $\mathcal{SO}(3)$ as the 3D rotation group. It is straightforward to show that objectivity is automatically satisfied if $W$ depends on $\F$ through $\C$, i.e. if $W = \tilde{W}(\C)$.

\paragraph{Material Symmetry} 
The strain energy density function should reflect the desired type of material symmetry, e.g. isotropy or a specific class of anisotropy, which can be formalized as follows

 \begin{equation*}
    W(\mathbf{Q} \, \F) = W(\F) \quad \forall \F \in \mathcal{GL}^+(3), \quad \mathbf{Q} \in G \subseteq \mathcal{O}(3), 
\end{equation*}
where $G$ is the symmetry group of the material. 

\paragraph{Polyconvexity}  The polyconvexity of the strain energy density function guarantees sequential weak lower semicontinuity, which along with coercivity is a sufficient condition for
the existence of minimizers, i.e. of solutions to boundary value problems under general boundary conditions and body forces \cite{ball1976convexity, hartmann2003polyconvexity,ebbing2010design}.  Also importantly, polyconvexity is sufficient
for quasiconvexity, which in turn implies rank-one convexity. For twice
differentiable and smooth elastic strain energy density functions, rank-one convexity is equivalent
to ellipticity,  \cite{schroder2010poly,ebbing2010design}.
$W(\F)$ is polyconvex if and only if there exists a function $\mathcal{P}$, \emph{convex} in its arguments, such that
%It can be shown to hold for strain energy expressions of the form:

\begin{equation*}
    \WF  = \mathcal{P} (\F, \rm{cof}\,\F, \rm{det} \,\F)
\end{equation*}
where 
%$\mathcal{P} (\F, \rm{cof}\,\F, \rm{det}\,\F)$ is convex in its arguments and 
$\rm{cof}\,\F = \rm{det}\,\F\,\F^{-T}$.

\paragraph{Normalization of Stress and Strain Energy Density} In an undeformed configuration, i.e. for $\F = \mathbf{I}$, it must be 
 \begin{equation*}
     W(\F= \mathbf{I}) = 0 \quad \text{and} \quad \P(\F= \mathbf{I}) = \mathbf{0}.
 \end{equation*}
In other words, an undeformed configuration implies no stresses and stores no energy. 

\paragraph{Growth Condition} The  volumetric  growth condition requires that the strain energy density grows to infinity as the volumetric deformation tends to zero or to infinity, as follows

\begin{equation*}
    \WF \rightarrow \infty \quad \text{as} \; J \rightarrow 0^+ \;\text{or}\; J \rightarrow \infty \quad \forall \,\F \in \mathcal{GL}^{+}(3).
\end{equation*}
Its physical meaning is that an infinitesimal material volume cannot grow to infinity or be compressed to a point.  Note that this is only one type of growth condition and probably the most widely used one. For a comprehensive discussion on growth conditions and coercivity we refer the readers to \cite{ebbing2010design}.

\paragraph{Non-Negativity of the Strain Energy Density} The strain energy density also needs to satisfy

\begin{equation*}
    \WF \geq 0,
\end{equation*}
as a negative strain energy density is physically meaningless. 

%%%%%%%%%%%%%%%%%%%%%%%%%%%%%%%%%%%%%%%%%%%%%%%%%%%%%%%%%%%%%%%%%%%%
\subsection{Enforcement of Constraints}
\label{sec:enforcement of constraints}

% In this section we propose creating a language $\curlyL(\curlyG_{CL})$ whose derived expressions satisfy constitutive law relations. The S-expressions derived by $\curlyL(\curlyG_{CL})$ needs to be syntactically and semantically valid. Given that the expression are produced by a set of rules the syntactic validity is implied by fact that the grammar uses re-write rules. The semantic validity in our case translates to whether or not an expression coming from $\curlyL(\curlyG)$ is a valid constitutive law or not. 
Now that the requirements for elastic strain energy densities have been defined, we discuss  how these constraints can be translated to language semantics.
The grammar $\curlyG_{NH}$ introduced in Section \ref{sec:formal grammars} leads to expressions that do not necessarily satisfy the constraints in Section \ref{sec:semantics for the language of materials}. If we consider e.g. $ W = (J - 0.5)^2 + 3 \cdot I_1^2$ (one of the examples in Figure \ref{fig:binary tree examples} with $a = 0.5$), 
%and $b=1.5$, 
it is evident that this expression does not satisfy the normalization condition. In this section, we propose a process to embed some of the constitutive law constraints directly in the definition of the grammar and to apply additional semantic checks to generated expressions in order to ensure that the remaining constraints are satisfied. Therefore, we propose creating a grammar, %$\curlyG_{CL}$ 
and the corresponding language, 
%$\curlyL(\curlyG_{CL})$ 
such that its derived expressions comprise automatically valid strain energy density functions. 

\paragraph{Intrinsic Constraints}
Some of the requirements in Section \ref{sec:semantics for the language of materials} can be easily fulfilled a priori in the grammar construction. Thermodynamic consistency for hyperelastic materials is trivially satisfied; symmetry of the stress tensor and frame indifference are automatically fulfilled by considering $W$ as a function of the deformation gradient through the right Cauchy-Green deformation tensor $\C$. In terms of material symmetry, from now on we focus on isotropic hyperelasticity; isotropy is automatically guaranteed by considering $W$ as a function of the invariants of $\C$. 

To satisfy normalization, we correct $W(\F)$ as follows:

\begin{equation}
\label{corr}
    \tilde{W}(\F) = W(\F) - W^0  - W^c
\end{equation}
where $W^0$ and $W^c$ are corrections to satisfy the strain energy density and the stress normalization, respectively, given by 

%The potential correction is computed as follows:
\begin{equation}
\label{en_corr}
     \tilde{W}(\F=\I)= \mathbf{0} \rightarrow W^0 = W|_{\F =\I},
\end{equation}
%and the stress correction $\mathbf{S}^c$:
\begin{equation}
\label{str_corr}
   \frac{\partial \tilde{W}}{ \partial \mathbf{C}}(\mathbf{C}=\mathbf{I})= \mathbf{0} \rightarrow W^c = n \cdot (J-1),
   % \rightarrow W^c =  - \frac{\partial W}{\partial J} |_{\mathbf{C} = \mathbf{I}} (J-1).
    % \tilde{\P}(\F=\I) = \frac{\partial \tilde{W}(\F)} {\partial \F} \Big |_{\F=\I} = \mathbf{0} \rightarrow \W^c = -  \dpsiF \Big |_{\F=\I}.
\end{equation}
where $n = 2 \cdot ( \frac{\partial W}{\partial I_1} + 2 \cdot \frac{\partial W}{\partial I_2} +\frac{\partial W}{\partial I_3} )|_{\mathbf{C} = \mathbf{I}}$ and $I_3 = J^2$. To derive the stress correction we follow \cite{linden2023neural}.

\paragraph{Extrinsic Constraints} The remaining requirements in Section \ref{sec:semantics for the language of materials} are not satisfied a priori, but verified on the final produced expression. If they are not fulfilled, the expression is discarded. In order to empirically check if the growth condition holds, we choose $\F$ as the diagonal matrix:

\begin{equation*}
    \F = \begin{bmatrix}
a & 0 & 0\\
0 & 1 & 0 \\
0 & 0 & 1
\end{bmatrix}
\end{equation*}
for which $J = a$. We set $a$ to a large positive real number $a_l$ to check the case $J \rightarrow \infty$ and to a small positive real number $a_s$ for the case $J \rightarrow 0^+$. Then, we choose a large positive real number $H$ and consider the growth condition satisfied if $W>H$ for $a=a_l$ and $a=a_s$. In our later numerical examples, we choose $a_l=10^4$, $a_s=10^{-4}$ and $H=10^3$. 

\begin{figure}
  \centering
    \includegraphics[width=0.5\textwidth]{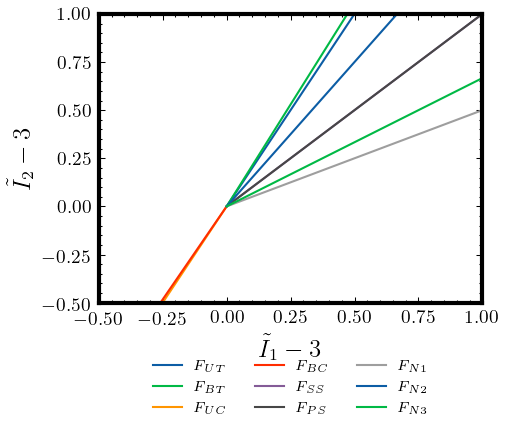}
  \caption{ Deformation paths used to check positivity and monotonicity of the elastic strain energy density plotted in the space of the isochoric invariants.}
  \label{fig:deformation paths}
\end{figure}

To check (again empirically) the non-negativity and the monotonicity of the strain energy density function, we consider one-parametric deformation gradients corresponding to several simple tests, namely Uni-axial Tension (UT), Bi-axial Tension (BT), Uni-axial Compression (UC), Bi-axial Compression (BC), Simple Shear (SS), and Pure Shear (PS) \cite{flaschel2021unsupervised},  as well as three additional loading paths denoted as $N_1$, $N_2$ and $N_3$, as follows 

\begin{equation}
\label{eq: deformation gradients}
\begin{split}
    \F_{UT} &= \begin{bmatrix}
1 + \gamma & 0 & 0\\
0 & 1 & 0 \\
0 & 0 & 1 
\end{bmatrix}, \quad \quad
\F_{UC} = \begin{bmatrix}
\frac{1}{1 + \gamma} & 0 & 0\\
0 & 1 & 0 \\
0 & 0 & 1 
\end{bmatrix}, \quad
    \F_{SS} = \begin{bmatrix}
1 & \gamma & 0\\
0 & 1 & 0 \\
0 & 0 & 1 
\end{bmatrix}, \\
\F_{BT} &= \begin{bmatrix}
1 + \gamma & 0 & 0\\
0 & 1 + \gamma & 0 \\
0 & 0 & 1 
\end{bmatrix},
\F_{BC} = \begin{bmatrix}
\frac{1}{1 + \gamma} & 0 & 0\\
0 & \frac{1}{1 + \gamma} & 0 \\
0 & 0 & 1 
\end{bmatrix}, 
\F_{PS} = \begin{bmatrix}
 1 + \gamma & 0 & 0\\
0 & \frac{1}{1 + \gamma} & 0 \\
0 & 0 & 1 
\end{bmatrix}, \\
\F_{N_1} &= \begin{bmatrix}
\sqrt{\gamma}& 0 & 0\\
0 & \sqrt{\frac{4 - \gamma}{1 + 2\gamma}}& 0 \\
0 & 0 & 1 
\end{bmatrix},  
\F_{N_2}  = \begin{bmatrix}
 \sqrt{\frac{\gamma}{2\gamma - 1}} & 0 & 0\\
0 & \sqrt{\gamma} & 0 \\
0 & 0 & 1 
\end{bmatrix}, 
\F_{N_3}  = \begin{bmatrix}
 \sqrt{\frac{5 -\gamma}{1 + 3\gamma}} & 0 & 0\\
0 & \sqrt{\gamma} & 0 \\
0 & 0 & 1
\end{bmatrix},
\end{split}
\end{equation}
with $\gamma$ as a real non-negative parameter (we give it values such that all the expressions are well defined). 
The deformation paths resulting from the parameterized deformation gradients in (\ref{eq: deformation gradients}) are visualized in Figure \ref{fig:deformation paths}, revealing a reasonable coverage of the invariant space at least in the first quadrant. For all cases,
%The monotonicity of the strain energy function changes at $\F = \I$; 
$W(\F(\gamma))$ should be a positive and monotonically increasing function of $\gamma$. We then consider parameters $\gamma_1$, $\gamma_2$ with $\gamma_2 > \gamma_1$ and check that
%and create a parameterization of %$\F$. Instead of checking if $\WF \geq 0, \forall \F \in GL_+(3)$ which is intractable, we check if

\begin{equation}
\label{eq: inequality of strain}
    0 < W (\F(\gamma_1)) <  W (\F(\gamma_2)), \quad \forall \gamma_1, \gamma_2 \in (0, \infty).
\end{equation}
%for different $\F$ tensors that correspond to the different loading conditions:
If Equation \ref{eq: inequality of strain} does not hold for a pair $(\gamma_1, \gamma_2)$ for any of the loading conditions in (\ref{eq: deformation gradients}), the mathematical expression is rejected.

\paragraph{Polyconvexity} The requisite of polyconvexity deserves a separate discussion. 
In view of the equivalence mentioned in Section \ref{sec:semantics for the language of materials} and of an additivity property \cite{hartmann2003polyconvexity},
%$W(\F)$ is polyconvex if and only if there exists a \emph{convex} function $P$ such that
%\begin{equation*}
%    W(\F) = P(\F, \rm{det} \ \F, \rm{cof} \ \F),
%\end{equation*}
a possible simple choice of a polyconvex strain energy density is the following:

\begin{equation*}
    W(\F) = W_1(\rm{det} \ \F)+ W_2(\tilde{I}_1(\F,\rm{det} \F)) + W_3(\tilde{I}^{3/2}_2(\rm{cof} \F,\rm{det} \F)),
\end{equation*}

with $W_1,W_2$ and $W_3$ convex and monotonically increasing with respect to their arguments. Since $\tilde{I}_1$ and $\tilde{I}^{3/2}_2$ are polyconvex \cite{hartmann2003polyconvexity}, the resulting strain energy density function is naturally polyconvex.  
%Even though the terms $\tilde{I}_1$ and $\tilde{I}^{3/2}_2$ are polyconvex their mixture, meaning $\tilde{I}_1 \cdot \tilde{I}^{3/2}_2$ is not. 
Note that this choice is quite restrictive, as we are not including any terms containing both $\tilde{I}_1$ and $\tilde{I}_2$ (since these mixed terms are not polyconvex \cite{hartmann2003polyconvexity}). Empirical tests on the convexity of each expression can be potentially performed, however one needs to consider deformation gradient tensors parameterized for different loading cases that also adequately cover the space of kinematic quantities. Designing the form of the deformation gradient tensors for such loading paths is not a trivial task and it is left as a subject of future research.

%For this reason, we can restrict the grammar to not produce any mixture terms using either multiplication or division between invariants.  %Considering, $W_1, W_2, W_3$ to be composed of convex functions that take the invariants as an argument and $W^{vol}=W_2$, $W^{iso}=W_1 + W_3$, a polyconvex strain energy $W$ is naturally derived. 

The approach we follow in this paper allows for more flexibility. As will become clear later, we work with a strain energy density function of the form

\begin{equation*}
    W = \hat{W}(J,\tilde{I}_1,\tilde{I}^{3/2}_2),
\end{equation*}

thus, we bias the grammar to derive polyconvex expressions through the choice of the exponent for $\tilde{I}_2$, however we do not strictly guarantee $\hat{W}$ convexity or monotonicity. Over polyconvexity, which may be regarded as an even too strong requirement for practical purposes, we prefer a more flexible grammar construction to allow for more exotic expressions to be generated. However, already at this point we would like to stress that the grammar construction fully controls the properties of the derived expressions; using mathematical tools from formal languages, these properties can be provably controlled in a rigorous manner \cite{linz2022introduction}, hence, in principle they provide the possibility to strictly enforce polyconvexity if desired.

\paragraph{Example}To exemplify the process of hyperelastic model generation including constraint enforcement, we refer back to $\curlyG_{\text{NH}}$ and apply a minor modification. Without loss of generality, we consider $a = 0.5$, $b = 1.5$ to define a specific material and the isochoric invariant $\tilde{I}_1$ in place of $I_1$. We present the grammar construction in two ways. First, we rewrite $\curlyG_{\text{NH}}$ upon implementation of the above modifications:
\begin{equation*}
\NORMAL{
    \begin{split}
        S = \{& \ C \  \}, \\
        \Phi  = \{& \ C, \ \Psi , L \  \}, \\
        \Sigma = \{& \ +, - ,  \cdot, ( )^2,  \tilde{I}_1, J, \  0.5 \ , 1, 1.5,  3 \  \}, \\
        R  = \{  &  \ C \rightarrow  \ \Psi + \Psi, \mathbf{(1)} \\
                &  \ \Psi \rightarrow L \cdot \Psi, \mathbf{(2)} \\
                &  \ \Psi \rightarrow \  (L - L), \mathbf{(3)} \\
                &  \ \Psi \rightarrow \  (\Psi)^2, \mathbf{(6)} \\
                &  \ \Psi \rightarrow \  L, \mathbf{(7)} \\
                 &  L \rightarrow 0.5, \mathbf{(8)} \\ 
                 & L \rightarrow  1,  \mathbf{(9)}\\ 
                 & L \rightarrow 1.5, \mathbf{(10)} \\
                 & L \rightarrow 3, \mathbf{(11)} \\
                 & L \rightarrow \tilde{I}_1, \mathbf{(12)} \\
                 & L \rightarrow J \mathbf{(13)} \ \}.
    \end{split}
    }
\end{equation*}
As we already did in Section \ref{CFG}, we generate an expression by consecutively applying a sequence of production rules, i.e.$r = [\mathbf{(1)}, \mathbf{(2)}, \mathbf{(2)},  \mathbf{(6)}, \mathbf{(3)}, \mathbf{(9)}, \mathbf{(6)}, \\
     \mathbf{(13)}, \mathbf{(8)}, \mathbf{(11)}, \mathbf{(2)}, \mathbf{(9)}, \mathbf{(7)}, \mathbf{(12)}]$, in a top-to-bottom-left-to-right fashion, as follows:

\begin{equation*}
\NORMALONE{
\begin{split}
    C & \xrightarrow{\mathbf{(1)}} \Psi + \Psi, \\
         & \xrightarrow{\mathbf{(2)}} L \cdot \Psi + \Psi, \\
         & \xrightarrow{\mathbf{(2)}} L \cdot \Psi +  L  \cdot \Psi, \\
         & \xrightarrow{\mathbf{(6)}} L \cdot (\Psi)^2 +  L  \cdot \Psi, \\
         & \xrightarrow{\mathbf{(3)}} L \cdot (L - L)^2 + L  \cdot \Psi, \\
         & \xrightarrow{\mathbf{(9)}} 1 \cdot (L - L)^2 + L  \cdot \Psi, \\
         & \xrightarrow{\mathbf{(6)}} 1 \cdot (L - L)^2 + L \cdot (\Psi)^2, \\
         & \xrightarrow{\mathbf{(13)}} 1 \cdot (J - L)^2 + L \cdot (\Psi)^2, \\
         & \xrightarrow{\mathbf{(8)}} 1 \cdot (J - 0.5)^2 + L \cdot (\Psi)^2, \\
         & \xrightarrow{\mathbf{(11)}} 1 \cdot (J - 0.5)^2 + 3 \cdot (\Psi)^2, \\
         & \xrightarrow{\mathbf{(2)}} 1 \cdot (J - 0.5)^2 + 3 \cdot (C \cdot \Psi )^2, \\
         & \xrightarrow{\mathbf{(9)}} 1 \cdot (J - 0.5)^2 + 3 \cdot (1 \cdot \Psi )^2, \\
         & \xrightarrow{\mathbf{(7)}} 1 \cdot (J - 0.5)^2 + 3 \cdot (1 \cdot L )^2, \\
         & \xrightarrow{\mathbf{(12)}} 1 \cdot (J - 0.5)^2 + 3 \cdot (1 \cdot \tilde{I}_1 )^2
\end{split}
}
\end{equation*}
and derive $\WF = (J - 0.5)^2 + 3 \cdot \tilde{I}_1^2$. 
As a second alternative, we introduce two new non-terminal symbols $\Psi^{iso}$ and $\Psi^{vol}$ to indicate the isochoric and volumetric parts of the strain energy density, and define production rules for each of these non-terminals:
\begin{equation*}
\NORMALONE{
    \begin{split}
        S = \{& \ C \  \}, \\
        \Phi  = \{& \ C, \ \Psi^{vol}, \ \Psi^{iso} , L,  L^{vol}, L^{iso} \  \}, \\
        \Sigma = \{& \ +, - ,  \cdot, ( )^2,  \tilde{I}_1, J, \  0.5 \ , 1, 1.5,  3 \  \}, \\
        R  = \{  &  \ C \rightarrow  \ \Psi^{iso} + \Psi^{vol}, \mathbf{(1)} \\
                &  \ \Psi^{vol} \rightarrow L \cdot \Psi^{vol}, \mathbf{(2)} \\
                &  \ \Psi^{vol} \rightarrow \  (L^{vol} - L^{vol}), \mathbf{(3)} \\
                &  \ \Psi^{vol} \rightarrow \  (\Psi^{vol})^2, \mathbf{(4)} \\
                &  \ \Psi^{vol} \rightarrow \  L^{vol}, \mathbf{(5)} \\
                &  \ \Psi^{iso} \rightarrow L \cdot \Psi^{iso}, \mathbf{(6)} \\
                &  \ \Psi^{iso} \rightarrow \  (L^{iso} - L^{iso}), \mathbf{(7)} \\
                &  \ \Psi^{iso} \rightarrow \  (\Psi^{iso})^2, \mathbf{(8)} \\
                &  \ \Psi^{iso} \rightarrow \  L^{iso}, \mathbf{(9)} \\
                 &  L^{vol} \rightarrow 0.5, \mathbf{(10)} \\ 
                 & L \rightarrow  1,  \mathbf{(11)}\\ 
                 & L \rightarrow 1.5, \mathbf{(12)} \\
                 & L^{iso} \rightarrow 3, \mathbf{(13)} \\
                 & L^{iso} \rightarrow \tilde{I}_1, \mathbf{(14)} \\
                 & L^{vol} \rightarrow J \mathbf{(15)} \ \},
    \end{split}
    }
\end{equation*}
With this grammar definition, it is possible to derive expressions which distinguish between volumetric and deviatoric (isochoric) contributions, which is a common choice in constitutive modeling of hyperelastic materials. From this grammar we again sample a sequence of production rules, i.e. $r = [\mathbf{(1)}, \mathbf{(2)}, \mathbf{(6)},  \mathbf{(4)}, \mathbf{(3)}, \mathbf{(11)}, \mathbf{(8)}, \mathbf{(15)}, \mathbf{(10)}, \mathbf{(13)}, \mathbf{(6)}, \mathbf{(11)}, \mathbf{(9)}, \mathbf{(14)}]$, and consecutively apply them in a top-to-bottom-left-to-right fashion to re-write the non-terminals until we derive a terminal expression, as follows:
\begin{equation*}
\NORMALONE{
\begin{split}
    C & \xrightarrow{\mathbf{(1)}} \Psi^{vol} + \Psi^{iso}, \\
         & \xrightarrow{\mathbf{(2)}} L \cdot \Psi^{vol} + \Psi^{iso}, \\
         & \xrightarrow{\mathbf{(6)}} L \cdot \Psi^{vol} +  L  \cdot \Psi^{iso}, \\
         & \xrightarrow{\mathbf{(4}} L \cdot (\Psi^{vol})^2 +  L  \cdot \Psi^{iso}, \\
         & \xrightarrow{\mathbf{(3)}} L \cdot (L^{vol} - L^{vol})^2 + L  \cdot \Psi^{iso}, \\
         & \xrightarrow{\mathbf{(11)}} 1 \cdot (L^{vol} - L^{vol})^2 + L  \cdot \Psi^{iso}, \\
         & \xrightarrow{\mathbf{(8)}} 1 \cdot (L^{vol} - L^{vol})^2 + L \cdot (\Psi^{iso})^2, \\
         & \xrightarrow{\mathbf{(15)}} 1 \cdot (J - L^{vol})^2 + L \cdot (\Psi^{iso})^2, \\
         & \xrightarrow{\mathbf{(10)}} 1 \cdot (J - 0.5)^2 + L \cdot (\Psi^{iso})^2, \\
         & \xrightarrow{\mathbf{(13)}} 1 \cdot (J - 0.5)^2 + 3 \cdot (\Psi^{iso})^2, \\
         & \xrightarrow{\mathbf{(6)}} 1 \cdot (J - 0.5)^2 + 3 \cdot (C \cdot \Psi^{iso} )^2, \\
         & \xrightarrow{\mathbf{(11)}} 1 \cdot (J - 0.5)^2 + 3 \cdot (1 \cdot \Psi^{iso} )^2, \\
         & \xrightarrow{\mathbf{(9)}} 1 \cdot (J - 0.5)^2 + 3 \cdot (1 \cdot L )^2, \\
         & \xrightarrow{\mathbf{(14)}} 1 \cdot (J - 0.5)^2 + 3 \cdot (1 \cdot \tilde{I}_1 )^2
\end{split}
}
\end{equation*}
obtaining again $\WF = (J - 0.5)^2 + 3 \cdot \tilde{I}_1^2$. Note that with this second version, where separate non-terminals are considered for the isochoric and volumetric parts of the strain energy density, the number of grammar rules is larger since we need rules for each non-terminal. For the sake of simplicity, in the following developments of this paper we will consider a single non-terminal $\Psi$ and not distinguish between volumetric and deviatoric contributions. However, considering both terms separately would be straightforward as shown above.

\begin{figure}
\centering
  \begin{subfigure}[!t]{0.3\textwidth}
  \vspace{1.0cm}
    \centering
    \begin{tikzpicture}[
      level distance=1cm,
      level 1/.style={sibling distance=1.25cm},
      level 2/.style={sibling distance=0.65cm},
      level 3/.style={sibling distance=0.65cm},
      every node/.style={draw, circle, minimum size=0.05em, font=\tiny},
      edge from parent/.style={draw, -, >=stealth},
    ]
      \node {$+$}
        child { node {$()^2$}
        child { node {$-$}
          child { node {$J$} }
          child { node {$0.5$} }
        }}
        child { node {$\cdot$}
          child { node {$3$} }
          child { node {$()^2$}
            child { node {$\tilde{I}_1$} }
          }
      };
    \end{tikzpicture}
  \end{subfigure}
  \begin{subfigure}[!t]{0.3\textwidth}
    \centering
    \begin{tikzpicture}[
      level distance=1cm,
      level 1/.style={sibling distance=2cm},
      level 2/.style={sibling distance=1.3cm},
      level 3/.style={sibling distance=0.65cm},
      every node/.style={draw, circle, minimum size=0.05em, font=\tiny},
      edge from parent/.style={draw, -, >=stealth},
    ]
    \node[color=red]{$-$}
      child { node {$+$}
        child { node {$()^2$}
        child { node {$-$}
          child { node {$J$} }
          child { node {$0.5$} }
        }}
        child { node {$\cdot$}
          child { node {$3$} }
          child { node {$()^2$}
            child { node {$\tilde{I}_1$} }
          }
      }}
      child{ node[red] {$-$}
      child { node[red] {$26.25$} }
      child {node[red] {$J$}}
      };
    \end{tikzpicture}
  \end{subfigure}
    \begin{subfigure}[!t]{0.3\textwidth}
    \vspace{1.0cm}
    \centering
        \includegraphics[width=0.7\textwidth]{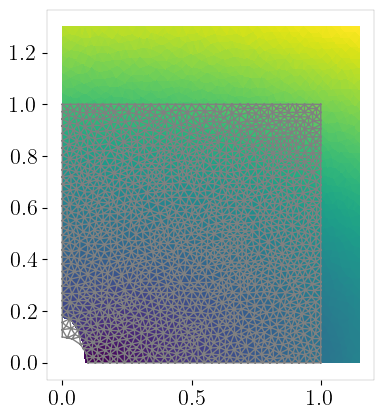}
  \end{subfigure}
  \caption{Left: The tree representation of $W = 3 \cdot \tilde{I}_1^2 + (J - 0.5)^2$. Middle: The tree representation of the expression  $ \tilde{W} = 3 \cdot \tilde{I}_1^2 + (J - 0.5)^2  - 26.25 - J$, with the correction as a sub-tree attached to the original expression. Right: The displacement plot with $\tilde{W} $ used as constitutive law for the boundary value problem in Figure \ref{fig:benchmark} with $\delta = 0.3$. }
    \label{fig:parse tree of example}
\end{figure}
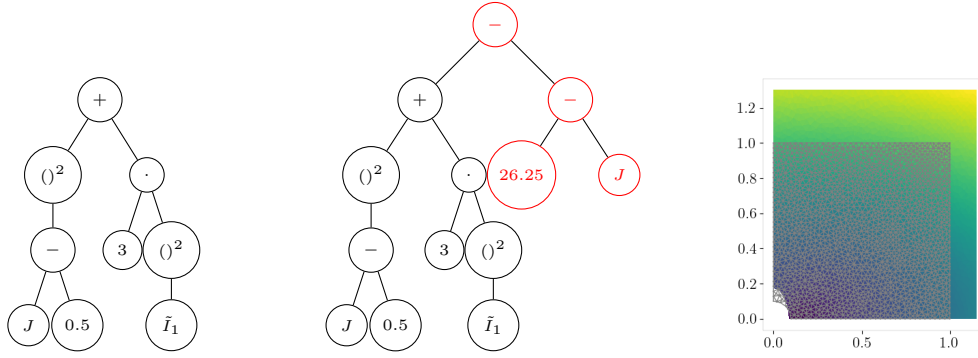

\begin{figure}
  \centering
    \includegraphics[width=0.5\textwidth]{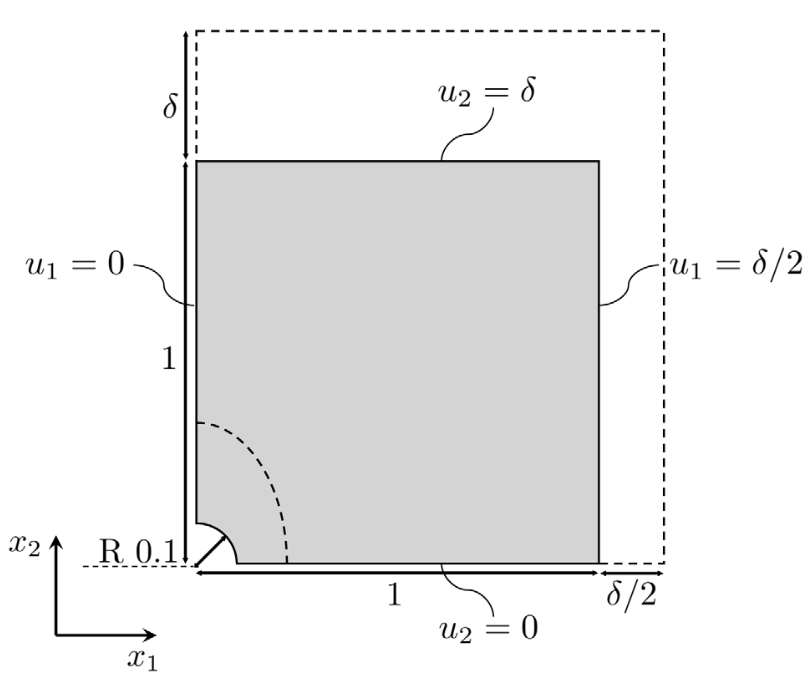}
  \caption{Benchmark boundary value problem considered in this work, adapted from \cite{flaschel2021unsupervised}.}
  \label{fig:benchmark}
\end{figure}

The obtained expression for $\WF$ does not satisfy the normalization conditions, because $W(\F =\I) = 27.25$ and $W^c (\F = \I) = (J - 1)$. Thus, we perform the corrections \eqref{en_corr} and \eqref{str_corr} and derive the expression:
\begin{equation}
\label{eq:corrected constitutive law sample}
    \tilde{W}(\F) = 3 \cdot \tilde{I}_1^2 + (J - 0.5)^2 - J - 26.25 .
\end{equation}
Performing the corrections is equivalent to adding a sub-tree to the original expression, see Figure \ref{fig:parse tree of example}. In Figure \ref{fig:parse tree of example} we also provide the predicted displacement field for the benchmark in \cite{flaschel2021unsupervised}, reproduced in Figure \ref{fig:benchmark}, considering $\tilde{W}(\F)$ from eq. \eqref{eq:corrected constitutive law sample}. 

We then execute the remaining empirical checks on volumetric growth and non-negativity, both of which hold in this case. Therefore, we conclude that the derived expression can indeed be deemed a constitutive law that can characterize a hyperelastic material. 
\subsection{Generating the Language of Hyperelastic Materials}
\label{sec:generating the language of hyperelastic materials}
So far we considered a simple grammar, based on which we could generate simple expressions which we guaranteed to be valid (simple) constitutive laws. In order to obtain more flexible expressions, we now consider a more complicated alphabet of terms, include production rules with more operations, and derive constants via recursive substitution of non-terminals instead of explicitly defining production rules that re-write non-terminals with reals. In the choice of the primitives we consider logarithmic, exponential, monomials and other operations that often appear in constitutive relations. To keep a reasonably compact grammar, we do not consider different non-terminals and production rules for the isochoric and volumetric parts of the strain energy density, but one non-terminal for both. However, if desired, this can be changed as exemplified earlier. Finally, we define the energy and stress corrections as production rules for the non-terminal $\Psi$. The resulting grammar of hyperelastic materials, $\curlyG_{\text{HM}}$, is then defined as follows: 
 % \begin{equation*}
 % \label{eq: hyperelastic grammar}
 % \NORMAL{
 {\scriptsize
 \begin{align*}
 \allowdisplaybreaks
    S = \{&   C  \}, \\
    \Phi = \{&  C ,  \Psi, \Psi^{0} ,  W^c , L, M, P, D \},\\
\Sigma = \{& +, -, /, \cdot, ()^2, ()^3, - \log, \exp, \tilde{I}_1, \tilde{I}_2, J, 0, 1, 2, 3, 4, 5, 6, 7, 8, 9,    \frac{ \partial ()  }{\partial \C }|_{\C= \I} : (J - 1),   |_{\F= \I}  \}, \\
         R  = \{&  C  \rightarrow   \Psi  +    \Psi^{0}  +   W^c , \mathbf{(1)} \\
         & \Psi   \rightarrow \Psi + \Psi , \mathbf{(2)} \\
         & \Psi^{0}  \rightarrow -  \Psi |_{\F=\I} ,  \mathbf{(3)} \\
         & W^c  \rightarrow - \frac{ \partial  \Psi  }{\partial \C }|_{\C= \I} (J - 1),  \mathbf{(4)}\\
         & \Psi   \rightarrow L , \mathbf{(5)} \\
         & L \rightarrow \  (L+L) \,  \mathbf{(6)}\\
         & L \rightarrow \  (L-L) \,  \mathbf{(7)}\\
         & L \rightarrow \  (L/M) \,  \mathbf{(8)}\\ %  L \rightarrow \  (L/L) \,  \mathbf{(9)}
         & L \rightarrow \  (M \cdot L) \,  \mathbf{(9)}\\ % L \rightarrow \  (L \cdot L) \,  \mathbf{(10)}
         & L \rightarrow \  (M \cdot L) \,  \mathbf{(9)}\\ % L \rightarrow \  (L \cdot L) \,  \mathbf{(10)}
         & L \rightarrow \  (L)^2 \ ,  \mathbf{(10)}\\
                  % \displaybreak
         & L \rightarrow \ (L)^3 \ ,  \mathbf{(11)}\\
         & L \rightarrow \  - \log(L) \ ,  \mathbf{(12)}\\ 
         & L \rightarrow \ \exp(L) \,  \mathbf{(13)}\\ 
         & L \rightarrow \  \tilde{I}_1 \,  \mathbf{(14)}\\
         & L \rightarrow \  \tilde{I}^{3/2}_2  \,  \mathbf{(15)}\\
         & L \rightarrow \  J \,  \mathbf{(16)}\\
         & L \rightarrow \ M ,  \mathbf{(17)}\\  
         & M \rightarrow \ P .  P ,  \mathbf{(18)}\\  
        & P  \rightarrow \ D \ ,  \mathbf{(19)}\\
        & P  \rightarrow \ D  P ,  \mathbf{(20)}\\
        & D  \rightarrow  \  0 \ ,  \mathbf{(21)}\\
        & D  \rightarrow  \  1 \ ,  \mathbf{(22)}\\
        & D  \rightarrow  \  2 \ ,  \mathbf{(23)}\\
        & D  \rightarrow  \  3 \ ,  \mathbf{(24)}\\
        & D  \rightarrow  \  4 \ ,  \mathbf{(25)}\\
        & D  \rightarrow  \  5 \ ,  \mathbf{(26)}\\
        & D  \rightarrow  \  6 \   \mathbf{(27)},\\
        & D  \rightarrow  \  7 \ ,  \mathbf{(28)}\\
        & D  \rightarrow  \  8 \ ,  \mathbf{(29)}\\
        & D  \rightarrow  \  9 \  \mathbf{(30)}\} 
\end{align*}
}%
% \end{equation*} 
In $\curlyG_{\text{HM}}$, we purposely define multiple production rules for the non-terminals $L, P, D$; this choice enforces recursions that increase the expressivity of the grammar. When two rules have the same left-hand side, we randomly select one to apply. In CFGs non-terminals should be treated independently in each production rule; however, for the purpose of the normalization correction (see eq. \eqref{corr}) we enforce that the same non-terminal $\Psi$ is substituted to the left-hand side in the production rules $\mathbf{(1)}, \mathbf{(3)}$, and $\mathbf{(4)}$. Now we return to eq. \eqref{eq:corrected constitutive law sample}, and show how this simple constitutive law can be derived also from the present more complex grammar via the sequence of production rules $r = [ \mathbf{(1)}, \mathbf{(3)}, \mathbf{(4)}, \mathbf{(2)}, \mathbf{(5)}, \mathbf{(10)}, \mathbf{(7)}, \mathbf{(16)}, \mathbf{(17)}, \mathbf{(18)},  \mathbf{(19)}, \mathbf{(21)}, \mathbf{(26)}, \mathbf{(5)}, \mathbf{(9)}, \mathbf{(18)}, \\\mathbf{(21)}, \mathbf{(10)}, \mathbf{(14)} ]$ with top-to-bottom-left-to-right ordering:
% \begin{equation*}
% \NORMAL{
{\scriptsize
\begin{align*}
\allowdisplaybreaks
    C & \xrightarrow{\mathbf{(1)}} \Psi + \Psi^0 + W^c, \\
         & \xrightarrow{\mathbf{(3)}}  \Psi - ( \Psi )|_{\F = \I} + W^c , \\
         & \xrightarrow{\mathbf{(4)}}  \Psi - ( \Psi )|_{\F = \I} - \frac{ \partial \Psi  }{\partial \C }|_{\C= \I} (J -1), \\
         & \xrightarrow{\mathbf{(2)}} \Psi  + \Psi - ( \Psi + \Psi )|_{\F = \I} - \frac{ \partial (\Psi  + \Psi)  }{\partial \C }|_{\C= \I} (J -1), \\
         & \xrightarrow{\mathbf{(5)}} L  + \Psi - ( L + \Psi )|_{\F = \I} - \frac{ \partial (L  + \Psi)  }{\partial \C }|_{\C= \I} (J -1), \\
         & \xrightarrow{\mathbf{(10)}} (L)^2  + \Psi - ( (L)^2 + \Psi )|_{\F = \I} - \frac{ \partial (L  + \Psi)  }{\partial \C }|_{\C= \I} (J -1), \\
         & \xrightarrow{\mathbf{(7)}} (L -L)^2  + \Psi - ( (L-L)^2 + \Psi )|_{\F = \I} - \frac{ \partial ( (L-L)  + \Psi)  }{\partial \C }|_{\C= \I} (J -1), \\
         & \xrightarrow{\mathbf{(16)}} (J -L)^2  + \Psi - ( (J-L)^2 + \Psi )|_{\F = \I} - \frac{ \partial ( (J-L)  + \Psi)  }{\partial \C }|_{\C= \I} (J -1), \\
         & \xrightarrow{\mathbf{(17)}} (J -M)^2  + \Psi - ( (J-M)^2 + \Psi )|_{\F = \I} - \frac{ \partial ( (J-M)  + \Psi)  }{\partial \C }|_{\C= \I} (J -1), \\
                  % \displaybreak
         & \xrightarrow{\mathbf{(18)}} (J - P.P )^2  + \Psi - ( (J-P.P)^2 + \Psi )|_{\F = \I} - \frac{ \partial ( (J-P.P)  + \Psi)  }{\partial \C }|_{\C= \I} (J -1), \\
         & \xrightarrow{\mathbf{(19)}} (J - D.D )^2  + \Psi - ( (J-D.D)^2 + \Psi )|_{\F = \I} - \frac{ \partial ( (J-D.D)  + \Psi)  }{\partial \C }|_{\C= \I} (J -1), \\
         & \xrightarrow{\mathbf{(21)}} (J - 0.D )^2  + \Psi - ( (J-0.D)^2 + \Psi )|_{\F = \I} - \frac{ \partial ( (J-0.D)  + \Psi)  }{\partial \C }|_{\C= \I}:(\F - \I), \\
         & \xrightarrow{\mathbf{(26)}} (J - 0.5 )^2  + \Psi - ( (J-0.5)^2 + \Psi )|_{\F = \I} - \frac{ \partial ( (J-0.5)  + \Psi)  }{\partial \C }|_{\C= \I} (J -1), \\
         & \xrightarrow{\mathbf{(5)}} (J - 0.5 )^2  + L - ( (J-0.5)^2 + L )|_{\F = \I} - \frac{ \partial ( (J-0.5)  + L)  }{\partial \C }|_{\C= \I} (J -1), \\
         & \xrightarrow{\mathbf{(9)}} (J - 0.5 )^2  + M \cdot L - ( (J-0.5) + M \cdot L )|_{\F = \I} - \frac{ \partial ( (J-0.5)^2  + M \cdot L)  }{\partial \C }|_{\C= \I}:(\F - \I), \\
         & \xrightarrow{\mathbf{(18)}} (J - 0.5 )^2  + P.P \cdot L - ( (J-0.5) + P.P \cdot L )|_{\F = \I} - \frac{ \partial ( (J-0.5)^2  + P.P \cdot L)  }{\partial \C }|_{\C= \I} (J -1), \\
         & \xrightarrow{\mathbf{(24)}} (J - 0.5 )^2  + 3.P \cdot L - ( (J-0.5) + 3.P \cdot L )|_{\F = \I} - \frac{ \partial ( (J-0.5)^2  + 3.P \cdot L)  }{\partial \C }|_{\C= \I} (J -1), \\
         & \xrightarrow{\mathbf{(21)}} (J - 0.5 )^2  + 3.0 \cdot L - ( (J-0.5) + 3.0 \cdot L )|_{\F = \I} - \frac{ \partial ( (J-0.5)^2  + 3.0 \cdot L)  }{\partial \C }|_{\C= \I} (J -1), \\
         & \xrightarrow{\mathbf{(10)}} (J - 0.5 )^2  + 3.0 \cdot (L)^2 - ( (J-0.5) + 3.0 \cdot (L)^2) )|_{\F = \I} - \frac{ \partial ( (J-0.5)^2  + 3.0 \cdot (L)^2)  }{\partial \C }|_{\C= \I} (J -1), \\
         & \xrightarrow{\mathbf{(14)}} (J - 0.5 )^2  + 3.0 \cdot (I_1)^2 - ( (J-0.5) + 3.0 \cdot (I_1)^2) )|_{\F = \I} - \frac{ \partial ( (J-0.5)^2  + 3.0 \cdot (I_1)^2)  }{\partial \C }|_{\C= \I} (J -1). 
\end{align*}
}
% \end{equation*}
By evaluating the production rules $r = [ \mathbf{(3)}, \mathbf{(4)} ]$ we derive eq. \eqref{eq:corrected constitutive law sample}. It takes more steps to complete the derivation and obtain the strain energy density expression because $\curlyG_{\text{HM}}$ contains more numerous and more general production rules than $\curlyG_{\text{NH}}$. As usual, the growth, non-negativity and monotonicity conditions are empirically checked a posteriori.  

To summarize, in the grammar of hyperelastic materials $\curlyG_{\text{HM}}$ the thermodynamic consistency, the symmetry of the stress tensor, the objectivity and material symmetry conditions are satisfied a priori, the normalization of the stress and the strain energy density are satisfied by construction and the growth, non-negativity and monotonicity conditions are empirically assessed a posteriori by testing the obtained expression on a number of simple one-parametric deformation paths. 

The proposed approach is highly versatile, because with minimal changes to the grammar one can describe different types of materials. For example in the case where anisotropy is present due to  fiber reinforcement, additional invariants need to be considered to describe the deformation along the direction of anisotropy. For example in a material reinforced with two fibers, $\curlyG_{\text{HM}}$ can be further adjusted by considering the production rules \cite{joshi2022bayesian}:
\begin{equation*}
\begin{split}
        &L \rightarrow \  J_4, \\
        &L \rightarrow  \ J_5, \\
        &L \rightarrow \ J_6, \\
        &L \rightarrow  \ J_7,
\end{split}
\end{equation*}
and adding $J_4, J_5, J_6, J_7$, i.e. the anisotropic invariants for the two fiber families, to the alphabet. The ensuing language $\curlyL(\curlyG_{\text{HM}})$ will include constitutive models for this anisotropy case. Clearly, analogous modifications can be applied for more complex anisotropy cases. 

We provided the derivation of one expression by choosing a sequence of production rules from $\curlyG_{\text{HM}}$. All the possible derivation trees that the grammar can produce comprise the Language of Hyperelastic Materials $\curlyL(\curlyG_{\text{HM}})$.
%and, as explained above, this can be adapted for different types of materials. 
In the next section, we elaborate on how this language can be used for the automated generation of a library of constitutive models, see the first block of Figure \ref{fig:master_figure_lhm}. 

%However, prior to proceeding with concrete outlining of use cases, we elaborate next on how such a tool can form part of an inverse problem formulation, as part of a pipeline for data-driven discovery of constitutive laws. 

\subsection{Deriving Parsimonious Expressions}
\label{sec:deriving parsimonious expressions}

%The tree generation initiates from the starting symbol; then, a successive application of production rules follows. 
The recursive substitution process in the grammar of hyperelastic materials $\curlyG_{\text{HM}}$ is not guaranteed to stop nor to provide a parsimonious expression, i.e. a shallow tree. To alleviate theses issues, the tree generation process can be manipulated so as to enforce or favor the generation of shallow trees. There are two main options. One option is to attribute different probabilities to the production rules that share the same left-hand side, favoring those that produce terminal strings. Another alternative is to specify a fixed maximum number of operations to obtain the final expression. In this subsection, we briefly discuss these two alternative procedures.
 
\paragraph{ Probabilistic Context-Free Grammars} As discussed earlier, in a CFG several production rules may have the same non-terminal on their left-hand side; as a result, a criterion is needed to select one of these production rules during the tree generation. A naive approach would consider a uniform probability over all production rules for a specific non-terminal. A greater flexibility and efficiency can be obtained using Probabilistic Context-Free Grammars (PCFGs) \cite{brence2021probabilistic, jelinek1992basic, chi1999statistical, geman2002probabilistic}. PCFGs can be constructed from CFGs by assigning probabilities to the grammar production rules with the same non-terminal on the left-hand side, with the sum of the probabilities summing up to one. The probability for each rule can be assigned by the user or learned from the data. As an example, a PCFG can be obtained from $\curlyG_{\text{NH}}$ (for $a=0.5, b=1.5$) by assigning probabilities to its production rules (given in square brackets) as follows
\begin{equation*}
\NORMAL{
    \begin{split}
        S = \{& \ C \  \}, \\
        \Phi  = \{& \ C,  \ \Psi , L \  \}, \\
        \Sigma = \{& \ +, - ,  \cdot, (\cdot)^2,  \tilde{I}_1, J, \  0.5 \ , 1, 1.5,  3 \  \}, \\
        R  = \{  &  \ C \rightarrow  [1] \ \ \Psi + \Psi, \mathbf{(1)} \\
                &  \ \Psi \rightarrow \ [0.3] \ C \cdot \Psi, \mathbf{(2)} \\
                &  \ \Psi \rightarrow \ [0.1] \ (C - C), \mathbf{(3)} \\
                &  \ \Psi \rightarrow \ [0.1] \ (\Psi)^2, \mathbf{(4)} \\
                &  \ \Psi \rightarrow \ [0.5] \  L, \mathbf{(5)} \\
                 &  L \rightarrow [0.1] \ 0.5, \mathbf{(6)} \\ 
                 & L \rightarrow  [0.1] \ 1,  \mathbf{(7)}\\ 
                 & L \rightarrow  [0.1] \ 1.5, \mathbf{(8)} \\
                 & L \rightarrow  [0.1] \ 3, \mathbf{(9)} \\
                 & L \rightarrow  [0.3] \ \tilde{I}_1, \mathbf{(10)} \\
                 & L \rightarrow  [0.3] \ J \mathbf{(11)} \ \}.
    \end{split}
    }
\end{equation*}
Clearly, the probability of a constitutive law is given by the product of all probabilities associated with the rules chosen for its derivation, i.e.
\begin{equation}
    P(W) = \prod_{i=1}^{N_p} P(r_i), 
\end{equation}
where $r= [r_1, ... ,r_{N_p}]$ is the sequence of production rules that derive $W$. It is also clear that the probabilities of all possible final expressions sum up to one \cite{chi1999statistical}. 
For deriving a deeper tree, i.e. a more complex expression, a longer sequence $r$ is required, which includes a larger number of production rules and is thus associated to a lower probability. Even though this is an important property that can be used to favor the generation of parsimonious expressions, it requires a careful design of the grammar, as unrealistically low probabilities may be obtained already for simple models. For example, with $\curlyG_{\text{NH}}$ the sequence of rules that produces  the Neo-Hookean model  $W = 0.5 \cdot (\tilde{I}_1 -3) + 1.5 \cdot (J-1)^2$ is $r = [\mathbf{(1)},\mathbf{(2)},\mathbf{(3)},\mathbf{(5)},\mathbf{(10)},\mathbf{(9)},\mathbf{(8)},\mathbf{(11)},\mathbf{(7)}]$, and the probability of this model being produced from the grammar (with the individual rule probabilities in the previous example)  is $P(W) = 1.35 \times 10^{-6}$. 

Clearly, it is possible to modify the grammar and/or the individual rule probabilities to influence the probability of the resulting model form, but such modifications would require hand-engineering, thereby introducing bias in the model generation. Thus, PCFGs are not well suited for generation of expressions. Moreover, PCFG are not efficient for performing data-driven discovery tasks for long expressions because these will be discovered with a low probability. They are frequently used for tasks involving parsing of expressions, where the probability of each rule is learned from the data and not manually assigned by a user. For example, we could consider a probabilistic grammar if we were given a library of constitutive laws and wishes to determine the probabilities of individual rules being used in deriving a valid expression. This could be useful in a scenario where there is uncertainty about the number and form of the rules used to derive expressions. In such a case, we could define a more general grammar than we believe is needed and then, by learning suitable probabilities, we could discover the rules used in generating the library. Probabilistic grammars are also useful in understanding the decoding process that we employ in Section \ref{sec:data driven constitutive law discovery}.  

\paragraph{ Constraining the Number of Operations in a Tree Derivation} The second alternative is to a priori decide the maximum number of operations that the grammar is allowed to perform in order to derive an expression. To realize this constraint, we construct the grammar rules such that they do not allow for long recursions. For this purpose, we introduce a non-terminal symbol for each operation, e.g. by transforming rule $\mathbf{(2)}$ from $\Psi \rightarrow \Psi + \Psi$ to $\Psi \rightarrow \Psi^1 + \Psi^2$, and for each of these non-terminal symbols we define unary and binary operations that in one step derive terminal nodes. In this way, we derive expressions of pre-set maximum length and complexity. We exemplify how this process works in practice in the next section. 
\subsection{Automatic Generation of an Hyperelastic Material Model Library}
\label{Library_generation}
In this section, we demonstrate that the grammar-based generation of elastic strain energy density functions proposed in the previous sections can be used for the automatic creation of a material model library, which can possibly serve as the basis for sparse regression approaches such as those in \cite{flaschel2021unsupervised, flaschel2023automated}. Moreover, we develop an automated computational pipeline which, once the mathematical expression for a valid constitutive law is generated, is used to produce the finite element code needed to solve boundary value problems where the material behavior obeys such a constitutive law. For the hyperelastic model library generation, we deploy a CFG for deriving mathematical expressions. As illustrated earlier, the grammar provides a systematic way of checking if the expressions are syntactically and semantically valid, and it is chosen to be as expressive as possible. 
For the purpose of obtaining parsimonious expressions, we no longer use the grammar $\curlyG_{\text{HM}}$ defined in Section \ref{sec:generating the language of hyperelastic materials}, but the following grammar, $\curlyG_{\text{HMp}}$:

\begin{equation}
\label{eq: hyperelastic grammar for library generation}
\NORMAL{
    \begin{split}
        S = \{& \ C \ \}, \\
    \Sigma = \{& +, -, /, \cdot, (\cdot)^2, (\cdot)^3, - \log, \exp, \tilde{I}_1, \tilde{I}_2, J, 0, 1, 2, 3, 4, 5, 6, 7, 8, 9, \ \frac{ \partial ()  }{\partial \C }|_{\C= \I}  : ( \F -\I), , \   |_{\F= \I}  \}\}, \\
         R  = \{ & C  \rightarrow \Psi  +  \Psi^{0} +   W^c, \mathbf{(1)} \\
         & \Psi  \rightarrow \  \Psi^1  +    \Psi^2 +    \Psi^3   +    \Psi^4, \mathbf{(2)}\\
         & \Psi^{0}  \rightarrow -  \Psi |_{\F=\I}, \mathbf{(3)}  \\
         & \Psi^{1}  \rightarrow  L  \  \mathbf{(4)} | \  U \mathbf{(5)} \ | \   Y \mathbf{(6)}  \ |  \     T \mathbf{(7)} \ , \\
         & \Psi^{2}  \rightarrow    L \mathbf{(8)} \  | \  U \mathbf{(9)} \ | \   Y \mathbf{(10)} \ |  \   T \mathbf{(11)}  \ ,  \\
         & \Psi^{3}  \rightarrow   L \mathbf{(12)} \  | \  U \mathbf{(13)} \ | \   Y \mathbf{(14)} \ |  \    T \mathbf{(15)} \ , \\
         & \Psi^{4}  \rightarrow   L \mathbf{(16)} \  | \  U \mathbf{(17)} \ | \   Y \mathbf{(18)} \ |  \    T \mathbf{(19)} \ , \\
         & W^c  \rightarrow - \frac{ \partial  \Psi  }{\partial \C }|_{\C= \I} : (\F - \I), \mathbf{(20)}\\
        & T  \rightarrow  (Y)^2 \mathbf{(21)}| (Y)^3 \mathbf{(22)} | \exp(Y) \mathbf{(23)}  | - \log(Y) \mathbf{(24)}, \\
        & Y  \rightarrow \  ( V + O ) \ \mathbf{(25)} | \   ( V -  O ) \mathbf{(26)} \ | \  ( V /  O )\mathbf{(27)} \  | \  ( V \cdot O ) \mathbf{(28)}\ | \ (V)^2  \mathbf{(29)} \  | \  (V)^3 \mathbf{(30)} \, \\
        & U  \rightarrow - \log( L )  \mathbf{(31)} \ | \exp( L )   \mathbf{(32)}\, \\
        & L  \rightarrow  V  \mathbf{(33)} \ |  O  \mathbf{(34)}, \\
         & V  \rightarrow  \tilde{I}_1  \mathbf{(35)}\  |   \ \tilde{I}_2^{3/2}  \mathbf{(36)} \  | \  J  \mathbf{(37)}, \\ 
         & O   \rightarrow  P  .  P ,  \mathbf{(38)} \\  
                & P  \rightarrow  D  \mathbf{(39)} \  |  D   P  \mathbf{(40)}, \\
                & D  \rightarrow  \  0 \mathbf{(41)}\  | \  1\mathbf{(42)}  \ | \ 2 \mathbf{(43)} \ | \ 3 \mathbf{(44)} \ | \ 4 \mathbf{(45)} \ | \ 5 \mathbf{(46)}\ | \ 6 \mathbf{(47)} \ | \ 7 \mathbf{(48)} \ | \ 8 \mathbf{(49)} \ | \ 9 \mathbf{(50)}\ 
                \}, \\
    \Phi = \{ \  & C ,   \Psi ,  \Psi^0 ,  \Psi^1 ,  \Psi^2 , \Psi^3 , \Psi^4 ,W^c, T, L, U, Y, O, D, P \ \}
\end{split}
}
\end{equation}

\begin{figure}
\centering
        \includegraphics[width=\textwidth]{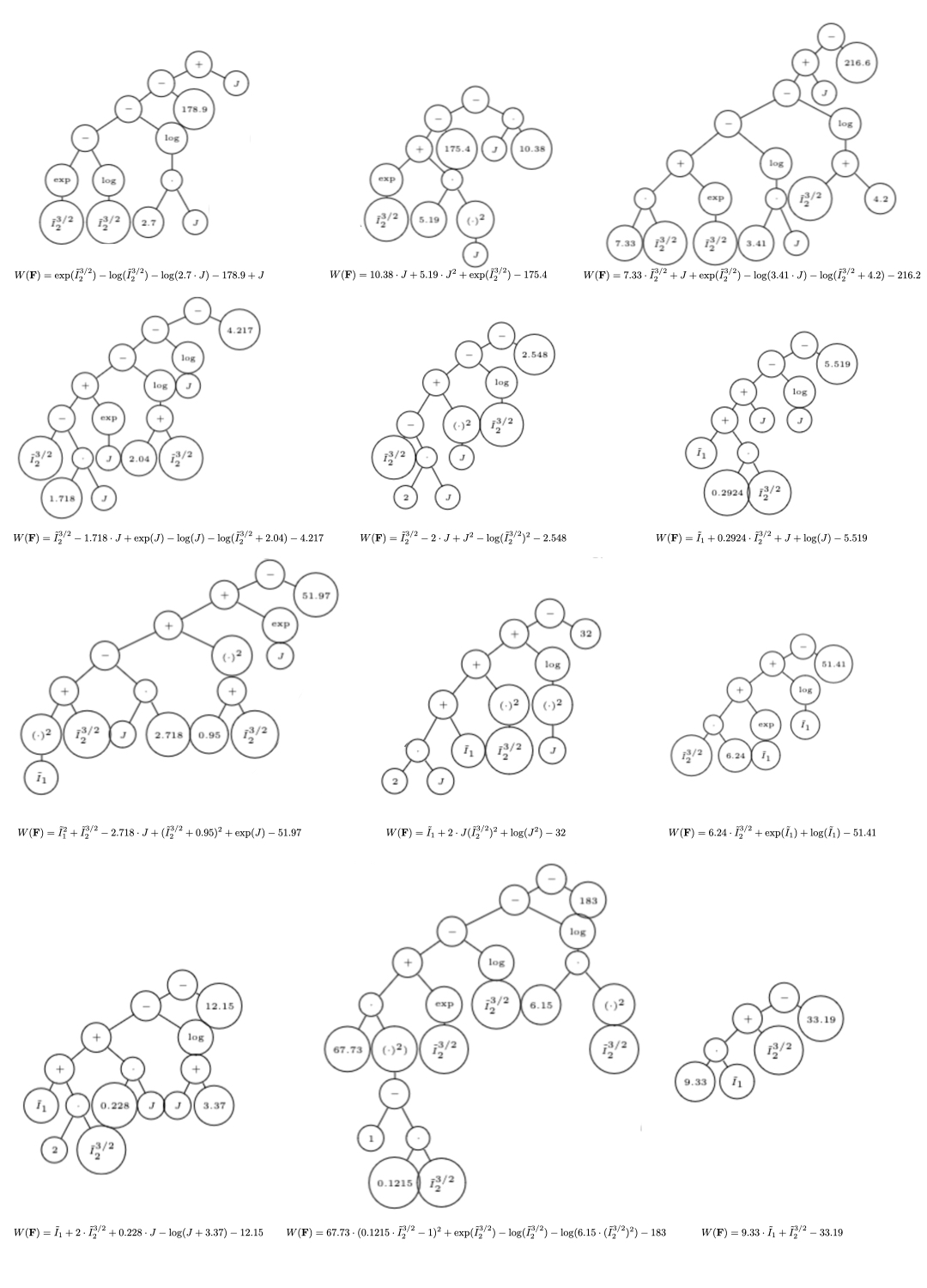}
  \caption{We present different constitutive laws from the language of hyperelastic materials $\curlyL(\curlyG_{\text{HMp}})$. These expressions satisfy all the constitutive law constraints.}
  \label{fig:constitutive laws from grammar}
\end{figure}

Here, $U$ stands for unary, $Y$ for binary, $T$ for combinations of unary and binary operations, $V$ for invariants, and $L$ for both invariants and constants. We also denote with $O$ real numbers, with $D$ integers and with $P$ parts of real numbers. We use the symbol "$|$" to separate rules with the same non-terminal on the left-hand side, for which we consider a uniform probability. As explained in Section \ref{sec:deriving parsimonious expressions}, in this new grammar we limit the left-hand side recursions in the grammar production rules in order to create shallow trees that correspond to parsimonious expressions by introducing non-terminal symbols and operations of predetermined depth. For this reason, we define the grammar rule  $\mathbf{(2)}$ of $\curlyG_{\text{HMp}}$ as $\Psi \rightarrow \Psi^1 + \Psi^2 + \Psi^3 + \Psi^4$ instead of $\Psi \rightarrow \Psi + \Psi$ and introduce the $L, U, Y, T$ to derive terminal nodes in less steps. Therefore, we enforce that the strain energy density is represented by a tree of maximum depth equal to seven. 
We use this grammar to produce syntactically valid expressions of elastic strain energy densities, which also satisfy intrinsic semantic constraints. The full semantic validity of these expressions is then assessed by checking if the growth and the non-negativity conditions hold, see Section \ref{sec:enforcement of constraints}. If these conditions hold, the expression is accepted as a valid constitutive law; if not, a new expression is constructed. Figure \ref{fig:constitutive laws from grammar} illustrates several examples of valid final expressions. 

% \begin{figure} 
% \centering
%     \begin{subfigure}{0.44\textwidth}
%     \centering
%         \includegraphics[width=0.5\textwidth]{Results_Library/example1.png}
%     \caption*{\tiny $\Tilde{I}^{3/2}_2 - 1.718 \cdot J + \exp(J) - \log(J) - \log(\Tilde{I}^{3/2}_2 + 2.04) - 4.217$}
%   \end{subfigure}
%     \begin{subfigure}{0.44\textwidth}
%     \centering
%         \includegraphics[width=0.5\textwidth]{Results_Library/example2.png}
%     \caption*{\tiny $ 67.73 \cdot (0.1215 \cdot \Tilde{I}^{3/2}_2 - 1.0)^2 + \exp(\Tilde{I}^{3/2}_2) - \log(\Tilde{I}^{3/2}_2) - \log(6.15 \cdot (\Tilde{I}^{3/2}_2)^2) - 183.0 $}
%   \end{subfigure}
%     \begin{subfigure}{0.44\textwidth}
%     \centering
%         \includegraphics[width=0.5\textwidth]{Results_Library/example3.png}
%     \caption*{\tiny $4.48 \cdot \Tilde{I}^{3/2}_2 - 148.1 \cdot J + 74.04 \cdot J^2 + (\Tilde{I}^{3/2}_2)^2 + \exp(I1) + 3.678$}
%   \end{subfigure}
%     \begin{subfigure}{0.44\textwidth}
%     \centering
%         \includegraphics[width=0.5\textwidth]{Results_Library/example4.png}
%     \caption*{\tiny $\Tilde{I}_1 + 0.2924 \cdot \Tilde{I}^{3/2}_2 + J - \log(J) - 5.519$}
%   \end{subfigure}
%   \caption{We choose four constitutive law relations out of the ones presented in Figure \ref{fig:constitutive laws from grammar} and perform a finite element simulation for the benchmark boundary value problem in Figure \ref{fig:benchmark}. We present the results for $\delta = 0.3$.}
%     \label{fig:examples from library}
% \end{figure}

\begin{figure}
\centering
        \includegraphics[width=0.8\textwidth]{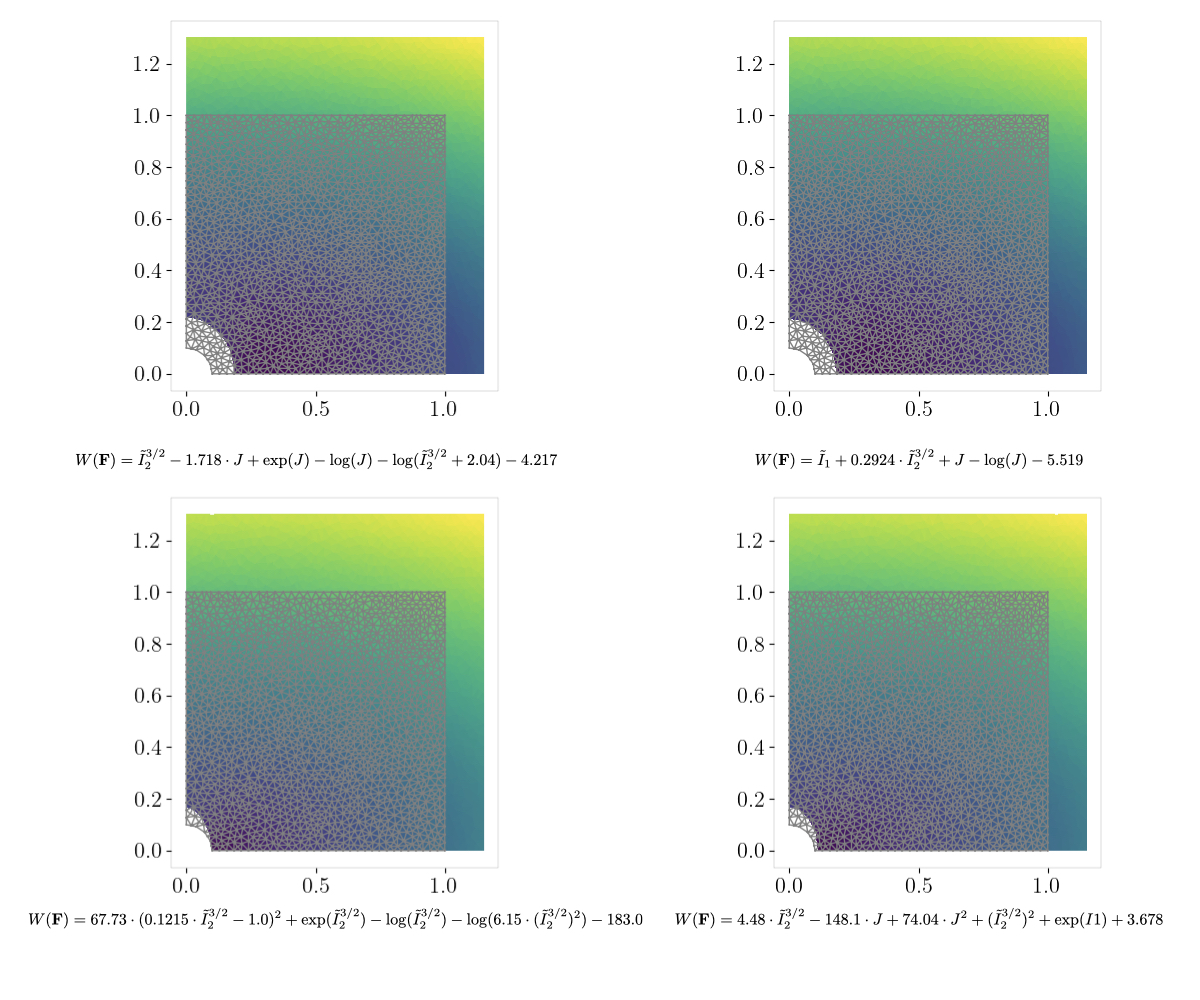}
  \caption{We choose four constitutive law relations out of the ones presented in Figure \ref{fig:constitutive laws from grammar} and perform a finite element simulation for the benchmark boundary value problem in Figure \ref{fig:benchmark}. We present the results for $\delta = 0.3$.}
    \label{fig:examples from library}
\end{figure}

Clearly, the total number of models (or trees) that a library can  potentially contain  depends on the grammar and can be computed using combinatorial calculus. In this work, we do not generate all possible models stemming from $\Hat{\curlyG}_{\text{HM}}$,  but fix upfront a desired number, and stop the generation process right after obtaining a number of valid models equal to the target. 
The automatic generation of the constitutive law library needs to be performed using symbolic computations, i.e. in SymPy \cite{meurer2017sympy}. The wall-clock time for the generation of a tree on a single processor is of order $O(10^{-4})$ seconds computed on an Alienware m16 Laptop with an Intel i9-13900HX CPU and 32GB RAM. In our experience, the probability of a generated tree to be accepted is about $33 \%$; the entirety of the discarded trees violates the growth condition of the strain energy density function, whereas about $33 \%$ violate both the growth and the non-negativity conditions. To further speed up the process we leverage the independence of the generation process for different trees and use multi-threading, whereby each tree is generated by a different core of the CPU in an asynchronous manner. Thus, the total time for a model library generation depends on the number of available CPU units. For example the process of generating $1,000,000$ valid trees, with 100 CPU units available, would take a few minutes. This numbers naturally depend on the chosen number of operations in the grammar rule $\mathbf{(2)}$, and we expect the acceptance rate of a tree to drop as the number increases, so the average time to produce a tree to increase. 

Once a material model library is created automatically, for its practical use it is of fundamental importance to automate the process leading from availability of a constitutive law to solution of a boundary value problem embedding such law. The goal should be to seamlessly integrate the library creation with finite element simulations without having to perform changes to a finite element code for every different strain energy density function. This requires the possibility to use automatic differentiation to derive the stress and the tangent stiffness tensors from the elastic strain energy density \cite{korelc2016automation}, and this possibility is provided by modern finite element tools such as Fenics \cite{LoggEtal2012, LoggWells2010, LoggEtal_10_2012} or AceGEN/AceFEM \cite{korelc2016automation}. In Figure \ref{fig:examples from library}, we present the displacement fields computed by solving the boundary value problem  in Figure \ref{fig:benchmark} with $\delta=0.3$, for four of the hyperelastic constitutive models in Figure \ref{fig:constitutive laws from grammar}, with the Fenics library. We discretize the domain using $1024$ quadratic triangular elements with a three-point Gauss quadrature rule.

The code for the generation of the hyperelastic material model library and the code for solving the boundary value problem in Figure \ref{fig:benchmark} will be made publicly available at the time of publication. 

\section{Data-Driven Discovery of Hyperelastic Constitutive Laws}
\label{sec:data driven constitutive law discovery}

One of the possible motivations for generating the Language of Hyperelastic Materials is to use it within the context of data-driven constitutive model discovery. As discussed already in the introduction, constitutive model identification is typically formulated as an inverse problem, where the unknown parameters in a chosen model are to be inferred on the basis of available experimental information from a tested system. Constitutive model discovery goes one step further, as it integrates model selection with parameter identification \cite{flaschel2023automated}. In Section \ref{sec:related work} we discussed the properties that a symbolic regression method needs to possess in order to be useful for constitutive law discovery. More specifically, we consider only algorithms that allow for generalization and therefore consider a structure composed from a map from the language $\mathcal{L}(\mathcal{G})$ to a latent vector $\mathbb{R}^{n_{\text{VAE}}}$ and another map to $\mathcal{L}(\mathcal{G})$. In this way, the latent space can be searched during the discovery stage for latent vectors that decode to expressions that fit the data. Moreover, the chosen dimensionality reduction method should accept trees as the data structure of the input, as well as, consider an encoder and decoder structure that can be combined with formal grammars. To our knowledge the RTGVAE algorithm is the only dimensionality reduction technique that fits these criteria. We have chosen to work with methods based on grammar due to their advantages: they overcome human bias, since they are not restricted to producing known models or combinations thereof, and the grammar construction imposes semantic and syntactic constraints to the constitutive law expressions. In this section, we aim at showcasing the solution of the inverse data-driven constitutive law discovery problem using a method that features reasonable complexity, accommodates different primitive operations without training instabilities, and generalizes across different materials. For this purpose, we consider a combination of the Recursive Tree Grammar Variational Autoencoder (RTGVAE) method proposed by \cite{paassen2022recursive}, applied to the Language of Hyperelastic Materials formulated in the previous section, and the Covariant Matrix Adaptation Evolutionary Strategy \cite{hansen2003reducing, hansen2001completely} (CMA-ES) for gradient-free optimization. Our purpose in choosing RTGVAE is to show that the proposed methodology can be seamlessly integrated with an already established symbolic regression method. However, in principle any other symbolic regression method could be considered, e.g. a LSPT model, in combination with our grammar-generated hyperelastic constitutive laws. In symbolic regression methods that consider grammar, the latent space is constructed by encoding sequences of grammar rules. Therefore, when performing the data-drive discovery, the models produced in the process of optimization are biased to satisfies the grammar and thus be semantically and syntactically valid. If we considered a LSPT model, for example, the latent space would not be constructed by encoding grammar rules, which means that there would be a higher probability of obtaining invalid expressions during the discovery process.

% In this section, leveraging the grammar-based generation of hyperelastic constitutive laws formulated in the previous section, we propose a framework for discovering constitutive laws from data that possesses all these properties. For this purpose, we consider the methodology based on VAEs on regular trees proposed in \cite{paassen2022recursive}.

% More specifically, its encoder extracts the production rules that derive different trees and encodes them to a reduced space. 
\subsection{Symbolic Regression Using Grammar and Model Discovery}
\label{model_discovery}
We discuss in Sections \ref{sec:related work} and \ref{sec:formal grammars} how trees can be characterized via grammar rules instead of their node labels and connectivity, see \ref{section: appendix regular tree grammars} for more information. In \ref{section: appendix regular tree grammars} we also discuss the operations of extracting a nested list of rules given a tree, called parsing, and deriving a tree given a list of rules, called generation. A \emph{tree VAE} is built on these two operations. More specifically, its encoder extracts the nested list of production rules that derive different trees (inputs) and produces its low-dimensional vectorial encoding. The decoder takes an encoding and provides the most probable nested list of rules that corresponds to the encoding (outputs). The VAE is trained to provide, with high probability, the correct list of rules given a latent vector. In the next paragraphs, we discuss in more detail how the encoder, the decoder and the loss function are defined. 

% For discovery of constitutive laws, we need to search the reduced space of the VAE and decode vectorial representations to find the model that best fits the data. In the next paragraphs we discuss the construction of tree VAEs in more detail. 

\begin{figure}
    \centering
        \includegraphics[width=\textwidth]{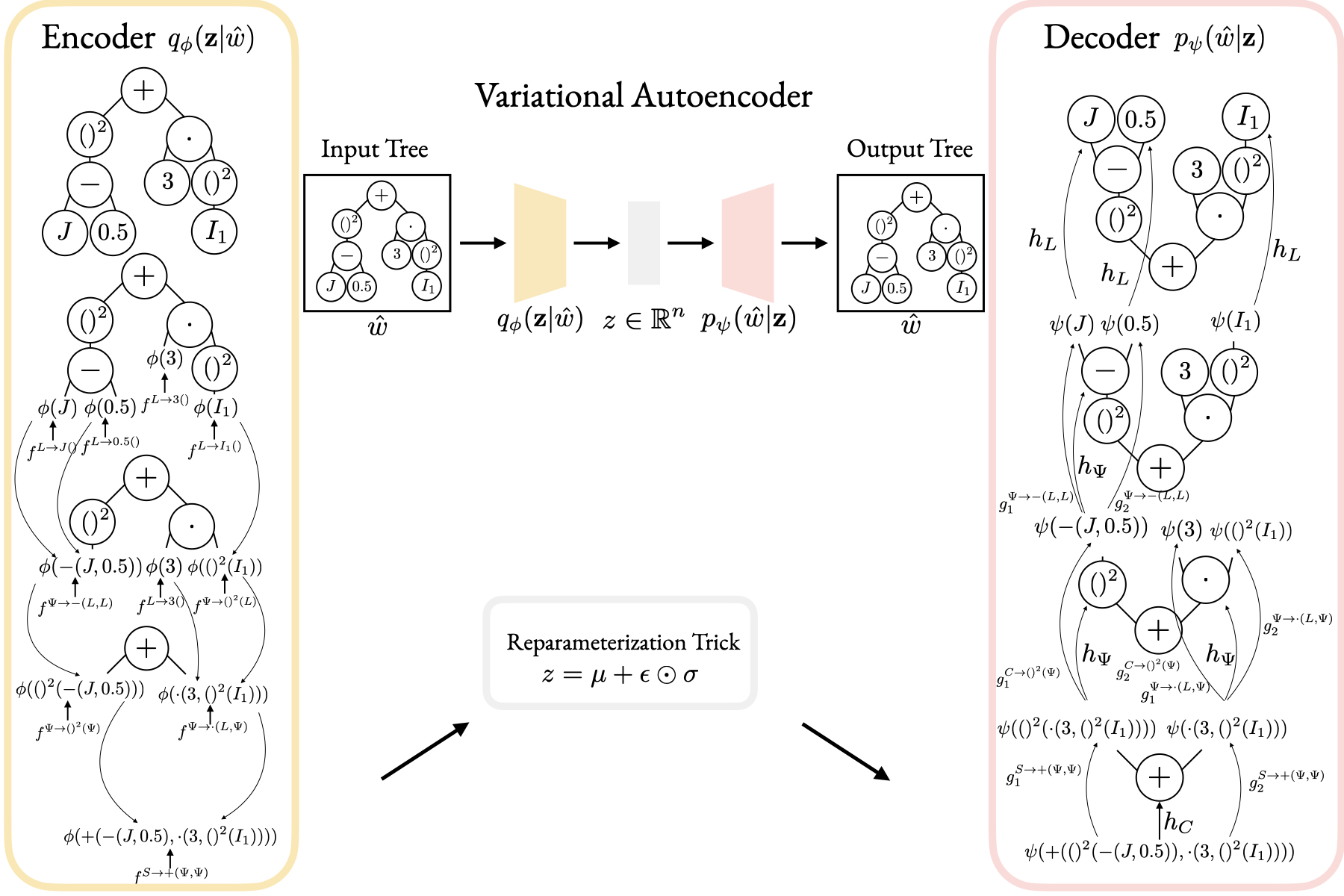}
\caption{A schematic representation of the recursive tree VAE process for the case of $\WF = (J - 0.5)^2 + 3 \cdot \tilde{I}_1^2$. Left: The encoder maps the representation of the children to the representation of the parent using $f^r$, where $r$ is the rule used to produce the children. The process starts from the leaf nodes until it reaches the root and encodes the whole tree. Middle: Perform the reparameterization trick to sample a latent vector $\mathbf{z}$. Right: The decoder generates a tree from a low-dimensional vector $\mathbf{z}$ by first using function $h_{\Psi}$ to decide the next rule to be applied, beginning from the starting symbol, and then function  $g^r_j$ to predict the vectorial representations of the children.}
\label{fig:treeVAE}
\end{figure}

\paragraph{Tree Encoder}The purpose of a tree encoder is to provide a systematic way to represent trees that adhere to a certain set of grammar rules using a low-dimensional vectorial representation. This encoding represents hierarchical or nested lists as a vector, which is a data structure that is easier to handle in computer programs. We explain how the encoder works by comparing encoding to the bottom-up parsing process, as they are closely connected. In parallel to parsing, the encoder assigns a representation to each child and recursively maps the representations of the children to the representation of the parent using a function $f^r$ for each rule $r$. In \cite{paassen2022recursive}, the $f^r$ functions are defined as Recursive Neural Networks \cite{socher2011parsing, tai2015improved,pollack1990recursive} to capture the hierarchical structure of the trees. This process begins from the leaf nodes of the tree and stops when it reaches the root, which means that the whole tree is encoded. Then two functions map the representation of the tree to the mean $\mu$ and the covariance $\sigma$ of the multivariate normal Gaussian distribution that models the latent variables of the VAE. Finally, the so-called reparameterization trick \cite{kingma2013auto} is performed to sample a latent vector of the VAE, see Figure \ref{fig:treeVAE} for a visual explanation of the encoding process. 

% The purpose of a tree encoder is to provide a systematic way of representing trees that adhere to a certain set of grammar rules using a low-dimensional vectorial representation. This encoding represents hierarchical or nested lists as a sequence, which is a data structure that is easier to handle in computer programs. A simple way to perform such encoding would be to list the nodes of a tree in an top-to-bottom - left-to-right ordering, assign one-hot vectors, dummy variable vectors \cite{draper1998applied}, to the node labels, and use a neural network architecture to encode this representation \cite{kusner2017grammar, dai2018syntaxdirected}. However, representing trees as lists can introduce long dependencies that can be hard to handle \cite{paassen2022recursive}. Considering that we work with RTGs, encoding trees via recursive neural networks \cite{socher2011parsing, tai2015improved,pollack1990recursive} to capture their hierarchical structure feels more natural. We explain how the encoder works by comparing it to the parsing processes, as they are closely connected.

More precisely, the tree encoding is defined as a function $\phi: \curlyL(\hat{\curlyG}) \rightarrow \mathbb{R}^{n_{\text{VAE}}}$ from a language $\curlyL(\hat{\curlyG})$ to a $n_{\text{VAE}}$-dimensional space. Consider a grammar rule of the form $\Psi \rightarrow s(L_1, ... L_k)$ that re-writes a non-terminal $\Psi$ with a $k$-ary operation and produces $L_1, ..., L_k$ children non-terminals. For each grammar rule, we introduce a function $f^r: \mathbb{R}^{k \times n} \rightarrow \mathbb{R}^n$ (with $n$ as a user-defined integer) that maps the representation of the $k$ children to the representation of the parent. If the rule re-writes the non-terminal with a nullary operation, i.e. a constant, then $f^r$ is a constant vector of size $n$. In practice $f^r$ is defined as a fully connected neural network with a single layer:

\begin{equation}
    f^r(t_1, ..., t_k) = \text{tanh} (U^{r,j} t_j + b^r), 
\end{equation} 
where $U^{r,j} \in \mathbb{R}^{n \times n}$ are the weights and $b \in \mathbb{R}^n$ the biases of the neural network. To obtain a representation of the whole tree, $f^r$ is applied during parsing. Therefore, the tree encoder is defined as a recursive equation of the form:

\begin{equation}
    \label{eq: recursive functions}
    \phi(s(\hat{t}_1, ... \hat{t}_k),\Psi) = f^r (\phi(\hat{t}_1, L_1), ..., \phi(\hat{t}_k, L_k) ),
\end{equation}
where $r \in R$ is the rule that derives $s(\hat{t}_1, ... \hat{t}_k)$ from $\Psi$. For the VAE, the encoding probability $q_{\phi}(\mathbf{z}| \hat{w})$, where $\hat{w}$ the derivation tree of an expression, is the Gaussian with mean $\mu: \mathbb{R}^{n} \rightarrow \mathbb{R}^{n_{\text{VAE}}}$ and the diagonal covariance $\sigma: \mathbb{R}^n \rightarrow \mathbb{R}^{n_{\text{VAE}}}$. Both $\mu$ and $\sigma$ are defined as linear fully connected networks with trainable parameters. The mean and the covariance are then used to perform the reparameterization $\mathbf{z} = \mathbf{\mu} + \mathbf{\epsilon} \odot \mathbf{\sigma}$, where $\epsilon \sim \mathcal{N}(0, \alpha \cdot \I)$ and $\alpha$ positive real number that scales the covariance of $\epsilon$. When the encoding finishes, the entire tree is represented by a vector $\mathbf{z} \in \mathbb{R}^{n_{\text{VAE}}}$.

\paragraph{Tree Decoder}
The purpose of a tree decoder is to provide a process to generate trees conditioned on a low-dimensional vectorial representation. Even though the decoder and the generation process discussed in \ref{section: appendix regular tree grammars} both work in a top-down manner, initiating from the starting symbol, they present a notable difference. For each step $l$ of the generation process, the generation algorithm recovers a non-terminal $\Psi$ from the tail of the list of non-terminals, applies the rule $r_l = \Psi \rightarrow s(L_1, ... L_k)$, removes $\Psi$ from the list and stores $L_1, ... L_k$, see \ref{section: appendix regular tree grammars}. In contrast to the generation process, the decoding process does not take a \emph{list of grammar rules} as an input but a \emph{vectorial encoding} $\mathbf{z}$. For this reason the rule $r_l$ to be applied at the $l$-th step is not known a priori, but needs to be chosen during the tree generation.

The tree decoding is defined as the opposite process of the encoding, or more precisely, as a function $\psi: \mathbb{R}^{n_{\text{VAE}}} \rightarrow \curlyL(\hat{\curlyG})$ from a $n_{\text{VAE}}$-dimensional space to the language $\curlyL(\hat{\curlyG})$. Mirroring the encoder, the first step of the decoding process is to lift the dimensions of the vector $\mathbf{z} \in \mathbb{R}^{n_{\text{VAE}}}$ using a linear fully connected network $\rho: \mathbb{R}^{n_{\text{VAE}}} \rightarrow \mathbb{R}^{n}$ to obtain the vector  $\mathbf{s} = \rho(\mathbf{z})$ that represents the entire tree. Then, at each step $l$ of the decoding process, the algorithm takes a vector $\mathbf{s}$ and a non-terminal $\Psi$ as inputs. The vectorial representation $\mathbf{s}$ is used to choose which rule $r_l$ will be applied as follows. Let $N_{\Psi}$ be the number of rules with $\Psi$ on their left-hand side. A linear fully connected network $h_\Psi: \mathbb{R}^n \rightarrow \mathbb{R}^{N_{\Psi}}$ is employed to obtain a score vector $\mathbf{\lambda} = h_\Psi(\mathbf{s})$ for each non-terminal $\Psi$. Then we choose a rule $r_l$, with $l \in [1,...,N_\Psi]$, based on the probability:
\begin{equation}
    p_{\Psi} (r_l | \mathbf{z}) = \frac{ \exp (\lambda_l) } {\sum^{N_\Psi}_{i=1} \exp(\lambda_i)}.
\end{equation}
For choosing the rules to be applied for non-terminals $L_1, ... L_k$, the representation of the parent needs to be mapped to the representation of the children. For this purpose, a neural network is defined with $k$ layers $g^r_1, ..., g^r_k$,  for each rule $r = \Psi \rightarrow s(L_1, ..., L_k)$, to map the vector representation $\mathbf{t}$ of each child, where $t_j = g_j^r(\mathbf{s})$ for $j \in [1, ...,k]$. At each step of the decoder, the tree representation $\mathbf{s}$ is updated by removing the representation at step $l$ from the overall tree representation, i.e. $\mathbf{s}$ is substituted by $\mathbf{s} - \mathbf{t}_l$. We repeat this process until no non-terminal symbol is left. The decoding function is then written in a recursive manner as follows:

\begin{equation}
    \psi(\mathbf{z}, \Psi) = s( \psi (\mathbf{t}_1, L_1) , ...., \psi(\mathbf{t}_k, L_k)).
\end{equation}
Using $\psi$ for $\Psi = C$ where $C$ is the starting symbol, we recursively generate a tree $\hat{w}$ given a vector $\mathbf{z}$. Even though the decoder learns probabilities of production rules conditioned to a library of valid material laws, which means that it is strongly biased to produce valid constitutive laws, it is not explicitly constrained to always satisfy the constitutive law constraints. This means that there exists a probability that the discovery process results to an expression that is not semantically valid.

\paragraph{Loss Function}The VAE introduces a probability density function $q_{\phi}(\mathbf{z} | \hat{w})$, parameterized by the encoder, and the conditional probability $p_{\psi}(\hat{w} | \mathbf{z})$ of decoding the input using the latent vector $\mathbf{z}$. The autoencoder is trained to minimize the loss:

\begin{equation}
\label{eq:training loss}
    \mathcal{L}(\phi, \psi) = \sum_{i=1}^m \mathbb{E}_{q_{\phi}(z_i|\hat{w}_i)} [  - \log(p_{\psi}(\hat{w}_i | \mathbf{z}))  ]  + \beta \mathcal{D}_{KL} (q_{\phi}|| \mathcal{N}),
\end{equation}
where $\mathcal{D}_{KL} (q_{\phi}|| \mathcal{N})$ is the Kullback–Leibler divergence between the distribution parameterized by the encoder and the standard normal distribution $\mathcal{N}$ and $\beta$ is a user-defined parameter that controls the effect of this term on the overall loss \cite{burda2015importance}. Eq. \eqref{eq:training loss} trains the model to maximize the probability of choosing the correct rule for a step in the decoding process given that all the choices for the previous steps have been correct \cite{paassen2022recursive}.

\paragraph{Inverse Problem Solution}The data-driven discovery stage of the process consists of searching the reduced space of the VAE for an encoding, which when decoded provides a model whose evaluation  matches well the measured data. At this stage, we do not focus on the optimal solution of this problem, but rather opt for use of an evlutionary approach, which is a popular choice for the solution of inverse formulations linked to multi-step model evaluation processes \cite{Agathos2018}. We consider a global instead of a gradient based optimization method, because we expect the landscape of the latent space to present multiple local minima. The data-driven discovery is performed on displacement measurements from only one type of loading (Biaxial Tension), which is not necessarily information to constrain the discovery to one material model, thus a global minimum. In this case, we opt for the adoption of a covariance matrix adaptation evolutionary strategy (CMA-ES) \cite{hansen2003reducing, hansen2001completely} for generating candidate solutions from a probability  distribution whose parameters are iteratively updated. The CMA-ES method fits naturally for performing global optimization in the latent space of VAEs, as they both consider a multivariate normal distribution over the latent vectors $\mathbf{z}$. A difference lies in the fact that the CMA-ES uses a full covariance matrix during the adaptation process, while the VAE considers a diagonal covariance matrix. For measuring the goodness of fit, a root mean square error metric is adopted reflecting the discrepancy between the model evaluation and the available measurements. The CMA-ES reflects one of several possible optimization tools one can opt from the class of evolutionary optimization tools, which allow for a more flexible formulation of the objective function. We opt for use of the CMA-ES as it comprises a stronger mathematical foundation, based on the maximum-likelihood principle, in comparison to metaheuristic schemes, such as Genetic Algorithms \cite{grandidier2006identification} and Particle Swarm Optimization \cite{hardt2021application}. 

%%%%%%%%%%%%%%%%%%%%%%%%%%%%%%%%%%%%%%%%%%%%%%%%%%%%%%%%%%%%%%%%%%%%%%%%%
% \section{The Language of Hyperelastic Materials: Use Cases}
% % \section{Examples}
% \label{sec:the language of hyperelastic materials use cases}

\subsection{Constitutive Law Discovery Based on Data}
\label{sec:constitutive law discovery based on data}

In this section, we demonstrate how the proposed grammar-based approach can be used to perform efficient data-driven model discovery, thus covering the spectrum of tasks in Figure \ref{fig:master_figure_lhm}. The model discovery is \emph{geometry and load agnostic}, which means that the same process without any change would work for any geometry, loading or boundary condition. We consider a simple RTG and generate a language for hyperelastic materials, which we then use to train a RTGVAE \cite{paassen2022recursive}. As a final step of the process we employ the CMA-ES \cite{hansen2003reducing, hansen2001completely} to discover the model that best fits synthetic data that are contaminated with different levels of noise.  In real applications, these measurements may come from experiments or from higher fidelity (e.g., multiscale) finite element computations. 

We consider two different setups for the discovery process, one based on supervised learning, and one based on unsupervised learning. We denote as \textit{supervised learning} the case in which the experimental values of the elastic strain energy density are assumed to be known. This is an unrealistic assumption since the elastic strain energy density cannot be measured experimentally. However, this setup is initially used as it is the simplest option and it is able to demonstrate the useful properties of the grammar-based model discovery approach. One such property is that when the baseline model cannot be represented using the grammar rules defined in the grammar, the discovery process recovers a constitutive law expression that represents similar physics. We demonstrate this property by evaluating the generalization performance of the discovered expression to different loading conditions.  Another useful property of the approach is that it is robust to noise in the displacement data. We demonstrate this property by introducing different noise levels in the measurements. To combine these useful properties with a realistic loss function, we adopt the \textit{unsupervised learning} approach, i.e. we assume that only displacement fields and the related values of global applied force are available from experimental measurements, which is a realistic scenario as discussed in \cite{flaschel2021unsupervised}. In the unsupervised case, we demonstrate that the proposed approach retains the properties of robustness and generalization and provides accurate results. For both discovery scenarios we consider one loading step in the discovery process for $\delta=0.3$.

In the cases of EQL and SR, see  Section \ref{sec:related work}, the data-driven discovery process is designed such that the learning algorithm chooses the combination that best fits the data out of a library of primitives, expressions or operations. The underlying assumption of these methods is that the constitutive laws can be described by a weighted sum of an priori chosen basis. The selection of such a basis biases the discovery process to recover combinations of already known relations and it is conditioned by the expertise of the person designing the library. However, in real applications the form of the constitutive law is not always known a priori such that an informed choice of the basis can be made. For such cases, a weighted sum of the library elements may not accurately describe the constitutive law, because the library construction may be missing important information. With our grammar-based approach, on one hand we can fully automatically generate a much wider library of material models than possible by manual construction, as already demonstrated in Section \ref{Library_generation}; on the other hand, we can discover an interpretable material model which well interprets the data even though the true model underlying the data is not present in the library. While the capability of finding a good approximation for a model not present in the library was already demonstrated with SR \cite{flaschel2021unsupervised}, the final model still had to be a combination of the library terms. This is no longer the case in the present setting, as we demonstrate in this section.

%we are interested in demonstrating discovery for a scenario where such information, which is rooted in a priori knowledge, is missing. To this end, we define a constitutive law library that is generated so that it is missing information on the complexity of the models we seek and we explore the capability of the proposed symbolic regression pipeline to discover models that have similar physical behavior. 

\paragraph{Data Generation}For generating the data that we are going to use to perform the discovery, we consider the benchmark proposed in \cite{flaschel2021unsupervised, thakolkaran2022nn} that emulates digital image correlation measurements using computational data generated by the finite element method. The domain is a plate with a central hole which undergoes a symmetric bi-axial loading controlled by a displacement parameter $\delta$. Due to symmetry, only a quarter of the plate is studied with symmetry boundary conditions on the left and bottom boundaries \cite{flaschel2021unsupervised}, see Figure \ref{fig:benchmark}.
Plane-strain conditions are assumed. In the data generation, we consider four different (baseline) hyperelastic constitutive models: 

\begin{itemize}
    \item A Neo-Hookean (NH) model containing a quadratic volumetric term
     \begin{equation*}
        W = 0.5\cdot (\tilde{I}_1 - 3) + 1.5 \cdot (J -1)^2;
     \end{equation*}
    \item An Isihara (IS) model \cite{isihara1951statistical} containing a quadratic deviatoric term
    \begin{equation*}
        W = 0.5\cdot (\tilde{I}_1 - 3) + (\tilde{I}_2 - 3) + (\tilde{I}_1 - 3)^2 + 1.5 \cdot (J -1)^2;
    \end{equation*}
    \item A Haines-Wilson (HW) model \cite{haines1979strain} containing a cubic deviatoric term
    \begin{equation*}
        W = 0.5 \cdot (\tilde{I}_1-3)+(\tilde{I}_2-3)+0.7 \cdot (\tilde{I}_1-3) \cdot (\tilde{I}_2-3) + 0.2 \cdot (\tilde{I}_1-3)^3 + 1.5 \cdot (J-1)^2;
    \end{equation*}
    \item A Gent-Thomas (GT) model \cite{gent1958forms} containing a logarithmic deviatoric term 
    \begin{equation*}
        W = 0.5 \cdot ( \tilde{I}_1-3) + \log(\tilde{I}_2/3)  + 1.5 \cdot (J-1)^2,
    \end{equation*}
    \item An Ogden (OG) model \cite{ogden1972large} depending on the principal stretches $\lambda_i$
    \begin{equation*}
        W = \frac{\mu}{\eta} ( \lambda_1^\eta + \lambda_2^\eta  + \lambda_3^\eta - 3),
    \end{equation*}with $\mu = \eta = 1.3$.
\end{itemize}
We set the applied displacement parameter to $\delta=0.3$ and, for each constitutive model, we obtain the displacement field $\mathbf{u}$ which we then use to compute the deformation gradient $\mathbf{F}$ and the volumetric and isochoric invariants of the right Cauchy-Green deformation tensor $J, \tilde{I}_1, \tilde{I}_2$. We add artificial noise to the displacement data, as follows:

\begin{equation}
     \tilde{u}^a_i = u^a_i + e^a_i, \quad e^a_i \sim \mathcal{N}(0, \sigma^*)\quad \forall i=1,2, \forall a=1,\dots,n_{dofs},
\end{equation}
with $n_{dofs}$ as the number of degrees of freedom in the finite element discretization. We adopt two levels of noise, namely $\sigma^* =10^{-4}$, and $\sigma^* =10^{-3}$, as in \cite{flaschel2021unsupervised}. Denoising of the resulting displacement data $\tilde{\bf{u}}$ is performed as in \cite{thakolkaran2022nn}.
%We perform no denoising on the resulting displacement data $\tilde{\bf{u}}$. 

\paragraph{Choosing the Grammar}For the reasons explained in Section \ref{sec: Properties of tree grammars}, we adopt a RTG for the data-driven discovery of the material model. With respect to the grammars we used so far (e.g. the CFG  $\curlyG_{\text{HMp}}$ in eq. \eqref{eq: hyperelastic grammar for library generation}), we introduce some simplifications; namely, we consider the terminals $J-1$,  $\tilde{I}_1-3$, and $\tilde{I}_2-3$ to promote expressions where the normalization holds a priori. However the normalization conditions are not strictly enforced, as the final expression could contain operations that violate them, e.g. additions between scalars. Moreover, we rewrite $C$ as $\Psi^1 + \Psi^2 + \Psi^3$ to derive final expressions that contain a smaller number of operations such that the grammar derives parsimonious expressions. The designed RTG, $\hat{\curlyG}_{\text{HMs}}$, is defined as follows:

\begin{equation}
\NORMAL{
\begin{split}
    \label{eq:hyperelastic grammar for discovery}
    S = \{ & \ C   \ \}, \\
    \tilde{\Sigma} = \{ & \  +:2, \   /:2, \  \cdot:2, (\cdot)^2:1, \  (\cdot)^3:1, \  \log:1, \ (\tilde{I}_1 - 3):0, \  (\tilde{I}_2-3):0, \\ &  (J-1):0, \  0.2 : 0, \  0.5:0, \  0.7:0, \   1.5 : 0, \   2: 0, \ 3 :0 \}, \\
         R  = \{ & C  \rightarrow \  \Psi^1 +    \Psi^{2}  +    \Psi^{3} , \\
         & \Psi^{1}  \rightarrow  L | U | Y | W | E, \\
         & \Psi^{2}  \rightarrow  L |  U | Y | W | E, \\
         & \Psi^{3}  \rightarrow   L |  U | Y | W | E , \\
        & Y  \rightarrow \   +( L , L ) \  | \  /( L ,  L ) \  | \  \cdot ( L , L ) \ , \\
        & U  \rightarrow \log( L ), \\
       & W  \rightarrow \ ()^2( Y ) \  | \  ()^3( Y )  , \\
        & E  \rightarrow \ \cdot (Y, Y), \\
        & L  \rightarrow  V  \ |  O , \\
         & V  \rightarrow  (\tilde{I}_1 - 3) () \  |   \ (\tilde{I}_2-3) () \  | \  (J-1) (), \\ 
           & O  \rightarrow \ 0.2 () \ | \ 0.5 () \ | \ 0.7 () \ | \ 1.5 ()  \}, \\
    \Phi = \{ \  & C,  \Psi^1 ,  \Psi^{2} ,  \Psi^3 ,   Y  ,  U ,  E, W ,  L ,  V ,  O       \ \}.
    \end{split}
    }
\end{equation}
In eq. \eqref{eq:hyperelastic grammar for discovery}, as done previously, we use the symbol "$|$"  to separate grammar rules with the same left-hand side. The grammar $\hat{\curlyG}_{\text{HMs}}$ contains additions and multiplications between terminals and thus it can potentially derive the NH model. However, the $\log(\Tilde{I}_2 /3)$ operation of the GT model or additional $+$ operations for the IS and HW models cannot be used when deriving the library of constitutive laws, hence these constitutive laws cannot be recovered exactly. Thus, for the data obtained using these laws our goal is to discover parsimonious expressions that deliver a physical behavior similar to the baseline using the available grammar $\hat{\curlyG}_{\text{HMs}}$. Moreover, for the case of the OG model the principal stretches  are not included in the grammar. Therefore, the symbolic regression pipeline is forced to discover a surrogate that performs closely to the OG model using the invariants.

\paragraph{Training Setup}We derive expressions from the language $\curlyL(\hat{\curlyG}_{\text{HMs}})$ and construct a library of $N_{\text{train}}=600,000$ valid constitutive models for different hyperelastic materials. We then train the recursive tree VAE model on this library. For the VAE, we set $\beta=0.0001$ in eq. \eqref{eq:training loss} and $\alpha = 0.0001$ for the reparameterization trick, see Section \ref{model_discovery}. We choose the latent dimension of the VAE as $n=8$, the learning rate as $0.001$ and the width of the recursive encoder and decoder as $128$. We check the performance of the model against $N_{\text{test}} = 100$ test expressions using a root mean tree edit distance metric \cite{zhang1989simple}.

\paragraph{Data-driven Discovery} Once we train the model on the library, we perform gradient-free optimization on the latent space of the RTGVAE to discover different hyperelastic material models using the CMA-ES method. For  CMA-ES in the supervised case we consider the number of new solutions per iteration as $1,000$ trees, a zero mean, a $0.1$ variance and $10$ maximum iterations per adaptation. In the unsupervised case, we consider $100$ new solutions per iteration, and $15$ maximum iterations per adaptation. 

In the supervised case, we assess the goodness of fit  using a root mean square error metric:

\begin{equation}
    \text{RMSE}_{W} = \sqrt{ \frac{1}{N_n} \sum_{i=1}^{N_n} (W_i - \Tilde{W}_i )^2},
\end{equation}
where $W_i$ and $\Tilde{W}_i$ are respectively the baseline and the predicted strain energies for node $i$, and $N_n=2,752$ is the number of nodes in the non-uniform finite element mesh used for data generation.  The baseline strain energy density is computed based on the available  dense (full-field) displacement measurements using the baseline expression of the constitutive law. The computation of the root mean square error is performed using the SymEngine \cite{sympy2016symengine} SymPy interface to decrease the wall-clock time. 
For the assessment of the prediction accuracy, we employ the relative $\mathcal{L}^2$ error metric referred to the entire domain:

\begin{equation*}
    \text{Relative}\, \mathcal{L}^2\,\text{Error} =  \frac{\| W - \Tilde{W} \|_2}{\| W \|_2}
\end{equation*}
and analogous for other quantities, such as the first component of the first Piola-Kirchhoff stress tensor. 

For the unsupervised learning scenario, we realistically assume the displacement field and the reaction force to be known and we compensate for the unknown baseline strain energy densities by enforcing the balance of linear momentum, both locally in the interior of the domain and globally on the loaded boundary, through the loss function in \cite{thakolkaran2022nn}, see \cite{thakolkaran2022nn} for more details. We additionally apply a penalty term to approximately satisfy the stress normalization conditions.

In Table \ref{tab:efficiency}, we report on the wall-clock time to run one iteration of the CMA-ES method with respect to the new solutions generated at each adaptation step on an Alienware m16 Laptop with an Intel i9-13900HX CPU and 32GB RAM. 

\begin{table}[ht!]
    \centering
    \begin{tabular}{|c|c|c|c|c|}
    \hline
          New solutions per iteration & $100$ & $500$ &  $1,000$ &  $5,000$ \\
         \hline
         Supervised Case: Wall-clock time in sec  & 1  & 5 & 11 & 60\\
         Unsupervised Case: Wall-clock time in sec  & 60  & 600 & 1300 & 7000\\
         \hline
    \end{tabular}
    \caption{The approximate wall-clock time in seconds of the discovery process for different numbers of new solution per iteration that are used for the covariance adaptation.}
    \label{tab:efficiency}
\end{table}

\paragraph{Assessing the Discovery Process for the NH Model}First, we test the performance of the symbolic regression algorithm in discovering the NH model from displacement data with  $\sigma^*=0, 10^{-4}, 10^{-3}$. The best discovered expressions as well as the relative $\mathcal{L}^2$ errors between the baseline and the predicted strain energy density for both the supervised and the unsupervised cases are presented in Table \ref{tab:nh_l2_error} for different noise levels. Being contained in the language generated by the starting grammar, the NH model is discovered \emph{exactly} for all cases in the unsupervised setting. For the supervised setting, it is  discovered \emph{exactly} for the noise-free case and for the noisy case with $\sigma^*=10^{-4}$.  For the $\sigma^*=10^{-3}$ case, a very accurate approximation is discovered, with a relative $\mathcal{L}^2$ error of $1.3 \%$. 

\begin{table}[ht!]
    \centering
    \begin{tabular}{|c|c|c|}
    \hline
            & {\small Supervised Case}   &  \\
         \hline
         {\small Model, Noise Level} &  {\small Discovered Expression} &   {\small Relative $\mathcal{L}^2$ Error} \\
         \hline
        { \small NH, $\sigma^*$ = $0$} & { \small $\Tilde{W} = 0.5 \cdot (I_1 -3 ) + 1.5 \cdot (J-1)^2$ }  &  {\small 0 } \\
        { \small NH, $\sigma^*$ = $10^{-4}$} & { \small $\Tilde{W} = 0.5 \cdot (I_1 -3 ) + 1.5 \cdot (J-1)^2$ }   & {\small 0 } \\
        { \small NH, $\sigma^*$ = $10^{-3}$} & { \small $\Tilde{W} = 0.49 \cdot (I_1 -3 ) + 1.5 \cdot (J-1)^2$  } & {\small $0.013$ } \\
         \hline
                 & {\small Unsupervised Case}   &  \\
         \hline
         {\small Model, Noise Level} &  {\small Discovered Expression} &   {\small Relative $\mathcal{L}^2$ Error} \\
         \hline
        { \small NH, $\sigma^*$ = $0$} & { \small $\Tilde{W} = 0.5 \cdot (I_1 -3 ) + 1.5 \cdot (J-1)^2$ }  &  {\small $0$ } \\
        { \small NH, $\sigma^*$ = $10^{-4}$} & { \small $\Tilde{W} = 0.5 \cdot (I_1 -3 ) + 1.5 \cdot (J-1)^2$ }   & {\small $0$ } \\
        { \small NH, $\sigma^*$ = $10^{-3}$} & { \small $\Tilde{W} = 0.5 \cdot (I_1 -3 ) + 1.5 \cdot (J-1)^2$  } & {\small $0$ } \\
         \hline
    \end{tabular}
    \caption{Best discovered expressions and relative $\mathcal{L}^2$ errors between the best predicted and the baseline strain energy density for the NH model, obtained from data with different noise levels for both the supervised and the unsupervised cases}.
    \label{tab:nh_l2_error}
\end{table}

\paragraph{Assessing the Discovery Process for the IS, HW, GT  and OG Models} We then run the discovery algorithm for each of the IS, HW,  GT  and OG  models separately. Since none of these models is contained in the language generated by the grammar in use, we do not expect an exact discovery. The best discovered models for the noise-free, $\sigma^*=10^{-4}$, and $\sigma^*=10^{-3}$ noise cases as well as the relative $\mathcal{L}^2$ errors between the baseline and the predicted strain energy density are presented in Table \ref{tab:models_l2_error}. We observe that for the IS and the HW models the relative $\mathcal{L}^2$ error is less than $1.5 \%$ for all noise levels irrespective of supervision, while for the GT model the error increases to $4.2 \%$ for the $10^{-3}$ noise level  in the supervised case and to $4.7 \%$ in the unsupervised case. For the OG model the relative error is between 2 and 3 $\%$ for all cases in the supervised and $4.7 \%$ in the unsupervised cases. 

We note that the models discovered from the IS data are very close to the baseline ones. While in the grammar \eqref{eq:hyperelastic grammar for discovery} we defined the constitutive law as $\Psi^{1} + \Psi^{2} + \Psi^{3}$, which means that constructed expressions in the library contain two binary $+$ operations, the discovery process returns an expression containing three binary operations; this indicates that the decoding process can predict expressions that are not limited to the strict structure of those contained within the library. Thus, interestingly, the discovered models  for the noiseless and the lowest-noise case have the same structure of the baseline model, but different material constants. We attribute the discrepancy in the constants to inaccuracies of the symbolic regression algorithm and the inability of the CMA-ES method to determine the global minimum of the optimization problem. 
Also interestingly, the predicted expression for the HW model is less complicated than the baseline and yet it achieves an excellent accuracy, with an error less than $1 \%$ independently of the noise level. 
In the supervised case, the GT discovered expression  is a linearized version of the baseline, corresponding to a generalized Mooney-Rivlin model, which is reasonable for the relatively limited amount of deformation included in the data (a similar result is reported in \cite{flaschel2021unsupervised}). Interestingly, also the discovered constitutive laws for the OG model have the same structure of the linearized GT (or generalized Mooney-Rivlin) model, only with different constants. In the unsupervised case, the process discovers the exact same model for both the OG and the GT baselines. These results further reflect the dependence of the inference process on the richness of the available dataset, including the amount of non-linearity that is observed. 

% In Figure \ref{fig:contours methods high noise}, we also present the strain energy obtained by evaluating the baseline and the predicted models for all cases and noise level $\sigma^* = 10^{-3}$ and their relative error. 

% \begin{figure}[!ht]
%   \begin{subfigure}{0.5\textwidth}
%     \centering
%     \includegraphics[width=\linewidth]{Contour_Plots_Strain_Energy/NH_W_1e3_contour.png}
%   \end{subfigure}%
%   \begin{subfigure}{0.5\textwidth}
%     \centering
%     \includegraphics[width=\linewidth]{Contour_Plots_Strain_Energy/IS_W_1e3_contour.png}
%   \end{subfigure}
%   \begin{subfigure}{0.5\textwidth}
%     \centering
%     \includegraphics[width=\linewidth]{Contour_Plots_Strain_Energy/HW_W_1e3_contour.png}
%   \end{subfigure}%
%   \begin{subfigure}{0.5\textwidth}
%     \centering
%     \includegraphics[width=\linewidth]{Contour_Plots_Strain_Energy/GT_W_1e3_contour.png}
%   \end{subfigure}
%   \caption{A comparison of the strain energy obtained by evaluating the baseline (Left) and predicted (Center) models for all case and noise level $\sigma^* = 10^{-3}$, as well as, their absolute error (Right).}
%   \label{fig:contours methods high noise}
% \end{figure}

\begin{table}[ht!]
    \centering
    \resizebox{\columnwidth}{!}{%
    \begin{tabular}{|c|c|c|}
    \hline
              &  {\small Supervised Case} &    \\
         \hline         
         {\small Model, Noise Level} &  {\small Discovered Expression} &    {\small Relative $\mathcal{L}^2$ Error} \\
         \hline
        { \small IS, $\sigma^*$ = $0$} & { \small $\Tilde{W} = \Tilde{I}_1 - 3 + 0.5\cdot(\Tilde{I}_2 - 3) + (\Tilde{I}_1 - 3)^2 + 1.5\cdot(J - 1)^2$ }  & {\small  0.011 } \\
        { \small IS, $\sigma^*$ = $10^{-4}$} & { \small $\Tilde{W} = \Tilde{I}_1 - 3 + 0.5\cdot(\Tilde{I}_2 - 3) + (\Tilde{I}_1 - 3)^2 + 1.5\cdot(J - 1)^2$ } & {\small 0.011  } \\
        { \small IS, $\sigma^*$ = $10^{-3}$} & { \small  $W = (\Tilde{I}_1 - 3)\cdot(\Tilde{I}_2 - 3) + 1.5\cdot(\Tilde{I}_1 - 3) + 1.5\cdot(J - 1)^2$ }  & {\small 0.013 } \\
         \hline
        { \small HW, $\sigma^*$ = $0$} & { \small $\Tilde{W} = 1.5\cdot(\Tilde{I}_2 - 3) + (\Tilde{I}_1 - 3)^2 + 1.5\cdot(-1 + J)^2$  }  & {\small 0.01 } \\
        { \small HW, $\sigma^*$ = $10^{-4}$} & { \small $\Tilde{W} = (\Tilde{I}_1 - 3)\cdot(\Tilde{I}_2 - 3) + 1.5\cdot(\Tilde{I}_2 - 3) + 1.5\cdot(-1 + J)^2$ }  & {\small 0.007 } \\
        { \small HW, $\sigma^*$ = $10^{-3}$} & {  \small $\Tilde{W} = (\Tilde{I}_1 - 3)\cdot(\Tilde{I}_2 - 3) + 1.5\cdot(\Tilde{I}_2 - 3) + 1.5\cdot(-1 + J)^2$ }  & {\small 0.007 } \\
         \hline
        { \small GT, $\sigma^*$ = $0$} & { \small $\Tilde{W} = 0.5 \cdot (\Tilde{I}_1 -3) + 0.3 \cdot (\Tilde{I}_2 - 3) + 1.5\cdot(J - 1)^2$ }   & {\small  0.026 } \\
        { \small GT, $\sigma^*$ = $10^{-4}$} & { \small  $W = 0.5\cdot(\Tilde{I}_1 - 3) + 0.3\cdot(\Tilde{I}_2 - 3) + 1.5\cdot(J-1)^2$ }  & {\small 0.026} \\
        { \small GT, $\sigma^*$ = $10^{-3}$} & { \small $\Tilde{W} = 0.5 \cdot (\Tilde{I}_1 -3) + 0.2 \cdot (\Tilde{I}_2  - 3) + 1.5\cdot(J - 1)^2$ }  & {\small  0.042 } \\
         \hline
        { \small OG, $\sigma^*$ = $0$} & { \small $\Tilde{W} = 0.7 \cdot (\Tilde{I}_1 - 3) + 0.1\cdot(\Tilde{I}_2 - 3) + 1.5\cdot(J-1)^2$ }   & {\small  0.027 } \\
        { \small OG, $\sigma^*$ = $10^{-4}$} & { \small  $W = 0.75 \cdot (\Tilde{I}_1 - 3) + 0.04\cdot(\Tilde{I}_2 - 3) + 1.5\cdot(J-1)^2$ }  & {\small 0.031} \\
        { \small OG, $\sigma^*$ = $10^{-3}$} & { \small $\Tilde{W} = 0.5 \cdot (\Tilde{I}_1 -3) + 0.3 \cdot (\Tilde{I}_2  - 3) + 1.5\cdot(J - 1)^2$ }  & {\small  0.020 } \\
         \hline
         &  {\small Unsupervised Case} &    \\
         \hline
        {\small Model, Noise Level} &  {\small Discovered Expression} &    {\small Relative $\mathcal{L}^2$ Error} \\
         \hline
        { \small IS, $\sigma^*$ = $0$} & { \small $\Tilde{W} = \Tilde{I}_1 - 3 + 0.5\cdot(\Tilde{I}_2 - 3) + (\Tilde{I}_1 - 3)^2 + 1.5\cdot(J - 1)^2$ }  & {\small  0.011 } \\
        { \small IS, $\sigma^*$ = $10^{-4}$} & { \small $\Tilde{W} =\Tilde{I}_1 - 3 + 0.5\cdot(\Tilde{I}_2 - 3) + (\Tilde{I}_1 - 3)^2 + 1.5\cdot(J - 1)^2$ } & {\small 0.011  } \\
        { \small IS, $\sigma^*$ = $10^{-3}$} & { \small  $W = 1.5\cdot(\Tilde{I}_2 - 3) + (\Tilde{I}_1 - 3)^2 + 1.5\cdot(-1 + J)^2$ }  & {\small 0.011 } \\
         \hline
        { \small HW, $\sigma^*$ = $0$} & { \small $\Tilde{W} = 1.5\cdot(\Tilde{I}_2 - 3) + (\Tilde{I}_1 - 3)^2 + 1.5\cdot(-1 + J)^2$  }  & {\small 0.01 } \\
        { \small HW, $\sigma^*$ = $10^{-4}$} & { \small $\Tilde{W} = (\Tilde{I}_1 - 3)\cdot(\Tilde{I}_2 - 3) + 1.5\cdot(\Tilde{I}_2 - 3) + 1.5\cdot(-1 + J)^2$ }  & {\small 0.007 } \\
        { \small HW, $\sigma^*$ = $10^{-3}$} & {  \small $\Tilde{W} = (\Tilde{I}_1 - 3)\cdot(\Tilde{I}_2 - 3) + 1.5\cdot(\Tilde{I}_2 - 3) + 1.5\cdot(-1 + J)^2$ }  & {\small 0.007 } \\
         \hline
        { \small GT, $\sigma^*$ = $0$} & { \small $\Tilde{W} =0.75 \cdot (\Tilde{I}_1 -3) + 1.5\cdot(J - 1)^2$ }   & {\small  0.047} \\
        { \small GT, $\sigma^*$ = $10^{-4}$} & { \small  $W = 0.75 \cdot (\Tilde{I}_1 -3) + 1.5\cdot(J - 1)^2$ }  & {\small 0.047} \\
        { \small GT, $\sigma^*$ = $10^{-3}$} & { \small $\Tilde{W} = 0.75 \cdot (\Tilde{I}_1 -3) + 1.5\cdot(J - 1)^2$ }  & {\small  0.047 } \\
         \hline
        { \small OG, $\sigma^*$ = $0$} & { \small $\Tilde{W} =0.75 \cdot (\Tilde{I}_1 -3) + 1.5\cdot(J - 1)^2$ }   & {\small  0.047 } \\
        { \small OG, $\sigma^*$ = $10^{-4}$} & { \small  $W = 0.75 \cdot (\Tilde{I}_1 -3) + 1.5\cdot(J - 1)^2$ }  & {\small 0.047} \\
        { \small OG, $\sigma^*$ = $10^{-3}$} & { \small $\Tilde{W} = 0.75 \cdot (\Tilde{I}_1 -3) + 1.5\cdot(J - 1)^2$ }  & {\small  0.047 } \\
         \hline
    \end{tabular}
    }
    \caption{Best discovered expressions and relative $\mathcal{L}^2$ errors between the best predicted and the baseline strain energy density for the IS, HW and GT models, obtained from data with different noise levels for both the supervised and the unsupervised cases}.
    \label{tab:models_l2_error}
\end{table}

\paragraph{Generalization to Different Loading Conditions} 

\begin{figure}
\caption*{  \hspace{1cm} Supervised Case \hspace{3.5cm} Unsupervised Case} 
  \begin{subfigure}{0.25\textwidth}
    \centering
    \includegraphics[width=\linewidth]{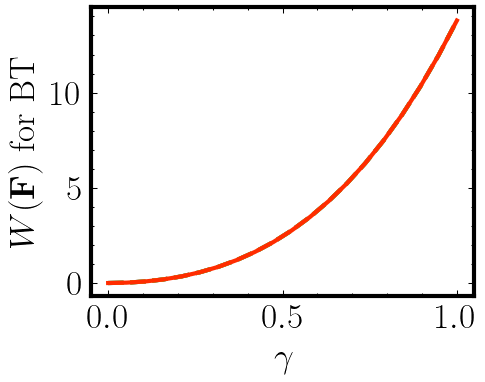}
  \end{subfigure}%
  \begin{subfigure}{0.25\textwidth}
    \centering
    \includegraphics[width=\linewidth]{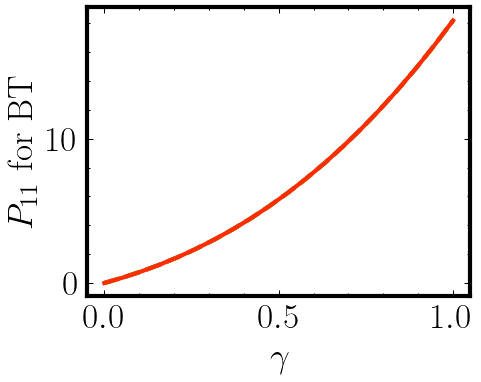}
  \end{subfigure}
    \begin{subfigure}{0.25\textwidth}
    \centering
    \includegraphics[width= \linewidth]{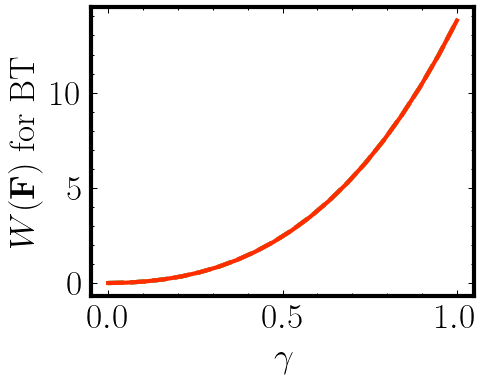}
  \end{subfigure}%
  \begin{subfigure}{0.25\textwidth}
    \centering
    \includegraphics[width=\linewidth]{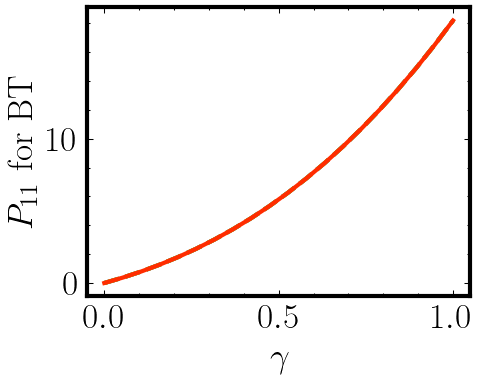}
  \end{subfigure}

  \begin{subfigure}{0.25\textwidth}
    \centering
    \includegraphics[width=\linewidth]{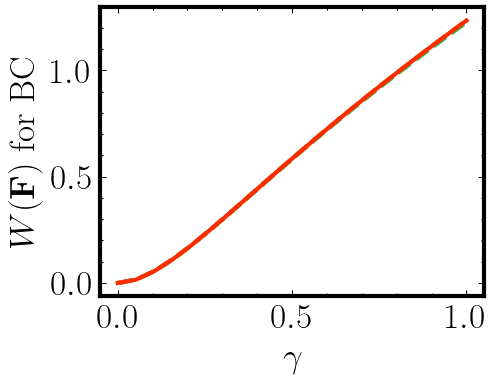}
  \end{subfigure}%
  \begin{subfigure}{0.25\textwidth}
    \centering
    \includegraphics[width=\linewidth]{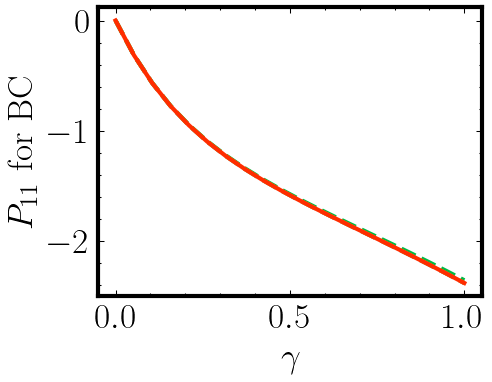}
  \end{subfigure}
    \begin{subfigure}{0.25\textwidth}
    \centering
    \includegraphics[width=\linewidth]{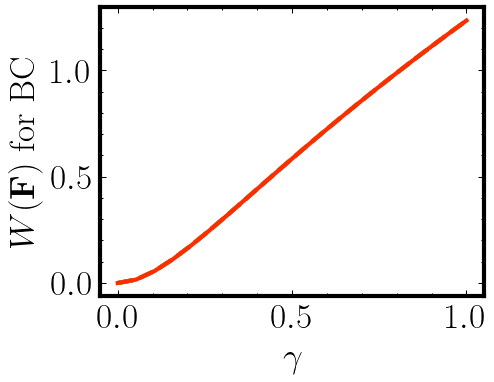}
  \end{subfigure}%
  \begin{subfigure}{0.25\textwidth}
    \centering
    \includegraphics[width=\linewidth]{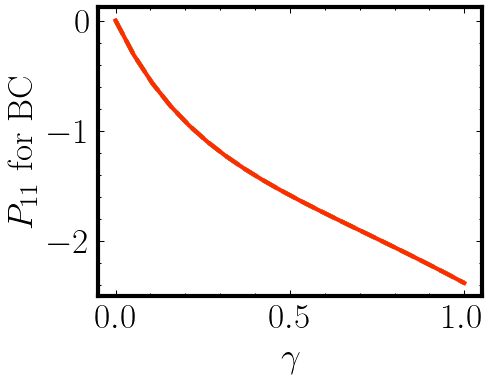}
  \end{subfigure}
  
  \begin{subfigure}{0.25\textwidth}
    \centering
    \includegraphics[width=\linewidth]{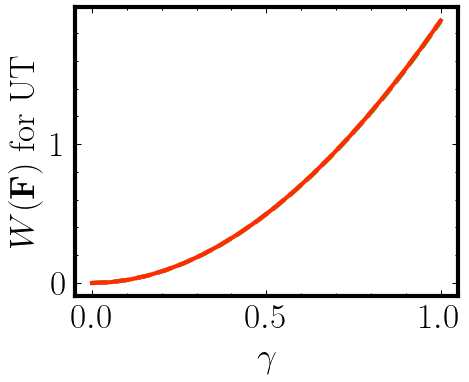}
  \end{subfigure}%
  \begin{subfigure}{0.25\textwidth}
    \centering
    \includegraphics[width=\linewidth]{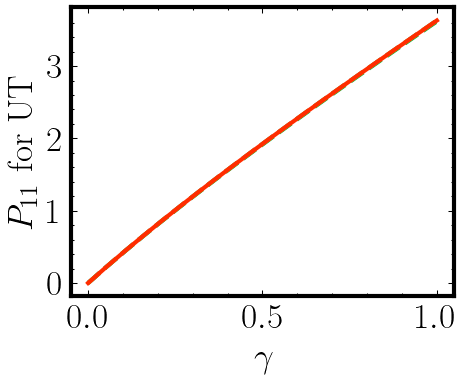}
  \end{subfigure}
    \begin{subfigure}{0.25\textwidth}
    \centering
    \includegraphics[width=\linewidth]{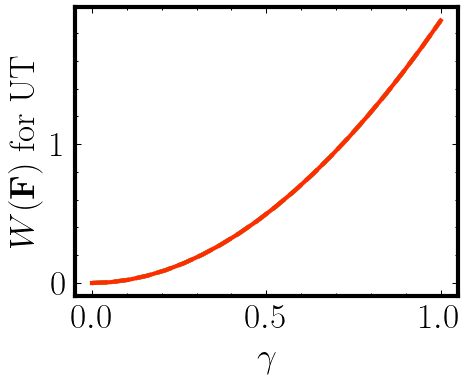}
  \end{subfigure}%
  \begin{subfigure}{0.25\textwidth}
    \centering
    \includegraphics[width=\linewidth]{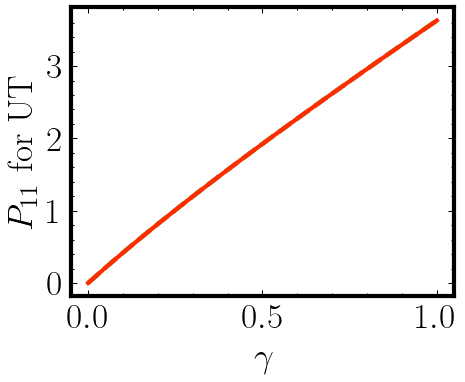}
  \end{subfigure}
  
  \begin{subfigure}{0.25\textwidth}
    \centering
    \includegraphics[width=\linewidth]{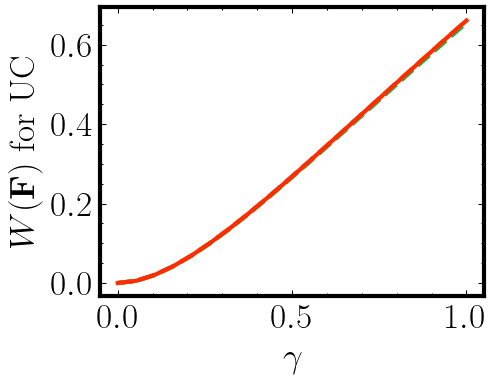}
  \end{subfigure}%
  \begin{subfigure}{0.25\textwidth}
    \centering
    \includegraphics[width=\linewidth]{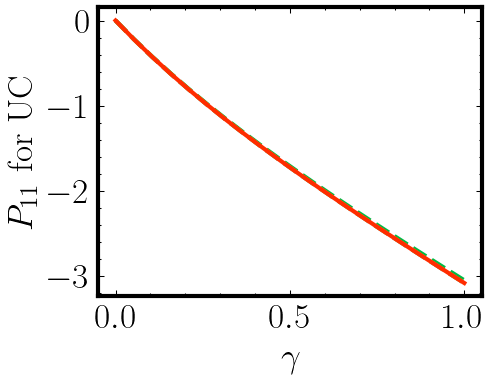}
  \end{subfigure}
  \begin{subfigure}{0.25\textwidth}
    \centering
    \includegraphics[width=\linewidth]{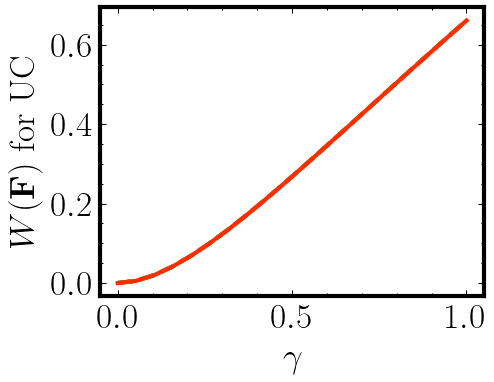}
  \end{subfigure}%
  \begin{subfigure}{0.25\textwidth}
    \centering
    \includegraphics[width=\linewidth]{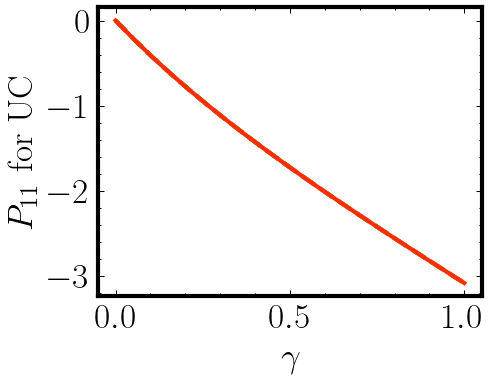}
  \end{subfigure}
  
   \begin{subfigure}{0.25\textwidth}
    \centering
    \includegraphics[width=\linewidth]{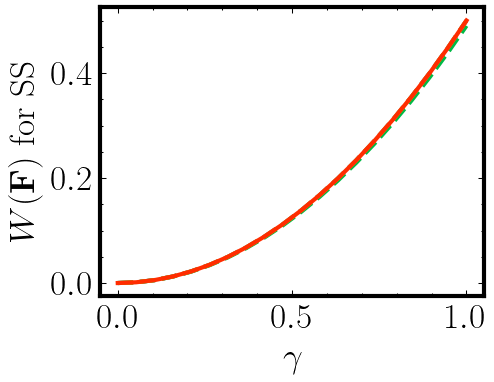}
  \end{subfigure}%
  \begin{subfigure}{0.25\textwidth}
    \centering
    \includegraphics[width=\linewidth]{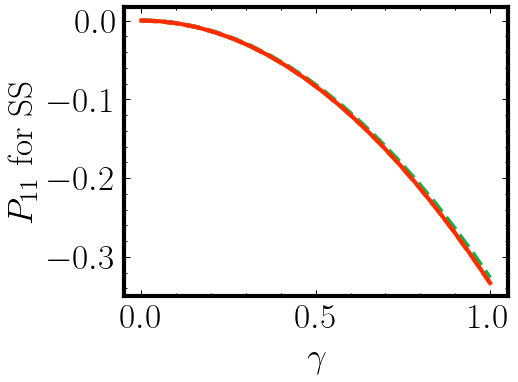}
  \end{subfigure}
    \begin{subfigure}{0.25\textwidth}
    \centering
    \includegraphics[width=\linewidth]{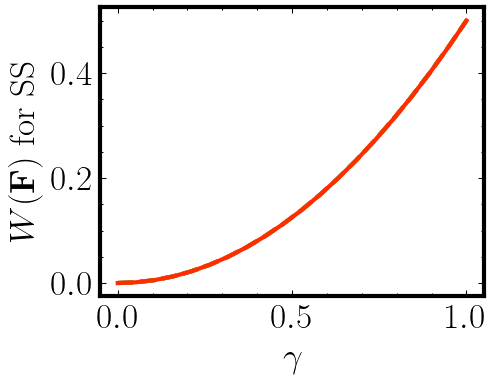}
  \end{subfigure}%
  \begin{subfigure}{0.25\textwidth}
    \centering
    \includegraphics[width=\linewidth]{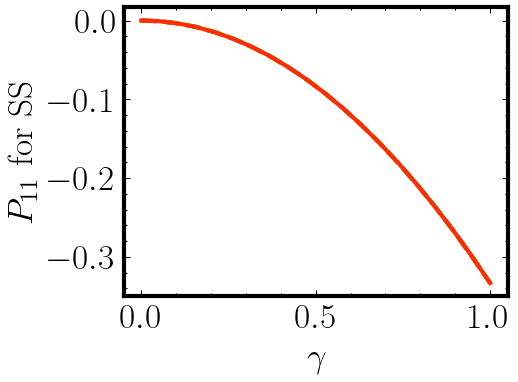}
  \end{subfigure}

  \begin{subfigure}{0.25\textwidth}
    \centering
    \includegraphics[width=\linewidth]{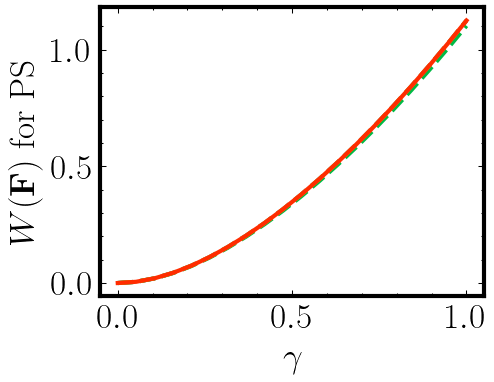}
  \end{subfigure}%
  \begin{subfigure}{0.25\textwidth}
    \centering
    \includegraphics[width=\linewidth]{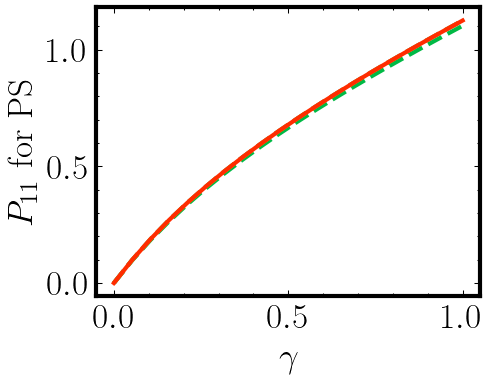}
  \end{subfigure}
    \begin{subfigure}{0.25\textwidth}
    \centering
    \includegraphics[width=\linewidth]{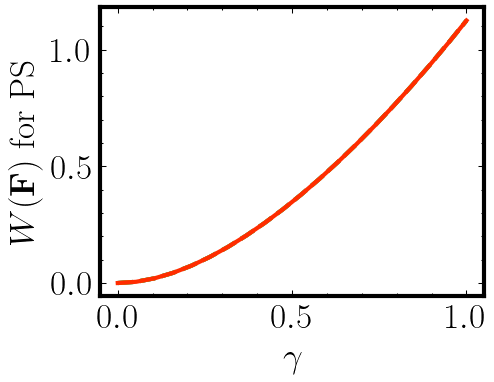}
  \end{subfigure}%
  \begin{subfigure}{0.25\textwidth}
    \centering
    \includegraphics[width=\linewidth]{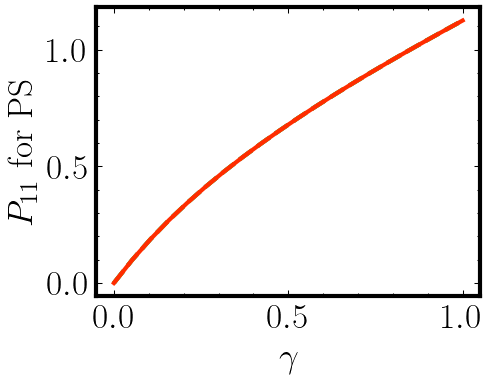}
  \end{subfigure}
  
  \caption{Neo-Hookean model: Left: comparison between the baseline and the predicted strain energy density, Right: comparison between the first component of the Piola-Kirchhoff stress tensor for the loading conditions not in the training set evaluated for different displacement magnitudes parameterized by $\gamma$. For all cases, the solid red line represents the baseline, the dashed blue line the prediction for the noise-free data, the dashed orange line the prediction for $\sigma^*=10^{-4}$ and the dashed green line the prediction for $\sigma^*=10^{-3}$.}
  \label{fig:results nh extrapolation}
\end{figure}

\begin{figure}
\caption*{  \hspace{1cm} Supervised Case \hspace{3.5cm} Unsupervised Case} 
  \begin{subfigure}{0.25\textwidth}
    \centering
    \includegraphics[width=\linewidth]{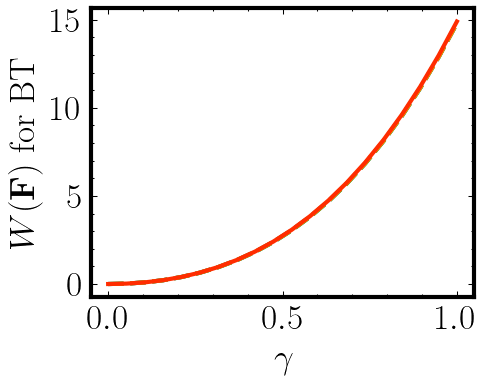}
  \end{subfigure}%
  \begin{subfigure}{0.25\textwidth}
    \centering
    \includegraphics[width=\linewidth]{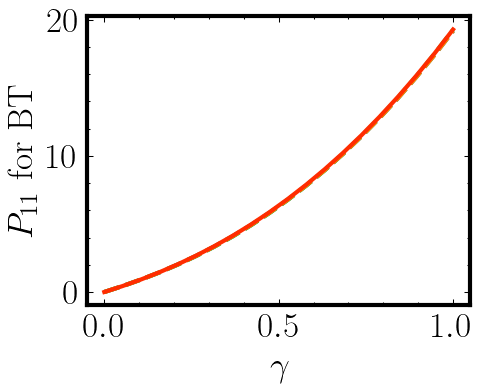}
  \end{subfigure}
  \begin{subfigure}{0.25\textwidth}
    \centering
    \includegraphics[width=\linewidth]{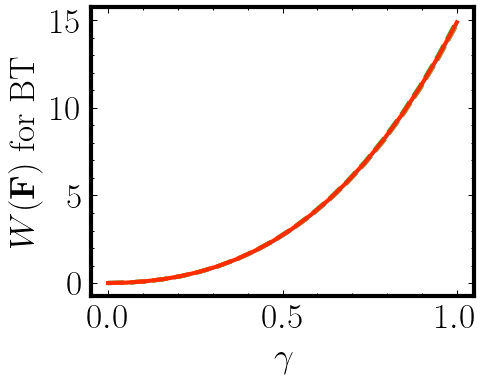}
  \end{subfigure}%
  \begin{subfigure}{0.25\textwidth}
    \centering
    \includegraphics[width=\linewidth]{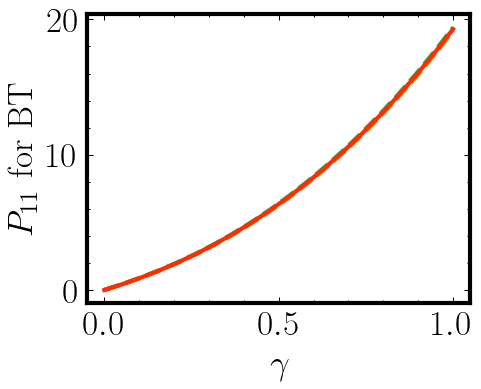}
  \end{subfigure}

  \begin{subfigure}{0.25\textwidth}
    \centering
    \includegraphics[width=\linewidth]{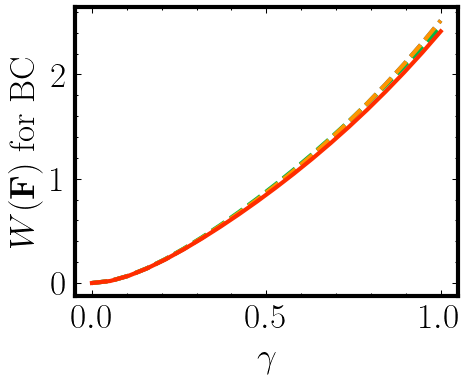}
  \end{subfigure}%
  \begin{subfigure}{0.25\textwidth}
    \centering
    \includegraphics[width=\linewidth]{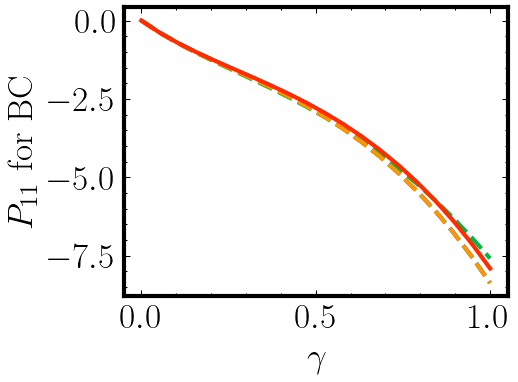}
  \end{subfigure}
  \begin{subfigure}{0.25\textwidth}
    \centering
    \includegraphics[width=\linewidth]{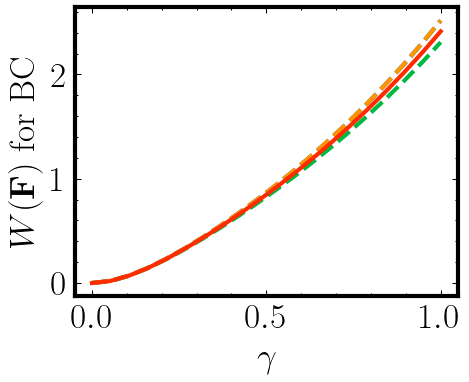}
  \end{subfigure}%
  \begin{subfigure}{0.25\textwidth}
    \centering
    \includegraphics[width=\linewidth]{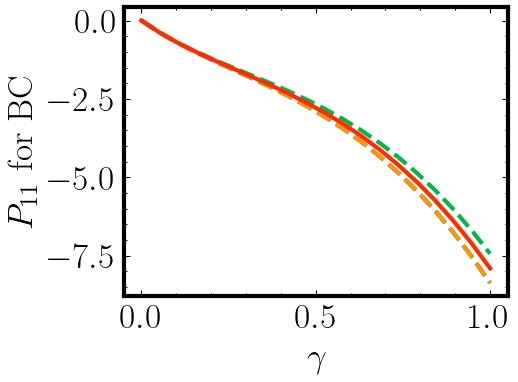}
  \end{subfigure}
  
  \begin{subfigure}{0.25\textwidth}
    \centering
    \includegraphics[width=\linewidth]{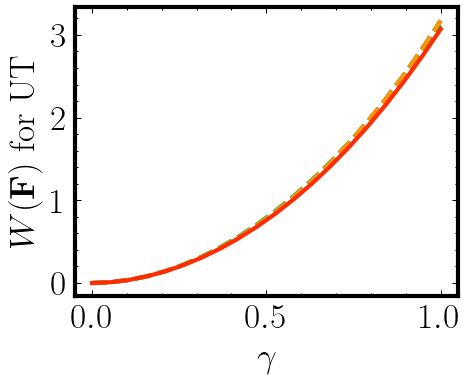}
  \end{subfigure}%
  \begin{subfigure}{0.25\textwidth}
    \centering
    \includegraphics[width=\linewidth]{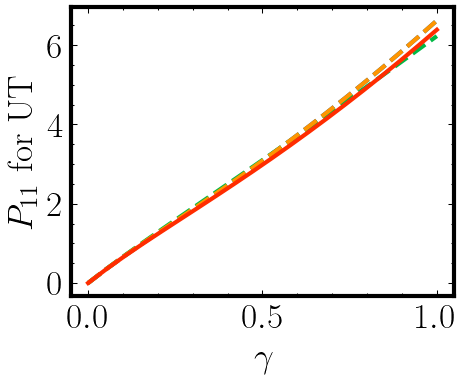}
  \end{subfigure}
    \begin{subfigure}{0.25\textwidth}
    \centering
    \includegraphics[width=\linewidth]{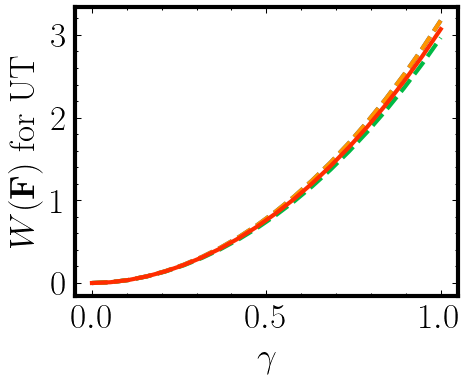}
  \end{subfigure}%
  \begin{subfigure}{0.25\textwidth}
    \centering
    \includegraphics[width=\linewidth]{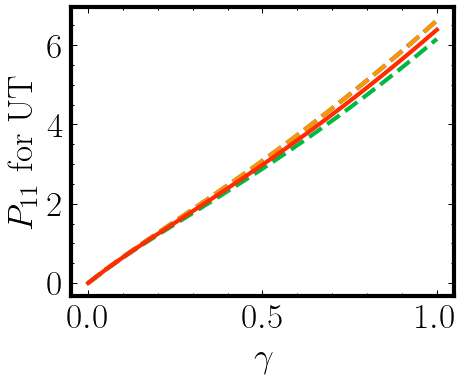}
  \end{subfigure}
  
  \begin{subfigure}{0.25\textwidth}
    \centering
    \includegraphics[width=\linewidth]{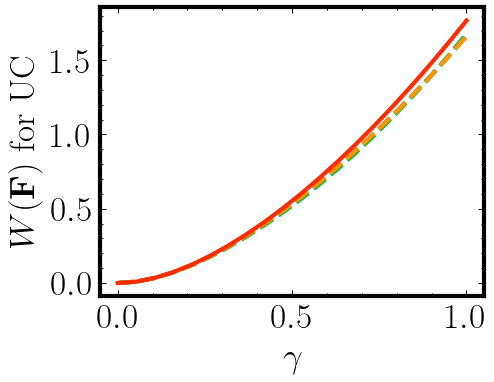}
  \end{subfigure}%
  \begin{subfigure}{0.25\textwidth}
    \centering
    \includegraphics[width=\linewidth]{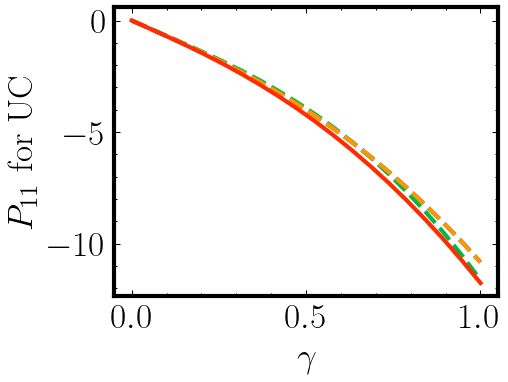}
  \end{subfigure}
  \begin{subfigure}{0.25\textwidth}
    \centering
    \includegraphics[width=\linewidth]{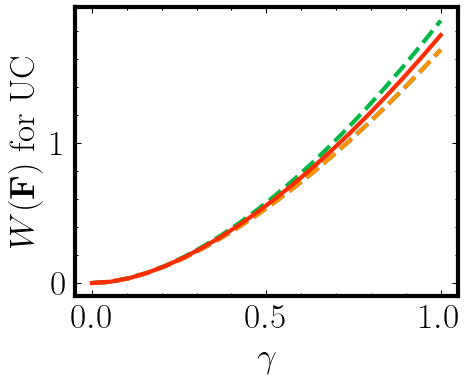}
  \end{subfigure}%
  \begin{subfigure}{0.25\textwidth}
    \centering
    \includegraphics[width=\linewidth]{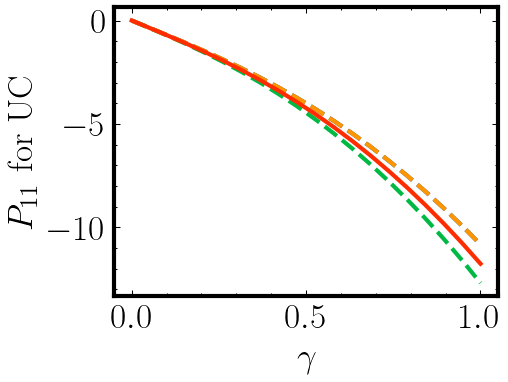}
  \end{subfigure}

   \begin{subfigure}{0.25\textwidth}
    \centering
    \includegraphics[width=\linewidth]{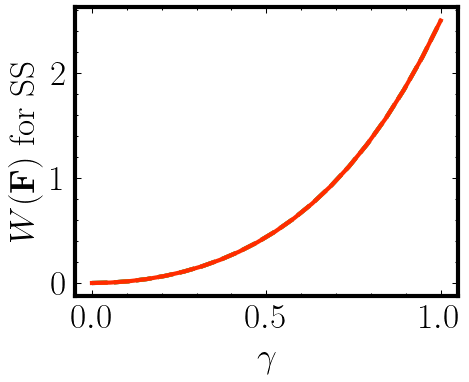}
  \end{subfigure}%
  \begin{subfigure}{0.25\textwidth}
    \centering
    \includegraphics[width=\linewidth]{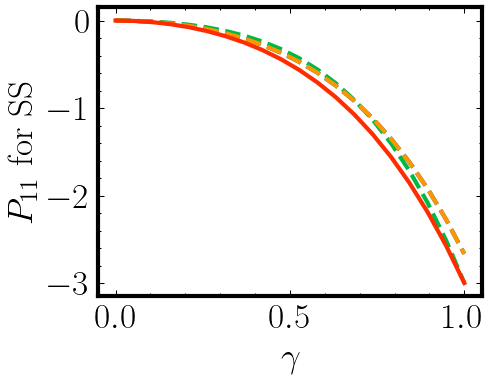}
  \end{subfigure}
     \begin{subfigure}{0.25\textwidth}
    \centering
    \includegraphics[width=\linewidth]{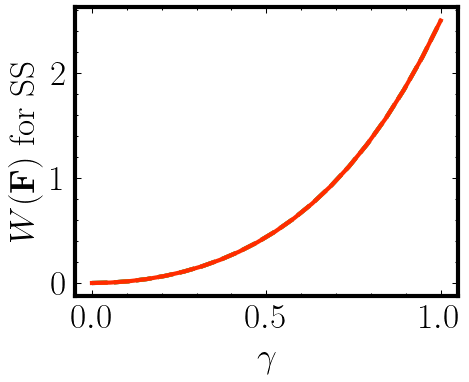}
  \end{subfigure}%
  \begin{subfigure}{0.25\textwidth}
    \centering
    \includegraphics[width=\linewidth]{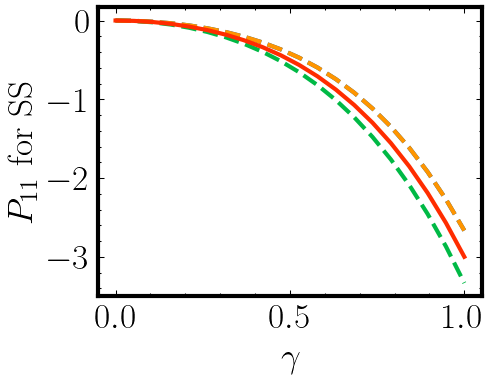}
  \end{subfigure}

  \begin{subfigure}{0.25\textwidth}
    \centering
    \includegraphics[width=\linewidth]{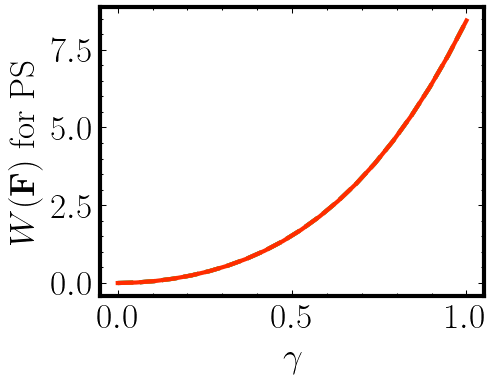}
  \end{subfigure}%
  \begin{subfigure}{0.25\textwidth}
    \centering
    \includegraphics[width=\linewidth]{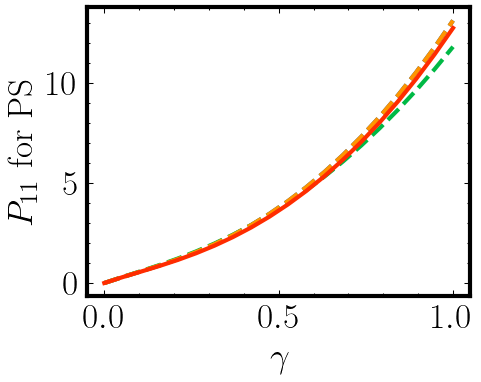}
  \end{subfigure}
  \begin{subfigure}{0.25\textwidth}
    \centering
    \includegraphics[width=\linewidth]{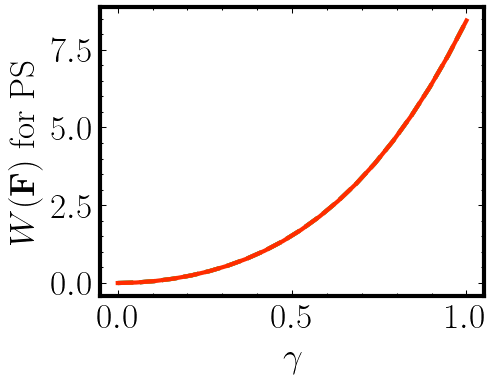}
  \end{subfigure}%
  \begin{subfigure}{0.25\textwidth}
    \centering
    \includegraphics[width=\linewidth]{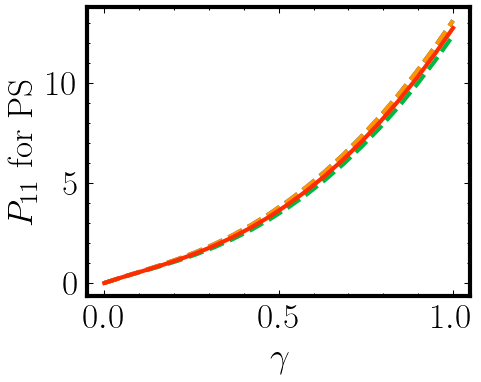}
  \end{subfigure}
  
  \caption{Isihara model: Left: comparison between the baseline and the predicted strain energy density, Right: comparison between the first component of the Piola-Kirchhoff stress tensor for the loading conditions not in the training set evaluated for different displacement magnitudes parameterized by $\gamma$. For all cases, the solid red line represents the baseline, the dashed blue line the prediction for the noise-free data, the dashed orange line the prediction for $\sigma^*=10^{-4}$ and the dashed green line the prediction for $\sigma^*=10^{-3}$.}
  \label{fig:results is extrapolation}
\end{figure}

\begin{figure}
\caption*{  \hspace{1cm} Supervised Case \hspace{3.5cm} Unsupervised Case} 
  \begin{subfigure}{0.25\textwidth}
    \centering
    \includegraphics[width=\linewidth]{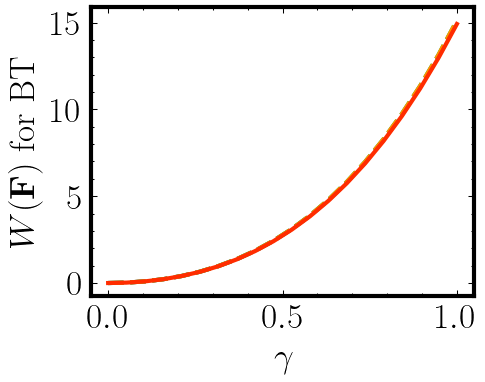}
  \end{subfigure}%
  \begin{subfigure}{0.25\textwidth}
    \centering
    \includegraphics[width=\linewidth]{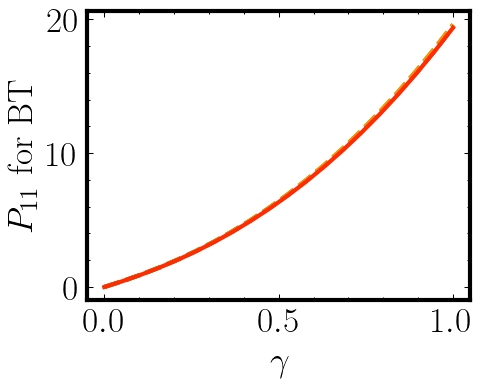}
  \end{subfigure}
  \begin{subfigure}{0.25\textwidth}
    \centering
    \includegraphics[width=\linewidth]{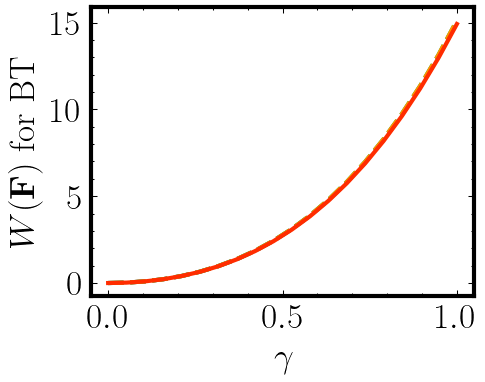}
  \end{subfigure}%
  \begin{subfigure}{0.25\textwidth}
    \centering
    \includegraphics[width=\linewidth]{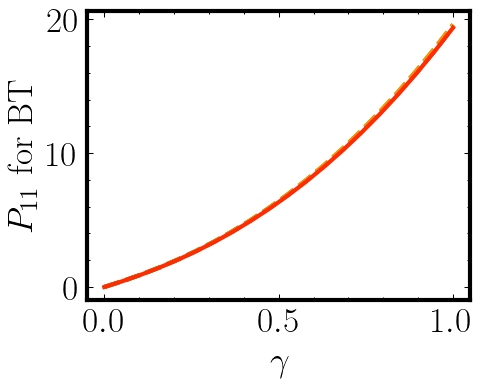}
  \end{subfigure}
  
  \begin{subfigure}{0.25\textwidth}
    \centering
    \includegraphics[width=\linewidth]{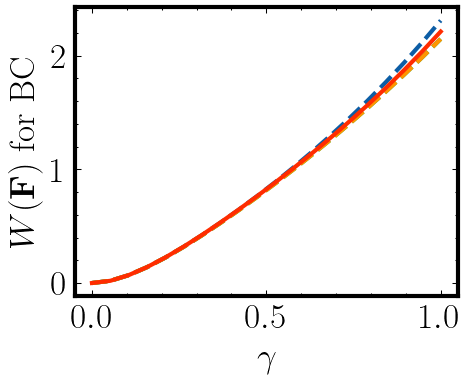}
  \end{subfigure}%
  \begin{subfigure}{0.25\textwidth}
    \centering
    \includegraphics[width=\linewidth]{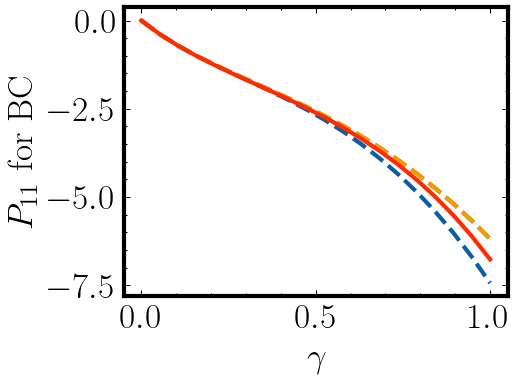}
  \end{subfigure}
    \begin{subfigure}{0.25\textwidth}
    \centering
    \includegraphics[width=\linewidth]{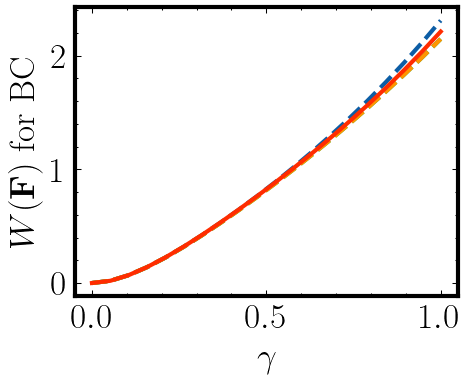}
  \end{subfigure}%
  \begin{subfigure}{0.25\textwidth}
    \centering
    \includegraphics[width=\linewidth]{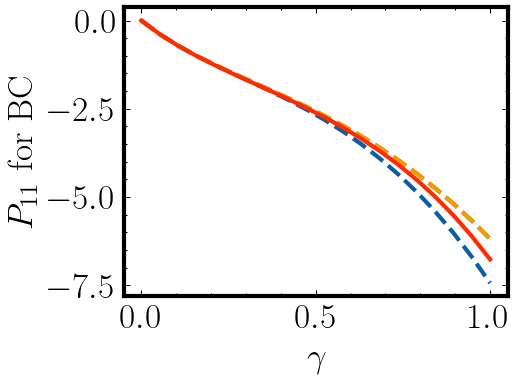}
  \end{subfigure}

  \begin{subfigure}{0.25\textwidth}
    \centering
    \includegraphics[width=\linewidth]{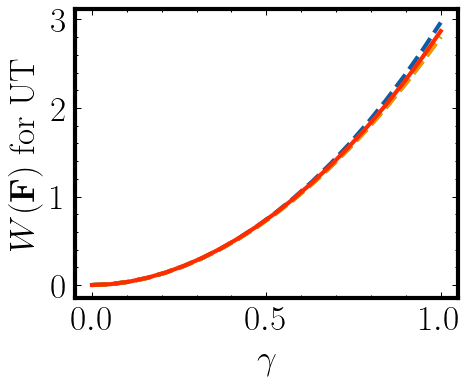}
  \end{subfigure}%
  \begin{subfigure}{0.25\textwidth}
    \centering
    \includegraphics[width=\linewidth]{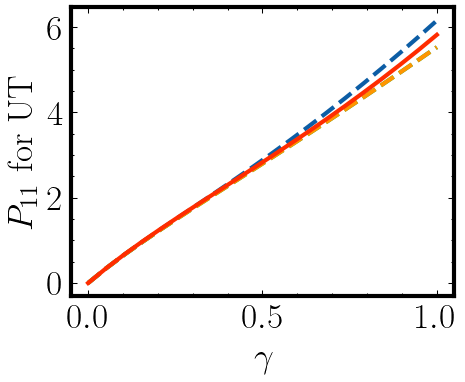}
  \end{subfigure}
  \begin{subfigure}{0.25\textwidth}
    \centering
    \includegraphics[width=\linewidth]{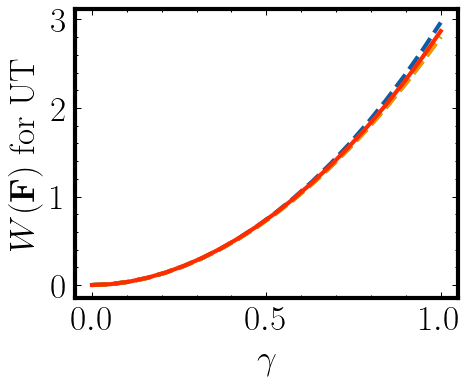}
  \end{subfigure}%
  \begin{subfigure}{0.25\textwidth}
    \centering
    \includegraphics[width=\linewidth]{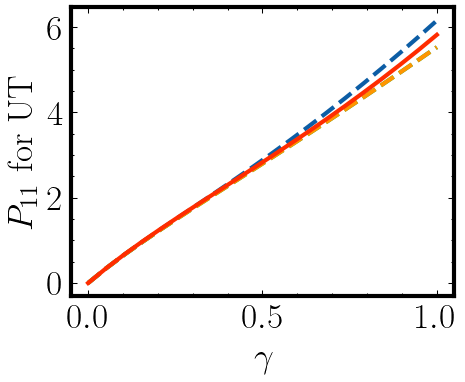}
  \end{subfigure}

  \begin{subfigure}{0.25\textwidth}
    \centering
    \includegraphics[width=\linewidth]{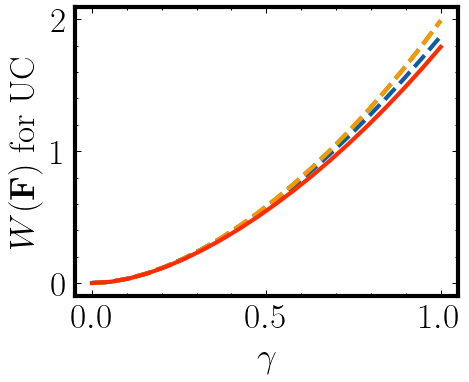}
  \end{subfigure}%
  \begin{subfigure}{0.25\textwidth}
    \centering
    \includegraphics[width=\linewidth]{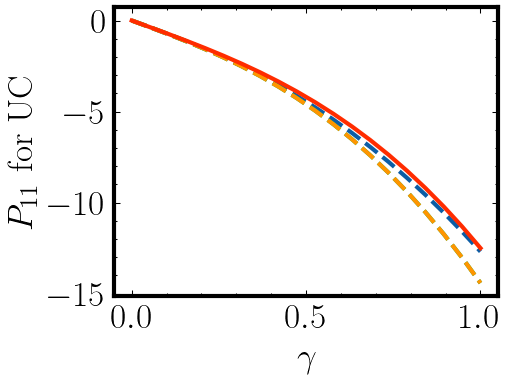}
  \end{subfigure}
  \begin{subfigure}{0.25\textwidth}
    \centering
    \includegraphics[width=\linewidth]{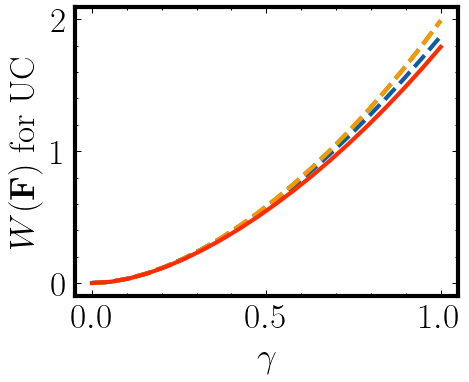}
  \end{subfigure}%
  \begin{subfigure}{0.25\textwidth}
    \centering
    \includegraphics[width=\linewidth]{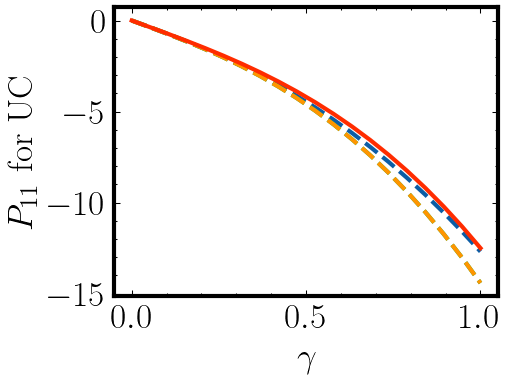}
  \end{subfigure}

   \begin{subfigure}{0.25\textwidth}
    \centering
    \includegraphics[width=\linewidth]{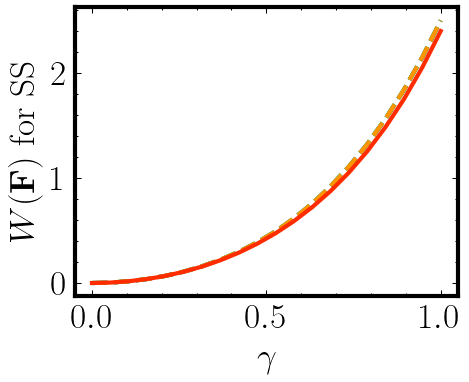}
  \end{subfigure}%
  \begin{subfigure}{0.25\textwidth}
    \centering
    \includegraphics[width=\linewidth]{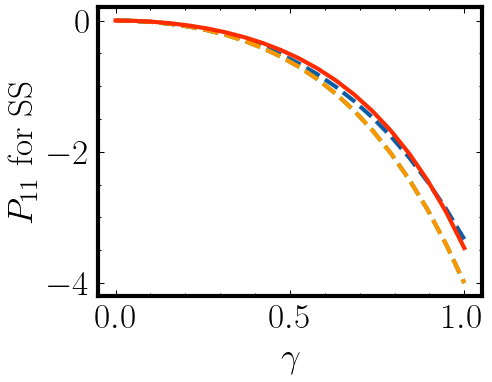}
  \end{subfigure}
   \begin{subfigure}{0.25\textwidth}
    \centering
    \includegraphics[width=\linewidth]{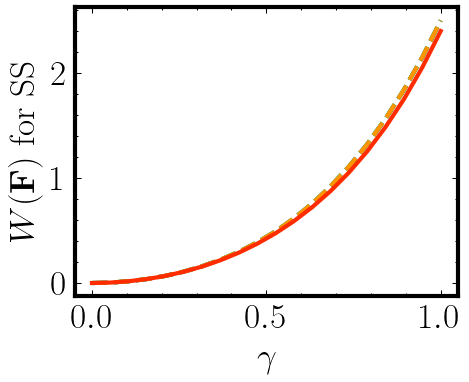}
  \end{subfigure}%
  \begin{subfigure}{0.25\textwidth}
    \centering
    \includegraphics[width=\linewidth]{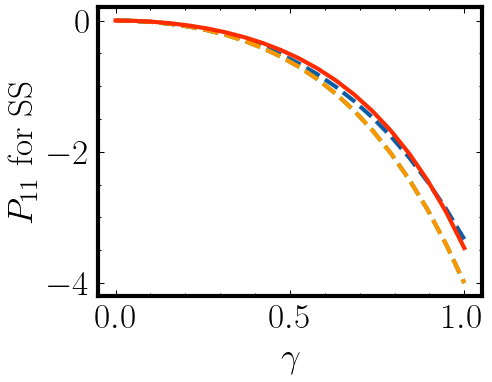}
  \end{subfigure}

  \begin{subfigure}{0.25\textwidth}
    \centering
    \includegraphics[width=\linewidth]{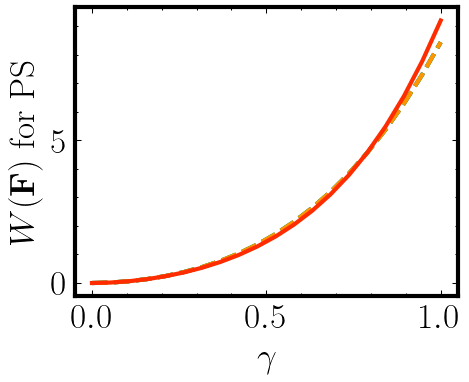}
  \end{subfigure}%
  \begin{subfigure}{0.25\textwidth}
    \centering
    \includegraphics[width=\linewidth]{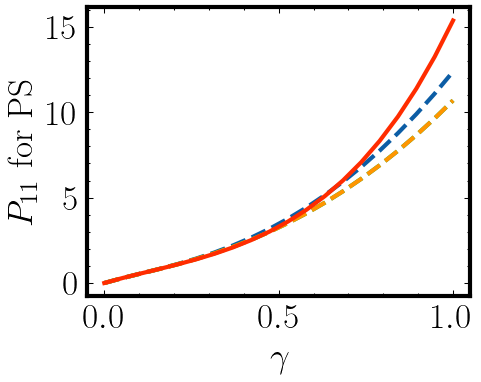}
  \end{subfigure}
  \begin{subfigure}{0.25\textwidth}
    \centering
    \includegraphics[width=\linewidth]{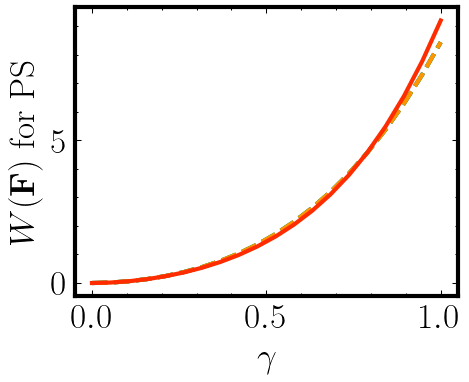}
  \end{subfigure}%
  \begin{subfigure}{0.25\textwidth}
    \centering
    \includegraphics[width=\linewidth]{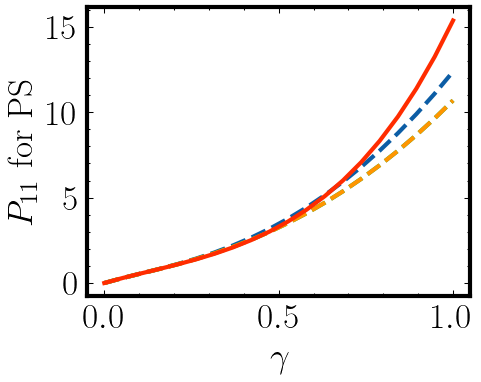}
  \end{subfigure}
  \caption{Haines-Wilson model: Left: comparison between the baseline and the predicted strain energy density, Right: comparison between the first component of the Piola-Kirchhoff stress tensor for the loading conditions not in the training set evaluated for different displacement magnitudes parameterized by $\gamma$. For all cases, the solid red line represents the baseline, the dashed blue line the prediction for the noise-free data, the dashed orange line the prediction for $\sigma^*=10^{-4}$ and the dashed green line the prediction for $\sigma^*=10^{-3}$.}
  \label{fig:results hw extrapolation}
\end{figure}

\begin{figure}
\caption*{  \hspace{1cm} Supervised Case \hspace{3.5cm} Unsupervised Case} 
  \begin{subfigure}{0.25\textwidth}
    \centering
    \includegraphics[width=\linewidth]{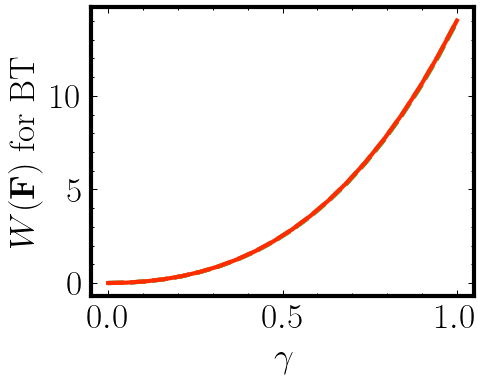}
  \end{subfigure}%
  \begin{subfigure}{0.25\textwidth}
    \centering
    \includegraphics[width=\linewidth]{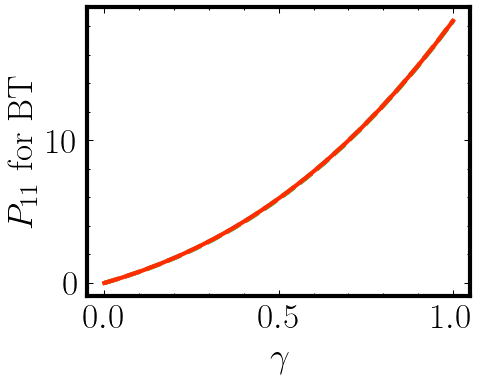}
  \end{subfigure}
  \begin{subfigure}{0.25\textwidth}
    \centering
    \includegraphics[width=\linewidth]{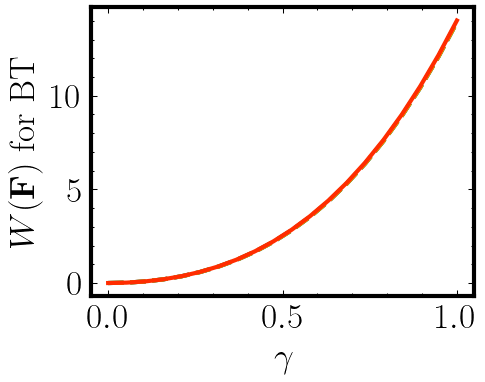}
  \end{subfigure}%
  \begin{subfigure}{0.25\textwidth}
    \centering
    \includegraphics[width=\linewidth]{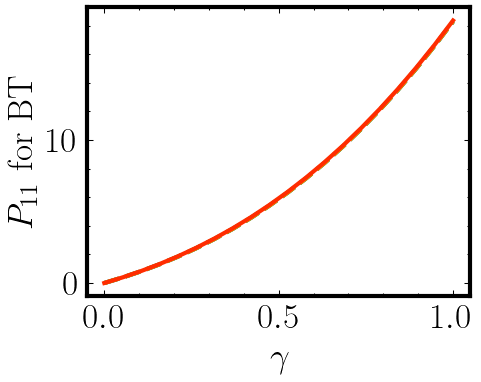}
  \end{subfigure}
  
  \begin{subfigure}{0.25\textwidth}
    \centering
    \includegraphics[width=\linewidth]{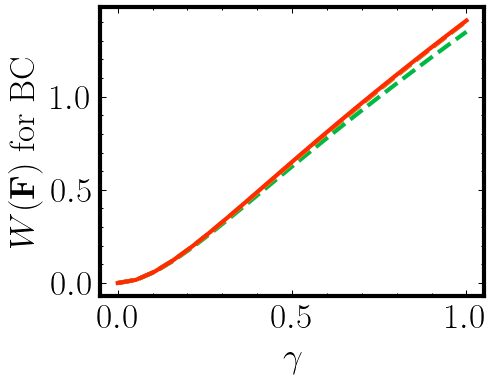}
  \end{subfigure}%
  \begin{subfigure}{0.25\textwidth}
    \centering
    \includegraphics[width=\linewidth]{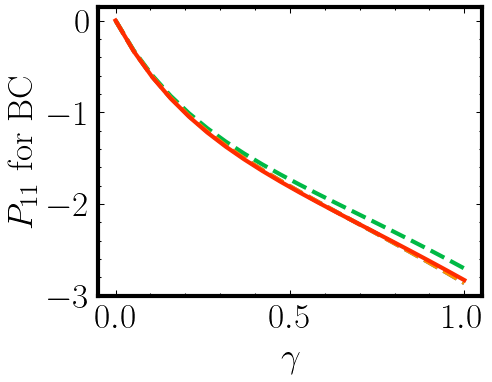}
  \end{subfigure}
  \begin{subfigure}{0.25\textwidth}
    \centering
    \includegraphics[width=\linewidth]{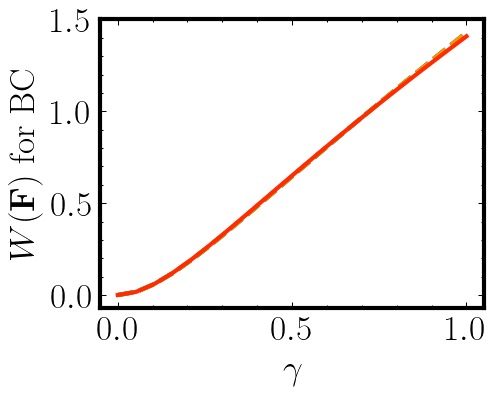}
  \end{subfigure}%
  \begin{subfigure}{0.25\textwidth}
    \centering
    \includegraphics[width=\linewidth]{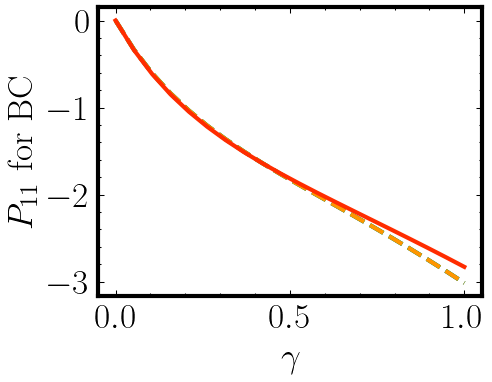}
  \end{subfigure}
  
  \begin{subfigure}{0.25\textwidth}
    \centering
    \includegraphics[width=\linewidth]{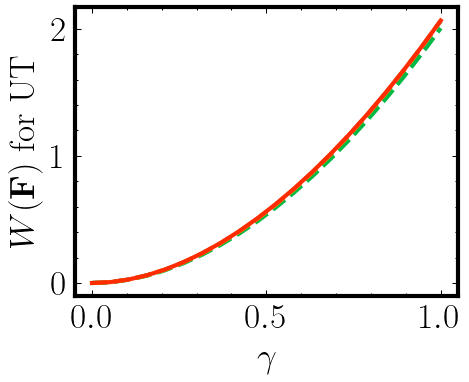}
  \end{subfigure}%
  \begin{subfigure}{0.25\textwidth}
    \centering
    \includegraphics[width=\linewidth]{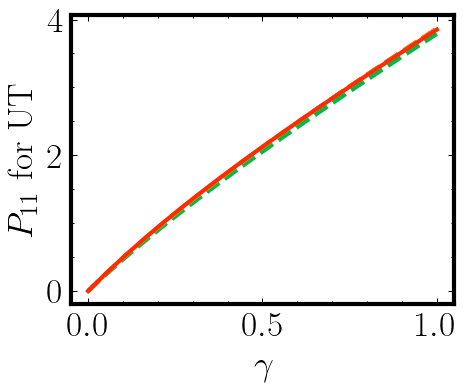}
  \end{subfigure}
  \begin{subfigure}{0.25\textwidth}
    \centering
    \includegraphics[width=\linewidth]{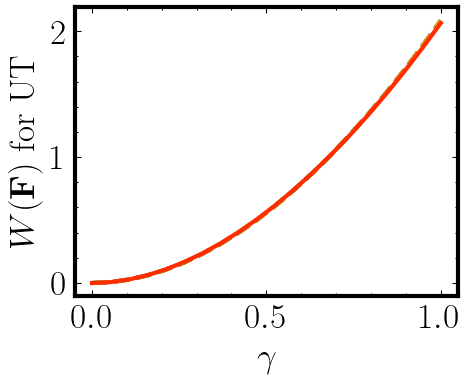}
  \end{subfigure}%
  \begin{subfigure}{0.25\textwidth}
    \centering
    \includegraphics[width=\linewidth]{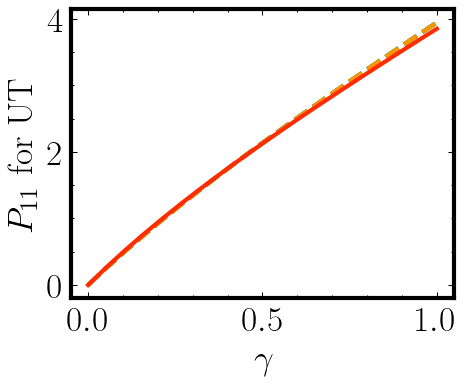}
  \end{subfigure}
  
  \begin{subfigure}{0.25\textwidth}
    \centering
    \includegraphics[width=\linewidth]{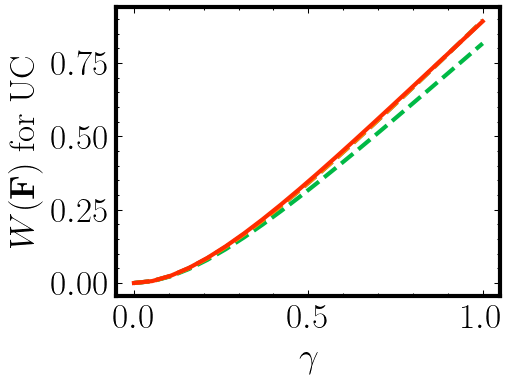}
  \end{subfigure}%
  \begin{subfigure}{0.25\textwidth}
    \centering
    \includegraphics[width=\linewidth]{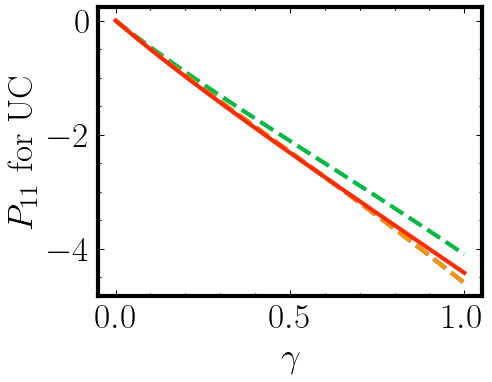}
  \end{subfigure}
  \begin{subfigure}{0.25\textwidth}
    \centering
    \includegraphics[width=\linewidth]{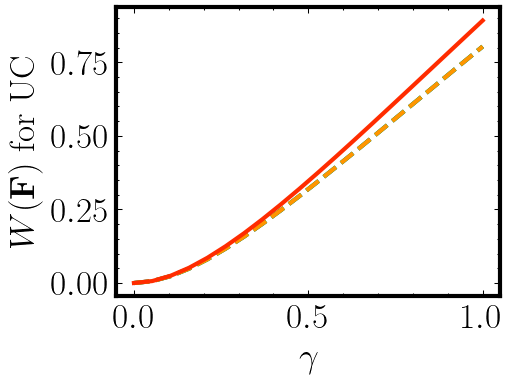}
  \end{subfigure}%
  \begin{subfigure}{0.25\textwidth}
    \centering
    \includegraphics[width=\linewidth]{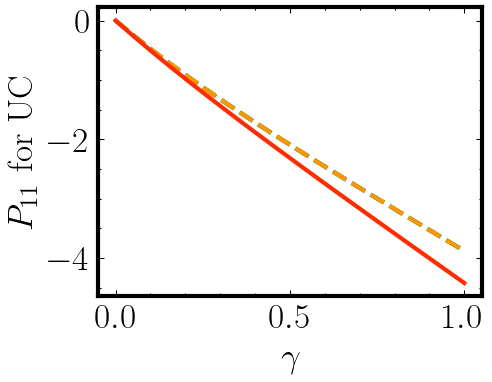}
  \end{subfigure}

   \begin{subfigure}{0.25\textwidth}
    \centering
    \includegraphics[width=\linewidth]{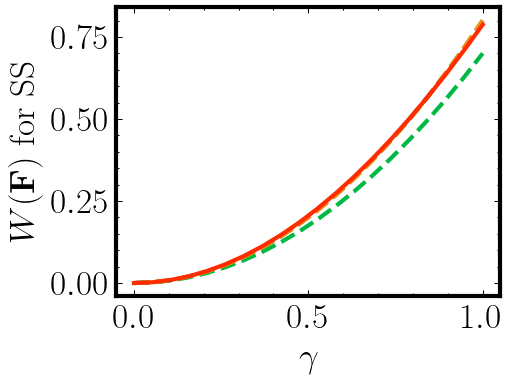}
  \end{subfigure}%
  \begin{subfigure}{0.25\textwidth}
    \centering
    \includegraphics[width=\linewidth]{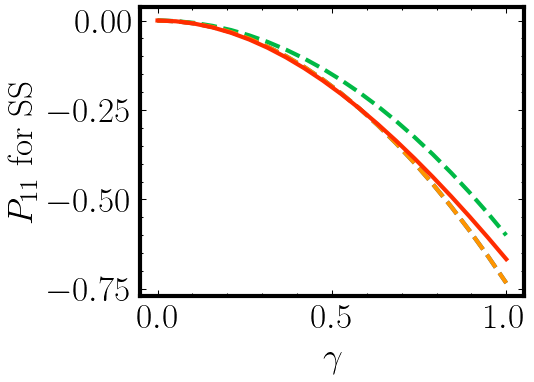}
  \end{subfigure}
   \begin{subfigure}{0.25\textwidth}
    \centering
    \includegraphics[width=\linewidth]{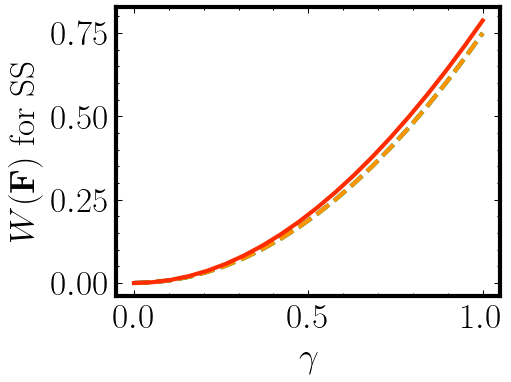}
  \end{subfigure}%
  \begin{subfigure}{0.25\textwidth}
    \centering
    \includegraphics[width=\linewidth]{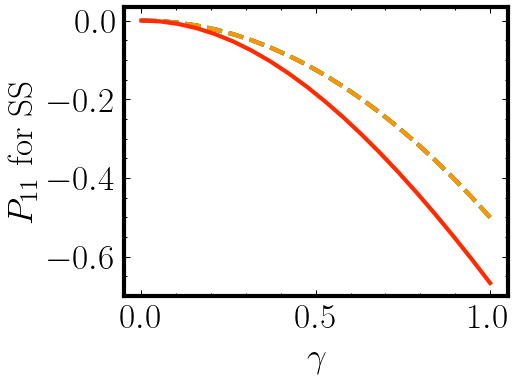}
  \end{subfigure}

  \begin{subfigure}{0.25\textwidth}
    \centering
    \includegraphics[width=\linewidth]{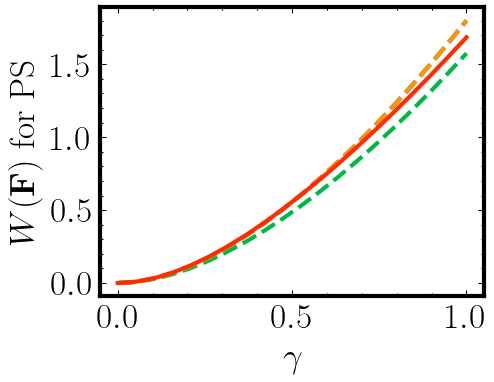}
  \end{subfigure}%
  \begin{subfigure}{0.25\textwidth}
    \centering
    \includegraphics[width=\linewidth]{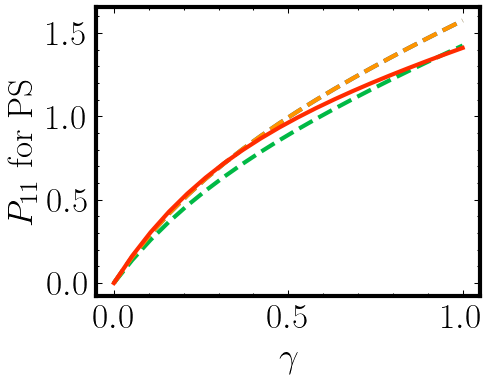}
  \end{subfigure}
  \begin{subfigure}{0.25\textwidth}
    \centering
    \includegraphics[width=\linewidth]{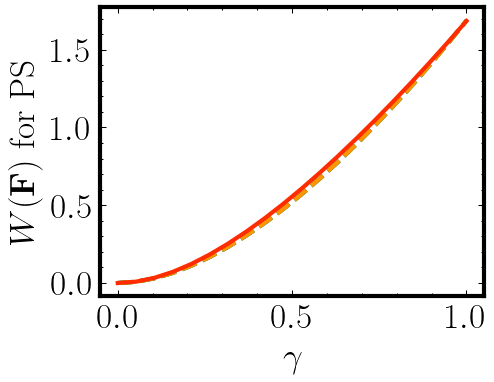}
  \end{subfigure}%
  \begin{subfigure}{0.25\textwidth}
    \centering
    \includegraphics[width=\linewidth]{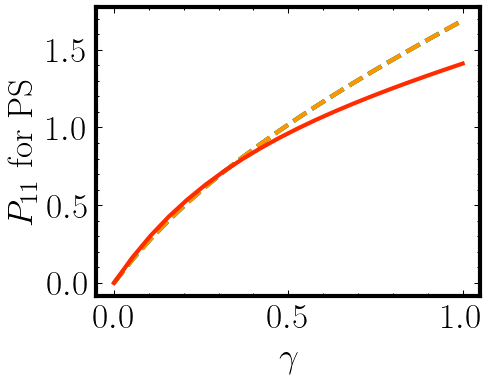}
  \end{subfigure}
  
  \caption{Gent-Thomas model: Left: comparison between the baseline and the predicted strain energy density, Right: comparison between the first component of the Piola-Kirchhoff stress tensor for loading conditions not in the training set evaluated for different displacement magnitudes parameterized by $\gamma$. For all cases, the solid red line represents the baseline, the dashed blue line the prediction for the noise-free data, the dashed orange line the prediction for $\sigma^*=10^{-4}$ and the dashed green line the prediction for $\sigma^*=10^{-3}$.}
  \label{fig:results gt extrapolation}
\end{figure}

In practice, we are interested in expressions that describe the behavior of a material under general loading conditions. For this reason, we now assess the performance of the discovered models in an extrapolation setup (i.e. for problems different from the one in Figure \ref{fig:benchmark}). For each discovered model, we perform the six simple loading tests in equation \ref{eq: deformation gradients} in the range $\gamma \in [0,1]$. We compare the baseline and the discovered constitutive laws in terms of strain energy density and first component of the first Piola-Kirchhoff stress tensor for all six loading conditions and noise levels. We present the results for the NH, IS, HW and GT models in Figures \ref{fig:results nh extrapolation}, \ref{fig:results is extrapolation}, \ref{fig:results hw extrapolation}, and \ref{fig:results gt extrapolation}, respectively.

As expected, the NH model predictions are identical to the baseline for the noise-free and $\sigma^* = 10^{-4}$ cases, while they are very close to the baseline for $\sigma^*=10^{-3}$. For the IS model, the predictions for the strain energy density in the noise-free case are very close to the baseline, while for the noisy cases we observe discrepancies especially in UC. The predictions for the first component of the first Piola-Kirchhoff stress tensor are very accurate for the noise-free case, but present discrepancies for the noisy cases especially for SS and UC loading. For the HW model, the predictions for the strain energy density are quite accurate for all loading conditions and noise levels, except for UC and BC, where we observe discrepancies for higher deformations. For the first component of the first Piola-Kirchhoff, we observe more pronounced discrepancies especially for the UC, SS and PS loading conditions and in the presence of noise. For the GT model, we observe a very good agreement between the baseline and the predicted model for all loading conditions for both strain energy density and stress in the noise-free and $\sigma^*=10^{-4}$ cases. However, for the noise level $\sigma^* =10^{-3}$ we observe high discrepancies for most loading conditions, which is reasonable considering the significant level of noise contamination and the usage of the raw noise-affected data with no denoising. For all the models and out of all loading cases,  slightly larger discrepancies are obtained for the loading conditions involving shear. This can be explained by the fact that the extent of shear in the training data is quite limited. Similar observations were made in previous works with model discovery approaches based on SR \cite{flaschel2021unsupervised}.

\begin{figure}
\caption*{  \hspace{1cm} Supervised Case \hspace{3.5cm} Unsupervised Case} 
  \begin{subfigure}{0.25\textwidth}
    \centering
    \includegraphics[width=\linewidth]{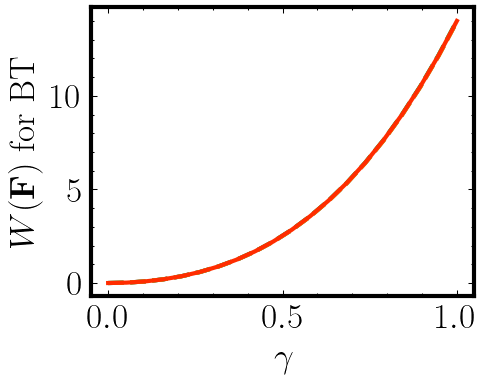}
  \end{subfigure}%
  \begin{subfigure}{0.25\textwidth}
    \centering
    \includegraphics[width=\linewidth]{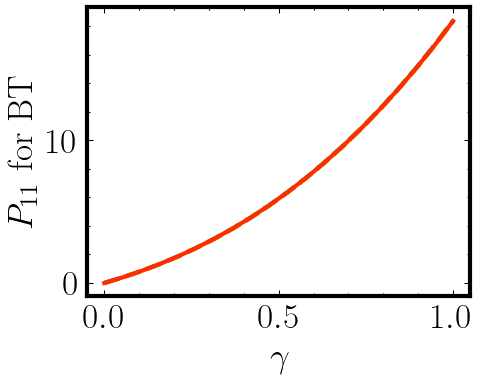}
  \end{subfigure}
  \begin{subfigure}{0.25\textwidth}
    \centering
    \includegraphics[width=\linewidth]{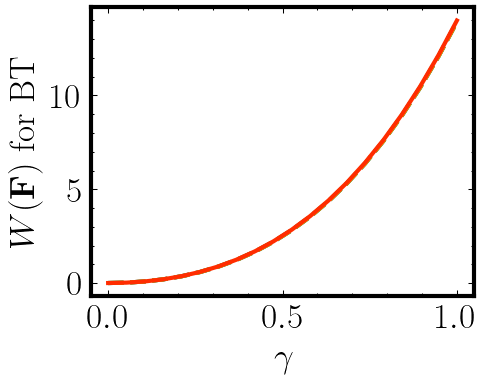}
  \end{subfigure}%
  \begin{subfigure}{0.25\textwidth}
    \centering
    \includegraphics[width=\linewidth]{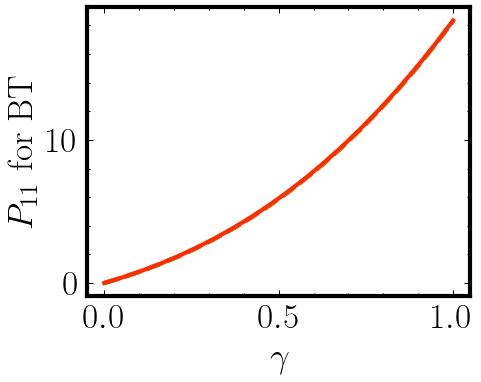}
  \end{subfigure}
  
  \begin{subfigure}{0.25\textwidth}
    \centering
    \includegraphics[width=\linewidth]{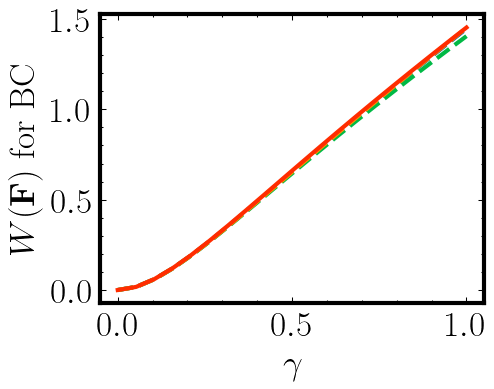}
  \end{subfigure}%
  \begin{subfigure}{0.25\textwidth}
    \centering
    \includegraphics[width=\linewidth]{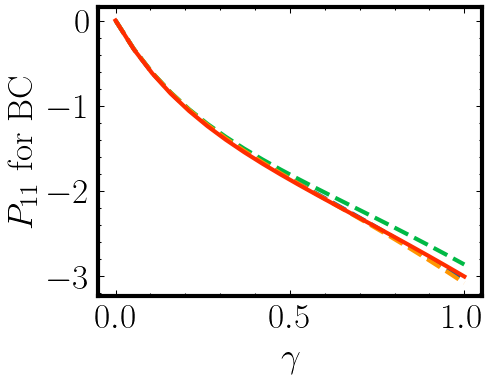}
  \end{subfigure}
  \begin{subfigure}{0.25\textwidth}
    \centering
    \includegraphics[width=\linewidth]{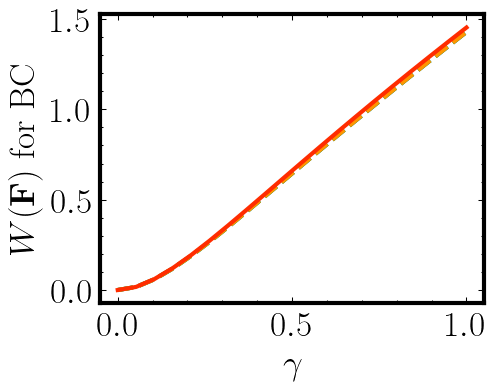}
  \end{subfigure}%
  \begin{subfigure}{0.25\textwidth}
    \centering
    \includegraphics[width=\linewidth]{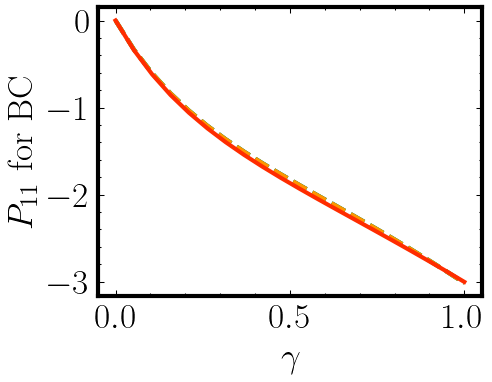}
  \end{subfigure}
  
  \begin{subfigure}{0.25\textwidth}
    \centering
    \includegraphics[width=\linewidth]{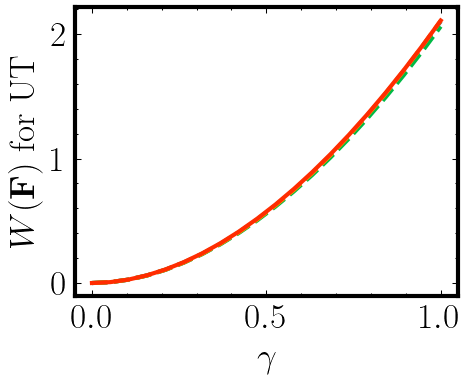}
  \end{subfigure}%
  \begin{subfigure}{0.25\textwidth}
    \centering
    \includegraphics[width=\linewidth]{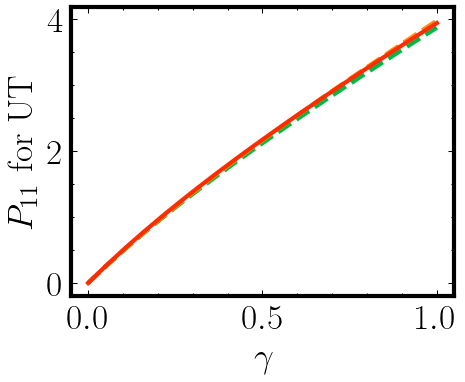}
  \end{subfigure}
  \begin{subfigure}{0.25\textwidth}
    \centering
    \includegraphics[width=\linewidth]{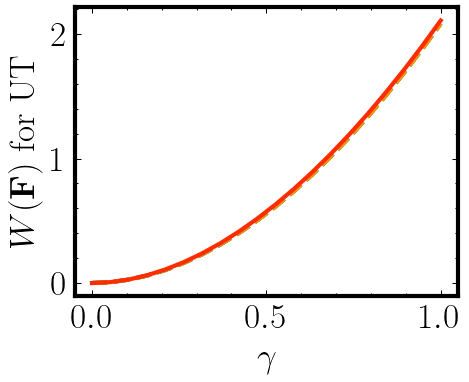}
  \end{subfigure}%
  \begin{subfigure}{0.25\textwidth}
    \centering
    \includegraphics[width=\linewidth]{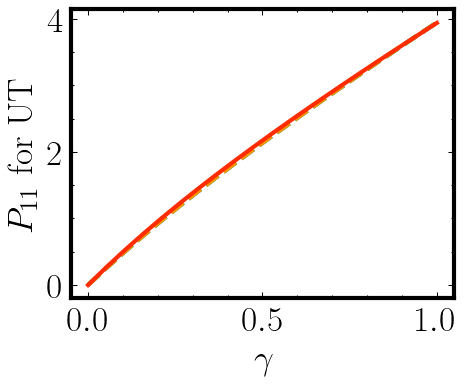}
  \end{subfigure}
  
  \begin{subfigure}{0.25\textwidth}
    \centering
    \includegraphics[width=\linewidth]{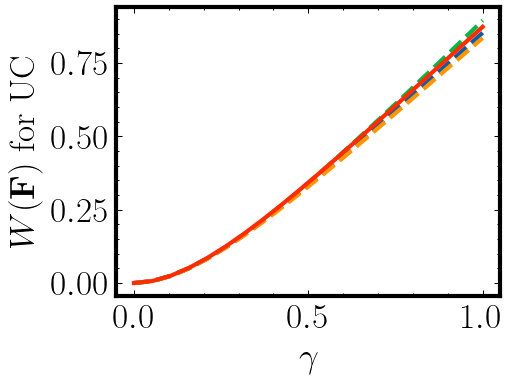}
  \end{subfigure}%
  \begin{subfigure}{0.25\textwidth}
    \centering
    \includegraphics[width=\linewidth]{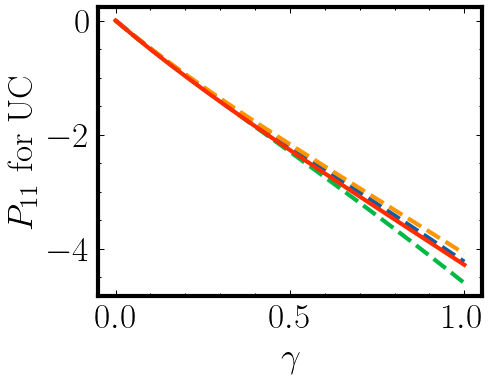}
  \end{subfigure}
  \begin{subfigure}{0.25\textwidth}
    \centering
    \includegraphics[width=\linewidth]{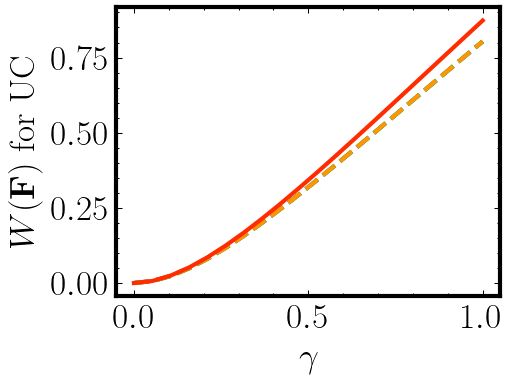}
  \end{subfigure}%
  \begin{subfigure}{0.25\textwidth}
    \centering
    \includegraphics[width=\linewidth]{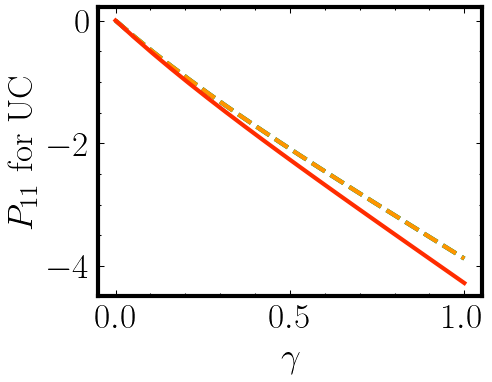}
  \end{subfigure}

   \begin{subfigure}{0.25\textwidth}
    \centering
    \includegraphics[width=\linewidth]{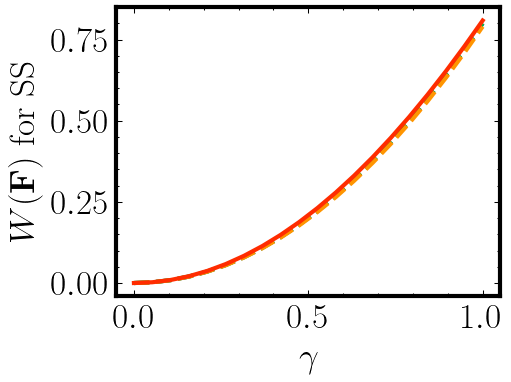}
  \end{subfigure}%
  \begin{subfigure}{0.25\textwidth}
    \centering
    \includegraphics[width=\linewidth]{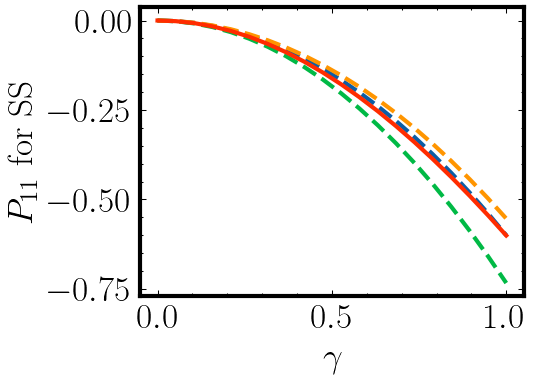}
  \end{subfigure}
   \begin{subfigure}{0.25\textwidth}
    \centering
    \includegraphics[width=\linewidth]{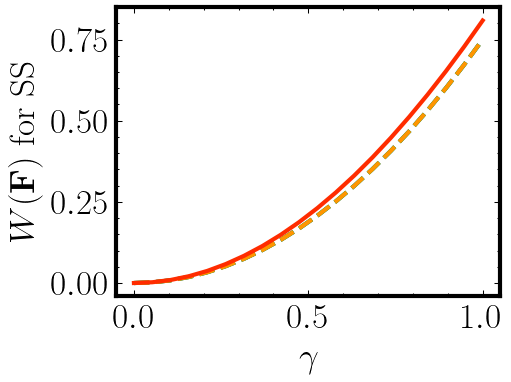}
  \end{subfigure}%
  \begin{subfigure}{0.25\textwidth}
    \centering
    \includegraphics[width=\linewidth]{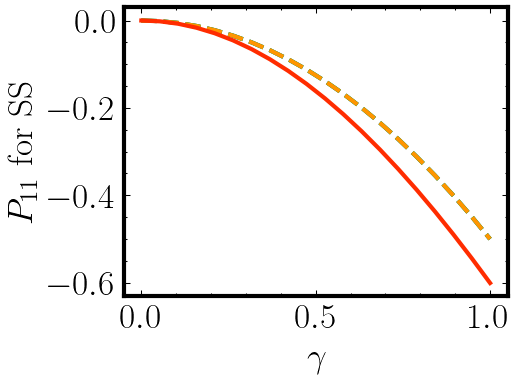}
  \end{subfigure}

  \begin{subfigure}{0.25\textwidth}
    \centering
    \includegraphics[width=\linewidth]{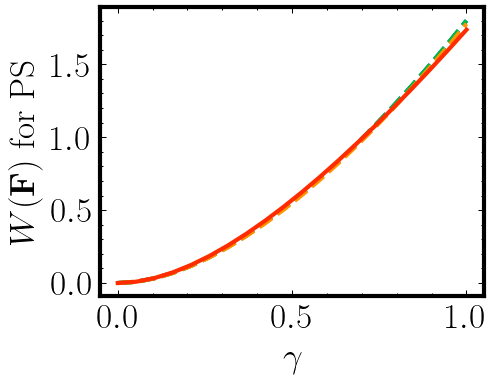}
  \end{subfigure}%
  \begin{subfigure}{0.25\textwidth}
    \centering
    \includegraphics[width=\linewidth]{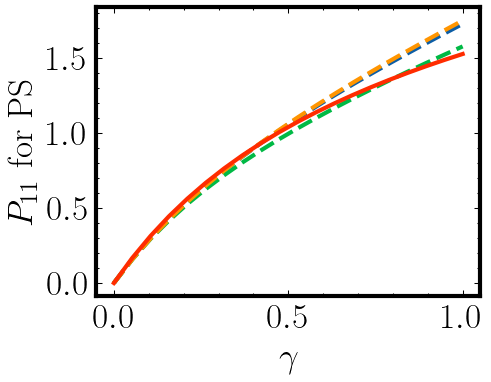}
  \end{subfigure}
  \begin{subfigure}{0.25\textwidth}
    \centering
    \includegraphics[width=\linewidth]{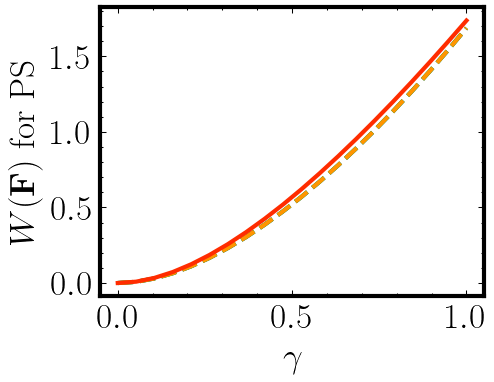}
  \end{subfigure}%
  \begin{subfigure}{0.25\textwidth}
    \centering
    \includegraphics[width=\linewidth]{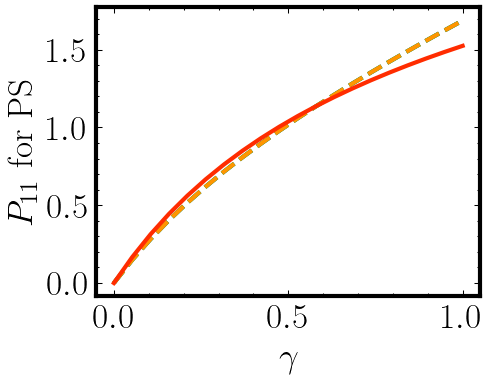}
  \end{subfigure}
  
  \caption{Ogden model: Left: comparison between the baseline and the predicted strain energy density, Right: comparison between the first component of the Piola-Kirchhoff stress tensor for loading conditions not in the training set evaluated for different displacement magnitudes parameterized by $\gamma$. For all cases, the solid red line represents the baseline, the dashed blue line the prediction for the noise-free data, the  dashed orange line the prediction for $\sigma^*=10^{-4}$ and the  dashed green line the prediction for $\sigma^*=10^{-3}$.}
  \label{fig:results og extrapolation}
\end{figure}

    % maxiter='100 + 150 * (N+3)**2 // popsize**0.5  #v maximum number of iterations',
    % popsize='4 + 3 * np.log(N)  # population size, AKA lambda, int(popsize) is the number of new solution per iteration',

\section{Conclusions}
\label{sec:discussion}

In this paper, we proposed formal grammars (specifically, Context-Free Grammars and Regular Tree Grammars) as conceptual and practical tools for the automated discovery of material models, featuring low complexity, ability to systematically embed constraints stemming from domain knowledge, and generalization capability. Leveraging formal grammars specifically designed for constitutive laws (focusing on hyperelasticity in this initial paper), we could achieve two important goals. On the one hand, we could automatically generate extensive libraries of hyperelastic material models including a desired, potentially very large number of terms which satisfy the relevant physics constraints, combine high expressivity with parsimony and can be deployed for model discovery based on sparse regression. On the other hand, we proposed and explored a grammar-based symbolic regression  pipeline for constitutive law discovery composed by a library construction step, a pre-training step, and a data-driven discovery step, and leveraging a combination of the Recursive Tree Grammar Variational Autoencoder method \cite{paassen2022recursive} and the Covariant Matrix Adaptation Evolutionary Strategy \cite{hansen2003reducing, hansen2001completely} for gradient-free optimization. We demonstrated that the approach is able to discover accurate and parsimonious hyperelastic models on four different benchmark cases. 

%\paragraph{Limitations:}
We here put forth an important first step toward establishing the language of material models, along with a number of possible useful downstream tasks and applications. However, the proposed method in its current form presents certain limitations. First, the efficiency and computational wall clock time required for the library generation is limited by the rate of rejections of the expressions, because some of the physics constraints are not imposed a priori but checked after the final expression is generated. Moreover, the decoder of the VAE does not allow to check the semantic validity of the generated expression during the process of decoding, which implies that expressions are yielded that may not satisfy the constitutive law constraints. 
A naive approach for imposing these constraints to the output of the decoder would be the following: Consider a strain energy function of the form $\tilde{W}(\mathbf{F}) = W(\mathbf{F}) + W^0  + W^c$. In this case, the symbolic regression model learns to predict $W(\mathbf{F})$ and the corrections, $W^0$ and $W^c$, are applied after the decoding takes place as physics-informed constraints. Even though this is possible, there exist two drawbacks for the particular approach: a) the evaluation of the corrections relies on symbolic computations, which makes the training prohibitively expensive, and b) as the physics informed constraints are not a natural part of the grammar (and therefore the encoding process) they are likely to result into disorienting the optimization process and moving it away from possibly suitable areas of the solution space. A more involved approach to add physics constraints would be through the definition of an attribute grammar \cite{knuth1968semantics}, where the attribute rules enforce the semantics, in this case the constitutive law constraints.
Furthermore, the CMA-ES method is an evolutionary strategy, which implies that different runs of CMA-ES can lead to different results and expressions offering different levels of approximation to the baseline. Additionally, the decoder is not suited to run on GPUs as it sequentially decides on the most probable grammar rule given a non-terminal symbol from a list of non-terminals. This limitation applies when choosing to work with tree structures. Finally, in this paper we considered grammars that only describe the behavior of isotropic hyperelastic materials, but the proposed methodology is extendable to more general and complex classes of material behavior. All these limitations will be addressed in forthcoming research.

%\paragraph{Future Work:}

% To overcome these limitations, we need to build a VAE approach that checks the semantic validity of the expressions during the generation. This can be done by constructing an auxiliary or attribute language and checking the attributes during decoding. The normalization and volumetric growth conditions can also be set as attributes of the language such that we do not need to check if these conditions are satisfied for the final expression, and thus increasing the efficiency of the automatic library creation. Finally, we need to consider different methods for searching the reduced space of the VAE to find the model that best fits that data in order to increase the robustness of the model discovery. An alternative to CMA-ES would be Bayesian Optimization as in \cite{kusner2017grammar}. 
%Finally, we need to design experiments for different amounts of deformation and containing different amounts of shear stress in order to increase the probability of getting a model that agrees with the baseline for different loading conditions. 

\section*{Code Availability}

The code for automated model discovery with the proposed procedure will be made available online at the time of publication.

\section*{Acknowledgments}

G.K. would like to acknowledge support from Asuera Stiftung via the ETH Zurich Foundation. G.K. would like to thank Anej Svete and Konstantinos Kallas for the fruitful discussions regarding formal languages. L.D.L. acknowledges funding by the Swiss National Science Foundation through grant N. 200021-204316 ‘‘Unsupervised data-driven discovery of material laws’’. S.M. acknowledges support in part by the DOE SEA-CROGS project (DE-SC0023191). 

\bibliographystyle{unsrt}  
\bibliography{references}

\begin{thebibliography}{10}

\bibitem{mahnken2004identification}
Rolf Mahnken.
\newblock Identification of material parameters for constitutive equations.
\newblock {\em Encyclopedia of computational mechanics}, 2004.

\bibitem{flaschel2021unsupervised}
Moritz Flaschel, Siddhant Kumar, and Laura De~Lorenzis.
\newblock Unsupervised discovery of interpretable hyperelastic constitutive
  laws.
\newblock {\em Computer Methods in Applied Mechanics and Engineering},
  381:113852, 2021.

\bibitem{oliveira2021numerical}
Hugo~Luiz Oliveira, Fran{\c{c}}ois Louf, and Fabrice Gatuingt.
\newblock Numerical study based on the constitutive relation error for
  identifying semi-rigid joint parameters between planar structural elements.
\newblock {\em Engineering Structures}, 236:112015, 2021.

\bibitem{mototake2020universal}
Yoh-ichi Mototake, Hitoshi Izuno, Kenji Nagata, Masahiko Demura, and Masato
  Okada.
\newblock A universal bayesian inference framework for complicated creep
  constitutive equations.
\newblock {\em Scientific Reports}, 10(1):10437, 2020.

\bibitem{pierron2021towards}
F~Pierron and M~Gr{\'e}diac.
\newblock Towards material testing 2.0. a review of test design for
  identification of constitutive parameters from full-field measurements.
\newblock {\em Strain}, 57(1):e12370, 2021.

\bibitem{li2023database}
Liang Li, Qian Shao, Yichen Yang, Zengtao Kuang, Wei Yan, Jie Yang, Ahmed
  Makradi, and Heng Hu.
\newblock A database construction method for data-driven computational
  mechanics of composites.
\newblock {\em International Journal of Mechanical Sciences}, 249:108232, 2023.

\bibitem{bock2019review}
Frederic~E Bock, Roland~C Aydin, Christian~J Cyron, Norbert Huber, Surya~R
  Kalidindi, and Benjamin Klusemann.
\newblock A review of the application of machine learning and data mining
  approaches in continuum materials mechanics.
\newblock {\em Frontiers in Materials}, 6:110, 2019.

\bibitem{kirchdoerfer2016data}
Trenton Kirchdoerfer and Michael Ortiz.
\newblock Data-driven computational mechanics.
\newblock {\em Computer Methods in Applied Mechanics and Engineering},
  304:81--101, 2016.

\bibitem{martius2016extrapolation}
Georg Martius and Christoph~H Lampert.
\newblock Extrapolation and learning equations.
\newblock {\em arXiv preprint arXiv:1610.02995}, 2016.

\bibitem{sahoo2018learning}
Subham Sahoo, Christoph Lampert, and Georg Martius.
\newblock Learning equations for extrapolation and control.
\newblock In {\em International Conference on Machine Learning}, pages
  4442--4450. PMLR, 2018.

\bibitem{costa2020fast}
Allan Costa, Rumen Dangovski, Owen Dugan, Samuel Kim, Pawan Goyal, Marin
  Solja{\v{c}}i{\'c}, and Joseph Jacobson.
\newblock Fast neural models for symbolic regression at scale.
\newblock {\em arXiv preprint arXiv:2007.10784}, 2020.

\bibitem{linka2023new}
Kevin Linka and Ellen Kuhl.
\newblock A new family of constitutive artificial neural networks towards
  automated model discovery.
\newblock {\em Computer Methods in Applied Mechanics and Engineering},
  403:115731, 2023.

\bibitem{tacc2023benchmarking}
Vahidullah Ta{\c{c}}, Kevin Linka, Francisco Sahli-Costabal, Ellen Kuhl, and
  Adrian~Buganza Tepole.
\newblock Benchmarking physics-informed frameworks for data-driven
  hyperelasticity.
\newblock {\em Computational Mechanics}, pages 1--17, 2023.

\bibitem{tibshirani1996regression}
Robert Tibshirani.
\newblock Regression shrinkage and selection via the lasso.
\newblock {\em Journal of the Royal Statistical Society Series B: Statistical
  Methodology}, 58(1):267--288, 1996.

\bibitem{tibshirani2013lasso}
Ryan~J Tibshirani.
\newblock The lasso problem and uniqueness.
\newblock 2013.

\bibitem{zou2007degrees}
Hui Zou, Trevor Hastie, and Robert Tibshirani.
\newblock On the "degrees of freedom" of the lasso.
\newblock {\em The Annals of Statistics}, 35(5):2173--2192, 2007.

\bibitem{landajuela2022unified}
Mikel Landajuela, Chak~Shing Lee, Jiachen Yang, Ruben Glatt, Claudio~P
  Santiago, Ignacio Aravena, Terrell Mundhenk, Garrett Mulcahy, and Brenden~K
  Petersen.
\newblock A unified framework for deep symbolic regression.
\newblock {\em Advances in Neural Information Processing Systems},
  35:33985--33998, 2022.

\bibitem{brunton2016discovering}
Steven~L Brunton, Joshua~L Proctor, and J~Nathan Kutz.
\newblock Discovering governing equations from data by sparse identification of
  nonlinear dynamical systems.
\newblock {\em Proceedings of the national academy of sciences},
  113(15):3932--3937, 2016.

\bibitem{JOSHI2022115225}
Akshay Joshi, Prakash Thakolkaran, Yiwen Zheng, Maxime Escande, Moritz
  Flaschel, Laura {De Lorenzis}, and Siddhant Kumar.
\newblock Bayesian-euclid: Discovering hyperelastic material laws with
  uncertainties.
\newblock {\em Computer Methods in Applied Mechanics and Engineering},
  398:115225, 2022.

\bibitem{FLASCHEL2023105404}
Moritz Flaschel, Huitian Yu, Nina Reiter, Jan Hinrichsen, Silvia Budday, Paul
  Steinmann, Siddhant Kumar, and Laura {De Lorenzis}.
\newblock Automated discovery of interpretable hyperelastic material models for
  human brain tissue with euclid.
\newblock {\em Journal of the Mechanics and Physics of Solids}, 180:105404,
  2023.

\bibitem{BODDAPATI2023105471}
Jagannadh Boddapati, Moritz Flaschel, Siddhant Kumar, Laura {De Lorenzis}, and
  Chiara Daraio.
\newblock Single-test evaluation of directional elastic properties of
  anisotropic structured materials.
\newblock {\em Journal of the Mechanics and Physics of Solids}, 181:105471,
  2023.

\bibitem{pierre2023principal}
Sarah R~St Pierre, Kevin Linka, and Ellen Kuhl.
\newblock Principal-stretch-based constitutive neural networks autonomously
  discover a subclass of ogden models for human brain tissue.
\newblock {\em Brain Multiphysics}, 4:100066, 2023.

\bibitem{pierre2023discovering}
Skyler R~St Pierre, Divya Rajasekharan, Ethan~C Darwin, Kevin Linka, Marc~E
  Levenston, and Ellen Kuhl.
\newblock Discovering the mechanics of artificial and real meat.
\newblock {\em Computer Methods in Applied Mechanics and Engineering},
  415:116236, 2023.

\bibitem{marino2023automated}
Enzo Marino, Moritz Flaschel, Siddhant Kumar, and Laura De~Lorenzis.
\newblock Automated identification of linear viscoelastic constitutive laws
  with euclid.
\newblock {\em Mechanics of Materials}, 181:104643, 2023.

\bibitem{flaschel2022discovering}
Moritz Flaschel, Siddhant Kumar, and Laura De~Lorenzis.
\newblock Discovering plasticity models without stress data.
\newblock {\em npj Computational Materials}, 8(1):91, 2022.

\bibitem{bahmani2023discovering}
Bahador Bahmani, Hyoung~Suk Suh, and WaiChing Sun.
\newblock Discovering interpretable elastoplasticity models via the neural
  polynomial method enabled symbolic regressions.
\newblock {\em arXiv preprint arXiv:2307.13149}, 2023.

\bibitem{flaschel2023automated}
Moritz Flaschel, Siddhant Kumar, and Laura De~Lorenzis.
\newblock Automated discovery of generalized standard material models with
  euclid.
\newblock {\em Computer Methods in Applied Mechanics and Engineering},
  405:115867, 2023.

\bibitem{koza1994genetic}
John~R Koza.
\newblock Genetic programming as a means for programming computers by natural
  selection.
\newblock {\em Statistics and computing}, 4:87--112, 1994.

\bibitem{mundhenk2021symbolic}
T~Nathan Mundhenk, Mikel Landajuela, Ruben Glatt, Claudio~P Santiago, Daniel~M
  Faissol, and Brenden~K Petersen.
\newblock Symbolic regression via neural-guided genetic programming population
  seeding.
\newblock {\em arXiv preprint arXiv:2111.00053}, 2021.

\bibitem{schmidt2009distilling}
Michael Schmidt and Hod Lipson.
\newblock Distilling free-form natural laws from experimental data.
\newblock {\em science}, 324(5923):81--85, 2009.

\bibitem{virgolin2022symbolic}
Marco Virgolin and Solon~P Pissis.
\newblock Symbolic regression is np-hard.
\newblock {\em arXiv preprint arXiv:2207.01018}, 2022.

\bibitem{pal1996calibration}
Surajit Pal, G~Wije Wathugala, and Sukhamay Kundu.
\newblock Calibration of a constitutive model using genetic algorithms.
\newblock {\em Computers and Geotechnics}, 19(4):325--348, 1996.

\bibitem{birky2023generalizing}
Donovan Birky, Karl Garbrecht, John Emery, Coleman Alleman, Geoffrey Bomarito,
  and Jacob Hochhalter.
\newblock Generalizing the gurson model using symbolic regression and transfer
  learning to relax inherent assumptions.
\newblock {\em Modelling and Simulation in Materials Science and Engineering},
  2023.

\bibitem{bahmani2023physics}
Bahador Bahmani and WaiChing Sun.
\newblock Physics-constrained symbolic model discovery for polyconvex
  incompressible hyperelastic materials.
\newblock {\em arXiv preprint arXiv:2310.04286}, 2023.

\bibitem{petersen2019deep}
Brenden~K Petersen, Mikel Landajuela, T~Nathan Mundhenk, Claudio~P Santiago,
  Soo~K Kim, and Joanne~T Kim.
\newblock Deep symbolic regression: Recovering mathematical expressions from
  data via risk-seeking policy gradients.
\newblock {\em arXiv preprint arXiv:1912.04871}, 2019.

\bibitem{petersen2021incorporating}
Brenden~K Petersen, Claudio~P Santiago, and Mikel Landajuela.
\newblock Incorporating domain knowledge into neural-guided search.
\newblock {\em arXiv preprint arXiv:2107.09182}, 2021.

\bibitem{kamienny2022endtoend}
Pierre-Alexandre Kamienny, St{\'e}phane d'Ascoli, Guillaume Lample, and
  Francois Charton.
\newblock End-to-end symbolic regression with transformers.
\newblock In Alice~H. Oh, Alekh Agarwal, Danielle Belgrave, and Kyunghyun Cho,
  editors, {\em Advances in Neural Information Processing Systems}, 2022.

\bibitem{biggio2021neural}
Luca Biggio, Tommaso Bendinelli, Alexander Neitz, Aurelien Lucchi, and
  Giambattista Parascandolo.
\newblock Neural symbolic regression that scales.
\newblock In {\em International Conference on Machine Learning}, pages
  936--945. PMLR, 2021.

\bibitem{vastl2022symformer}
Martin Vastl, Jon{\'a}{\v{s}} Kulh{\'a}nek, Ji{\v{r}}{\'\i} Kubal{\'\i}k, Erik
  Derner, and Robert Babu{\v{s}}ka.
\newblock Symformer: End-to-end symbolic regression using transformer-based
  architecture.
\newblock {\em arXiv preprint arXiv:2205.15764}, 2022.

\bibitem{kusner2017grammar}
Matt~J Kusner, Brooks Paige, and Jos{\'e}~Miguel Hern{\'a}ndez-Lobato.
\newblock Grammar variational autoencoder.
\newblock In {\em International conference on machine learning}, pages
  1945--1954. PMLR, 2017.

\bibitem{dai2018syntaxdirected}
Hanjun Dai, Yingtao Tian, Bo~Dai, Steven Skiena, and Le~Song.
\newblock Syntax-directed variational autoencoder for structured data.
\newblock In {\em International Conference on Learning Representations}, 2018.

\bibitem{paassen2022recursive}
Benjamin Paa{\ss}en, Irena Koprinska, and Kalina Yacef.
\newblock Recursive tree grammar autoencoders.
\newblock {\em Machine Learning}, 111(9):3393--3423, 2022.

\bibitem{hansen2003reducing}
Nikolaus Hansen, Sibylle~D M{\"u}ller, and Petros Koumoutsakos.
\newblock Reducing the time complexity of the derandomized evolution strategy
  with covariance matrix adaptation (cma-es).
\newblock {\em Evolutionary computation}, 11(1):1--18, 2003.

\bibitem{hansen2001completely}
Nikolaus Hansen and Andreas Ostermeier.
\newblock Completely derandomized self-adaptation in evolution strategies.
\newblock {\em Evolutionary computation}, 9(2):159--195, 2001.

\bibitem{hoogeboom2015undecidable}
Hendrik~Jan Hoogeboom.
\newblock Undecidable problems for context-free grammars.
\newblock {\em Preprint https://liacs. leidenuniv. nl/\~{}
  hoogeboomhj/second/codingcomputations. pdf}, 2015.

\bibitem{linz2022introduction}
Peter Linz and Susan~H Rodger.
\newblock {\em An introduction to formal languages and automata}.
\newblock Jones \& Bartlett Learning, 2022.

\bibitem{bonet1997nonlinear}
Javier Bonet and Richard~D Wood.
\newblock {\em Nonlinear continuum mechanics for finite element analysis}.
\newblock Cambridge university press, 1997.

\bibitem{holzapfel2002nonlinear}
Gerhard~A Holzapfel.
\newblock Nonlinear solid mechanics: a continuum approach for engineering
  science, 2002.

\bibitem{ball1976convexity}
John~M Ball.
\newblock Convexity conditions and existence theorems in nonlinear elasticity.
\newblock {\em Archive for rational mechanics and Analysis}, 63:337--403, 1976.

\bibitem{hartmann2003polyconvexity}
Stefan Hartmann and Patrizio Neff.
\newblock Polyconvexity of generalized polynomial-type hyperelastic strain
  energy functions for near-incompressibility.
\newblock {\em International journal of solids and structures},
  40(11):2767--2791, 2003.

\bibitem{ebbing2010design}
Vera Ebbing.
\newblock {\em Design of polyconvex energy functions for all anisotropy
  classes}.
\newblock Inst. f{\"u}r Mechanik, Abt. Bauwissenschaften, 2010.

\bibitem{schroder2010poly}
J{\"o}rg Schr{\"o}der and Patrizio Neff.
\newblock {\em Poly-, quasi-and rank-one convexity in applied mechanics},
  volume 516.
\newblock Springer Science \& Business Media, 2010.

\bibitem{linden2023neural}
Lennart Linden, Dominik~K Klein, Karl~A Kalina, J{\"o}rg Brummund, Oliver
  Weeger, and Markus K{\"a}stner.
\newblock Neural networks meet hyperelasticity: A guide to enforcing physics.
\newblock {\em Journal of the Mechanics and Physics of Solids}, page 105363,
  2023.

\bibitem{joshi2022bayesian}
Akshay Joshi, Prakash Thakolkaran, Yiwen Zheng, Maxime Escande, Moritz
  Flaschel, Laura De~Lorenzis, and Siddhant Kumar.
\newblock Bayesian-euclid: Discovering hyperelastic material laws with
  uncertainties.
\newblock {\em Computer Methods in Applied Mechanics and Engineering},
  398:115225, 2022.

\bibitem{brence2021probabilistic}
Jure Brence, Ljup{\v{c}}o Todorovski, and Sa{\v{s}}o D{\v{z}}eroski.
\newblock Probabilistic grammars for equation discovery.
\newblock {\em Knowledge-Based Systems}, 224:107077, 2021.

\bibitem{jelinek1992basic}
Frederick Jelinek, John~D Lafferty, and Robert~L Mercer.
\newblock {\em Basic methods of probabilistic context free grammars}.
\newblock Springer, 1992.

\bibitem{chi1999statistical}
Zhiyi Chi.
\newblock Statistical properties of probabilistic context-free grammars.
\newblock {\em Computational Linguistics}, 25(1):131--160, 1999.

\bibitem{geman2002probabilistic}
Stuart Geman and Mark Johnson.
\newblock Probabilistic grammars and their applications.
\newblock {\em International Encyclopedia of the Social \& Behavioral
  Sciences}, 2002:12075--12082, 2002.

\bibitem{meurer2017sympy}
Aaron Meurer, Christopher~P Smith, Mateusz Paprocki, Ond{\v{r}}ej
  {\v{C}}ert{\'\i}k, Sergey~B Kirpichev, Matthew Rocklin, AMiT Kumar, Sergiu
  Ivanov, Jason~K Moore, Sartaj Singh, et~al.
\newblock Sympy: symbolic computing in python.
\newblock {\em PeerJ Computer Science}, 3:e103, 2017.

\bibitem{korelc2016automation}
Joze Korelc and Peter Wriggers.
\newblock {\em Automation ofFinite Element Methods}.
\newblock Springer, 2016.

\bibitem{LoggEtal2012}
A.~Logg, {K.-A.} Mardal, G.~N. Wells, et~al.
\newblock {\em Automated Solution of Differential Equations by the Finite
  Element Method}.
\newblock Springer, 2012.

\bibitem{LoggWells2010}
A.~Logg and G.~N. Wells.
\newblock {DOLFIN:} automated finite element computing.
\newblock {\em {ACM} Transactions on Mathematical Software}, 37, 2010.

\bibitem{LoggEtal_10_2012}
A.~Logg, G.~N. Wells, and J.~Hake.
\newblock {DOLFIN:} a {C++/Python} finite element library.
\newblock In A.~Logg, {K.-A.} Mardal, and G.~N. Wells, editors, {\em Automated
  Solution of Differential Equations by the Finite Element Method}, volume~84
  of {\em Lecture Notes in Computational Science and Engineering}, chapter~10.
  Springer, 2012.

\bibitem{socher2011parsing}
Richard Socher, Cliff~C Lin, Chris Manning, and Andrew~Y Ng.
\newblock Parsing natural scenes and natural language with recursive neural
  networks.
\newblock In {\em Proceedings of the 28th international conference on machine
  learning (ICML-11)}, pages 129--136, 2011.

\bibitem{tai2015improved}
Kai~Sheng Tai, Richard Socher, and Christopher~D Manning.
\newblock Improved semantic representations from tree-structured long
  short-term memory networks.
\newblock {\em arXiv preprint arXiv:1503.00075}, 2015.

\bibitem{pollack1990recursive}
Jordan~B Pollack.
\newblock Recursive distributed representations.
\newblock {\em Artificial Intelligence}, 46(1-2):77--105, 1990.

\bibitem{kingma2013auto}
Diederik~P Kingma and Max Welling.
\newblock Auto-encoding variational bayes.
\newblock {\em arXiv preprint arXiv:1312.6114}, 2013.

\bibitem{burda2015importance}
Yuri Burda, Roger Grosse, and Ruslan Salakhutdinov.
\newblock Importance weighted autoencoders.
\newblock {\em arXiv preprint arXiv:1509.00519}, 2015.

\bibitem{Agathos2018}
Konstantinos Agathos, Eleni Chatzi, and Stéphane P.~A. Bordas.
\newblock Multiple crack detection in 3d using a stable xfem and global
  optimization.
\newblock {\em Computational mechanics}, 62:835--852, 2018.

\bibitem{grandidier2006identification}
JC~Grandidier and {\'E}~Lain{\'e}.
\newblock Identification by genetic algorithm of a constitutive law taking into
  account the effects of hydrostatic pressure and speeds.
\newblock {\em Oil \& Gas Science and Technology-Revue de l'IFP},
  61(6):781--787, 2006.

\bibitem{hardt2021application}
Marvin Hardt, Deepak Jayaramaiah, and Thomas Bergs.
\newblock On the application of the particle swarm optimization to the inverse
  determination of material model parameters for cutting simulations.
\newblock {\em Modelling}, 2(1):129--148, 2021.

\bibitem{thakolkaran2022nn}
Prakash Thakolkaran, Akshay Joshi, Yiwen Zheng, Moritz Flaschel, Laura
  De~Lorenzis, and Siddhant Kumar.
\newblock Nn-euclid: Deep-learning hyperelasticity without stress data.
\newblock {\em Journal of the Mechanics and Physics of Solids}, 169:105076,
  2022.

\bibitem{isihara1951statistical}
Akira Isihara, Natsuki Hashitsume, and Masao Tatibana.
\newblock Statistical theory of rubber-like elasticity. iv.(two-dimensional
  stretching).
\newblock {\em The Journal of Chemical Physics}, 19(12):1508--1512, 1951.

\bibitem{haines1979strain}
DW~Haines and WD~Wilson.
\newblock Strain-energy density function for rubberlike materials.
\newblock {\em Journal of the Mechanics and Physics of Solids}, 27(4):345--360,
  1979.

\bibitem{gent1958forms}
Alan~N Gent and AG~Thomas.
\newblock Forms for the stored (strain) energy function for vulcanized rubber.
\newblock {\em Journal of Polymer Science}, 28(118):625--628, 1958.

\bibitem{ogden1972large}
Raymond~William Ogden.
\newblock Large deformation isotropic elasticity--on the correlation of theory
  and experiment for incompressible rubberlike solids.
\newblock {\em Proceedings of the Royal Society of London. A. Mathematical and
  Physical Sciences}, 326(1567):565--584, 1972.

\bibitem{zhang1989simple}
Kaizhong Zhang and Dennis Shasha.
\newblock Simple fast algorithms for the editing distance between trees and
  related problems.
\newblock {\em SIAM journal on computing}, 18(6):1245--1262, 1989.

\bibitem{sympy2016symengine}
SymPy Developers.
\newblock Symengine, a fast symbolic manipulation library, written in c++,
  2016.

\bibitem{knuth1968semantics}
Donald~E Knuth.
\newblock Semantics of context-free languages.
\newblock {\em Mathematical systems theory}, 2(2):127--145, 1968.

\end{thebibliography}

\appendix

\section{Representation of Mathematical Expressions}
\label{section: appendix representation of expressions}

Mathematical expressions can be represented using graphs, which can then be used to perform downstream regression or classification tasks. Within a general approach, we can view mathematical operations, variables, or expressions, which we collectively call primitives, as graph node labels and define different expressions  as different ways to connect these nodes. To explain this concept we first introduce some basic terminology on graphs, S-expressions and Polish notation. We then summarize the construction of mathematical expression using graph representations and their utilization in the context of symbolic regression. 

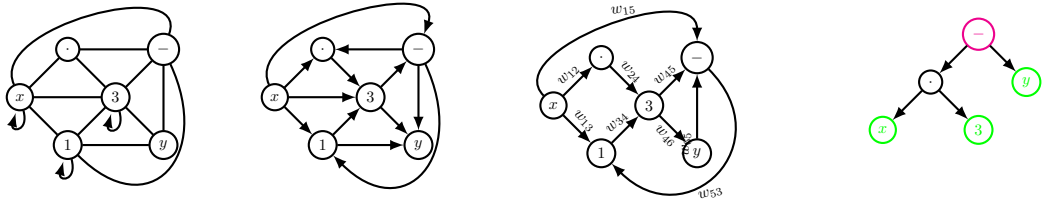
\begin{figure}
\begin{subfigure}[!t]{0.2\textwidth}
    \centering
    \vspace{-0.2cm}
\begin{tikzpicture}[node distance={15mm}, thick, scale=0.6, every node/.style={scale=0.6}, main/.style = {draw, circle}] 
\node[main] (1) {$x$}; 
\node[main] (2) [above right of=1] {$\cdot$}; 
\node[main] (3) [below right of=1] {$1$}; 
\node[main] (4) [above right of=3] {$3$}; 
\node[main] (5) [above right of=4] {$-$}; 
\node[main] (6) [below right of=4] {$y$}; 
\draw[-] (1) -- (2); 
\draw [-] (1) edge[loop below] (1);
\draw[-] (1) -- (3); 
\draw[-] (1) -- (4); 
\draw[-] (1) to [out=115,in=70,looseness=1] (5); 
\draw[-] (2) --  (4); 
\draw [-] (3) edge[loop below] (3);
\draw[-] (3) --  (4); 
\draw[-] (4) --  (5); 
\draw [-] (4) edge[loop below] (4);
\draw[-] (5) to [out=295, in=315, looseness=2.] (3); 
\draw[-] (4) -- (6); 
\draw[-] (3) -- (6); 
\draw[-] (5) -- (6); 
\draw[-] (5) -- (2); 
\end{tikzpicture} 
% \caption*{\small An example of a general graph. Each node contains a primitive, i.e. either a mathematical operation or a variable.}
\end{subfigure}
\begin{subfigure}[!t]{0.2\textwidth}
    \centering
    \vspace{-0.2cm}
\begin{tikzpicture}[node distance={15mm}, thick,scale=0.6, every node/.style={scale=0.6}, main/.style = {draw, circle}] 
\node[main] (1) {$x$}; 
\node[main] (2) [above right of=1] {$\cdot$}; 
\node[main] (3) [below right of=1] {$1$}; 
\node[main] (4) [above right of=3] {$3$}; 
\node[main] (5) [above right of=4] {$-$}; 
\node[main] (6) [below right of=4] {$y$}; 
\draw[->] (1) -- (2); 
\draw[->] (1) -- (3); 
\draw[->] (1) -- (4); 
\draw[->] (1) to [out=115,in=70,looseness=1] (5); 
\draw[->] (2) --  (4); 
\draw[->] (3) --  (4); 
\draw[->] (4) --  (5); 
\draw[->] (5) to [out=295, in=315, looseness=2.] (3); 
\draw[->] (4) -- (6); 
\draw[->] (3) -- (6); 
\draw[->] (5) -- (6); 
\draw[->] (5) -- (2); 
\end{tikzpicture} 
% \caption*{\small An example of a general graph. Each node contains a primitive, i.e. either a mathematical operation or a variable.}
\end{subfigure}
\begin{subfigure}[!c]{0.25\textwidth}
    \centering
\begin{tikzpicture}[node distance={15mm}, thick, scale=0.5, every node/.style={scale=0.6},  main/.style = {draw, circle}] 
\node[main] (1) {$x$}; 
\node[main] (2) [above right of=1] {$\cdot$}; 
\node[main] (3) [below right of=1] {$1$}; 
\node[main] (4) [above right of=3] {$3$}; 
\node[main] (5) [above right of=4] {$-$}; 
\node[main] (6) [below right of=4] {$y$}; 
\draw[->] (1) --  node[midway, above right, sloped, pos=0.01] {$w_{12}$} (2); 
\draw[->] (1) --  node[midway, above right, sloped, pos=0.01] {$w_{13}$} (3); 
\draw[->] (1) to [out=135,in=90,looseness=1]  node[midway, above left, sloped, pos=0.65] {$w_{15}$} (5); 
\draw[->] (2) -- node[midway, above right, sloped, pos=0.01] {$w_{24}$} (4); 
\draw[->] (3) -- node[midway, above right, sloped, pos=0.01] {$w_{34}$} (4); 
\draw[->] (4) -- node[midway, above right, sloped, pos=0.01] {$w_{45}$} (5); 
\draw[->] (5) to [out=315, in=315, looseness=2.] node[midway, below right, sloped, pos=0.65] {$w_{53}$} (3); 
\draw[->] (4) -- node[midway, below right, sloped, pos=0.01] {$w_{46}$} (6); 
\draw[->] (6) -- node[midway, above left, sloped, pos=0.2] {$w_{65}$} (5); 
\end{tikzpicture} 
% \caption*{\small An example of a directed acyclic graph.}
\end{subfigure}
\begin{subfigure}[!t]{0.25\textwidth}
    \centering
\begin{tikzpicture}[node distance={15mm}, thick, scale=0.6, every node/.style={scale=0.6},  main/.style = {draw, circle}] 
\node[main, magenta] (1) {$-$}; 
\node[main] (2) [below left of=1] {$\cdot$}; 
\node[main, green] (3) [below left of=2] {$x$}; 
\node[main, green] (4) [below right of=2] {$3$}; 
\node[main, green] (5) [below right of=1] {$y$}; 
\draw[->] (1) -- (2); 
\draw[->] (2) -- (3); 
\draw[->] (2) -- (4); 
\draw[->] (1) -- (5); 
\end{tikzpicture} 
\vspace{0.72cm}
% \caption*{\small Tree representation of $-(\cdot (x,3),y)$.}
\end{subfigure}
\caption{Examples of three graph topologies. On the far left, we have a general graph; In the middle left, a directed acyclic graph; In the middle right, a weighted directed acyclic graph; On the far right, a binary tree.}
\label{fig:glossary of graphs}
\end{figure}

\paragraph{Basic Glossary on Graphs} Throughout the manuscript, we consider nomenclature to describe different graphs and their nodes. For this purpose, we consider a short glossary accompanied by Figure \ref{fig:glossary of graphs} to explain these terms. A general graph is a topology whose edges are undirected, meaning that moving both from $x$ to $3$ and from $3$ to $x$ is allowed, all nodes can be connected with all other nodes without restrictions and also cycles, meaning a node is connected with itself, are allowed to exist. A directed acyclic graph is a topology where the edges are directed, meaning that moving only from $x$ to $3$ is allowed, all nodes can be connected with all nodes, and no cycles are allowed. A weighted directed acyclic is an directed acyclic graph whose edges have also weights. The weights can be thought of as the importance of edges or how strongly correlated two nodes are. A tree is a graph topology that nodes are connected to other nodes in a specific hierarchy. The node $-$ is called the parent of the nodes $\cdot$ and $y$ which are called the children of $-$. A tree takes its name from the number of children that appear at most in it. For example, a tree where the nodes have two children is called binary, with three ternary and with $k$ k-ary. The nodes that have no children are called leaf nodes and the node with no parent a root node in a tree, see the green and magenta nodes in Figure \ref{fig:glossary of graphs} respectively.  

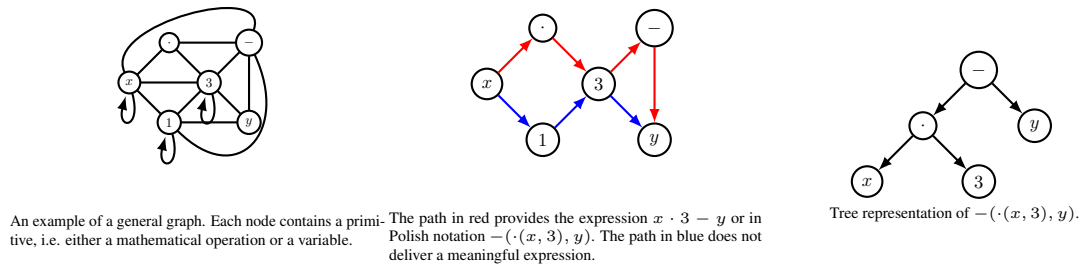
\begin{figure}
\begin{subfigure}[!t]{0.3\textwidth}
    \centering
    \vspace{-0.6cm}
\begin{tikzpicture}[node distance={15mm}, every node/.style={scale=0.5}, thick, main/.style = {draw, circle}] 
\node[main] (1) {$x$}; 
\node[main] (2) [above right of=1] {$\cdot$}; 
\node[main] (3) [below right of=1] {$1$}; 
\node[main] (4) [above right of=3] {$3$}; 
\node[main] (5) [above right of=4] {$-$}; 
\node[main] (6) [below right of=4] {$y$}; 
\draw[-] (1) -- (2); 
\draw [-] (1) edge[loop below] (1);
\draw[-] (1) -- (3); 
\draw[-] (1) -- (4); 
\draw[-] (1) to [out=115,in=70,looseness=1] (5); 
\draw[-] (2) --  (4); 
\draw [-] (3) edge[loop below] (3);
\draw[-] (3) --  (4); 
\draw[-] (4) --  (5); 
\draw [-] (4) edge[loop below] (4);
\draw[-] (5) to [out=295, in=315, looseness=2.] (3); 
\draw[-] (4) -- (6); 
\draw[-] (3) -- (6); 
\draw[-] (5) -- (6); 
\draw[-] (5) -- (2); 
\end{tikzpicture} 
\caption*{\tiny An example of a general graph. Each node contains a primitive, i.e. either a mathematical operation or a variable.}
\end{subfigure}
\begin{subfigure}[!t]{0.3\textwidth}
    \centering
\begin{tikzpicture}[node distance={15mm}, every node/.style={scale=0.7}, thick, main/.style = {draw, circle}] 
\node[main] (1) {$x$}; 
\node[main] (2) [above right of=1] {$\cdot$}; 
\node[main] (3) [below right of=1] {$1$}; 
\node[main] (4) [above right of=3] {$3$}; 
\node[main] (5) [above right of=4] {$-$}; 
\node[main] (6) [below right of=4] {$y$}; 
\draw[->, red] (1) -- (2); 
\draw[->, blue] (1) --  (3); 
\draw[->, red] (2) -- (4); 
\draw[->, blue] (3) -- (4); 
\draw[->, red] (4) -- (5); 
\draw[->, blue] (4) -- (6); 
\draw[->, red] (5) -- (6); 
\end{tikzpicture} 
\vspace{0.42cm}
\caption*{\tiny The path in red provides the expression $x \cdot 3 - y$ or in Polish notation $-(\cdot (x,3), y)$. The path in blue does not deliver a meaningful expression.}
\end{subfigure}
\begin{subfigure}[!t]{0.3\textwidth}
    \centering
\begin{tikzpicture}[node distance={15mm}, every node/.style={scale=0.7}, thick, main/.style = {draw, circle}] 
\node[main] (1) {$-$}; 
\node[main] (2) [below left of=1] {$\cdot$}; 
\node[main] (3) [below left of=2] {$x$}; 
\node[main] (4) [below right of=2] {$3$}; 
\node[main] (5) [below right of=1] {$y$}; 
\draw[->] (1) -- (2); 
\draw[->] (2) -- (3); 
\draw[->] (2) -- (4); 
\draw[->] (1) -- (5); 
\end{tikzpicture} 
\vspace{-0.2cm}
\caption*{\tiny Tree representation of $-(\cdot (x,3),y)$.}
\end{subfigure}
\caption{Left: a possible representation of mathematical expressions using an acyclic graph; middle: examples of paths on a directed acyclic graph; right: an example of a tree.}
\label{fig:explanatory figures}
\end{figure}

\paragraph{Graph and Tree Representations, S-expressions}A graph representation of a mathematical expression is one where the graph nodes represent primitives, see Figure \ref{fig:explanatory figures} left, and the edges of the graph describe different possible orders of operations. By considering orders of operations between the primitives, implying a direction on the graph edges, and different paths on the graph, possible expressions are defined, see Figure \ref{fig:explanatory figures} middle. A more simplified, yet also more restricted, representation that a priori assumes the order of the operations is the tree representation, see Figure \ref{fig:explanatory figures} right. Any tree can be represented using a recursive structure of nested lists called symbolic or S-expressions.

\paragraph{Polish Notation}To better clarify the relation between a mathematical formula and a tree, we introduce the so-called Polish notation. This is a mathematical notation where operators proceed operands. Mathematical operations on any number of arguments written in Polish notation can be easily translated into trees because their nested operation format is compatible with S-expressions. For example, let us consider the red path in Figure \ref{fig:explanatory figures}, which provides the expression $x \cdot 3 - y$. We can write this expression in Polish notation as $- (\cdot (x,3), y)$, and then translate it into a tree structure, as shown in Figure \ref{fig:explanatory figures} right. 

% \begin{subfigure}[!c]{0.3\textwidth}
%     \centering
% \begin{tikzpicture}[node distance={15mm}, thick, main/.style = {draw, circle}] 
% \node[main] (1) {$x_1$}; 
% \node[main] (2) [above right of=1] {$x_2$}; 
% \node[main] (3) [below right of=1] {$x_3$}; 
% \node[main] (4) [above right of=3] {$x_4$}; 
% \node[main] (5) [above right of=4] {$x_5$}; 
% \node[main] (6) [below right of=4] {$x_6$}; 
% \draw[->] (1) --  node[midway, above right, sloped, pos=0.01] {$w_{12}$} (2); 
% \draw[->] (1) --  node[midway, above right, sloped, pos=0.01] {$w_{13}$} (3); 
% \draw[->] (1) to [out=135,in=90,looseness=1]  node[midway, above left, sloped, pos=0.65] {$w_{15}$} (5); 
% \draw[->] (2) -- node[midway, above right, sloped, pos=0.01] {$w_{24}$} (4); 
% \draw[->] (3) -- node[midway, above right, sloped, pos=0.01] {$w_{34}$} (4); 
% \draw[->] (4) -- node[midway, above right, sloped, pos=0.01] {$w_{45}$} (5); 
% \draw[->] (5) to [out=315, in=315, looseness=2.] node[midway, below right, sloped, pos=0.65] {$w_{53}$} (3); 
% \draw[->] (4) -- node[midway, below right, sloped, pos=0.01] {$w_{46}$} (6); 
% \draw[->] (6) -- node[midway, above right, sloped, pos=0.2] {$w_{65}$} (5); 
% \end{tikzpicture} 
% % \caption*{\small An example of a directed acyclic graph.}
% \end{subfigure}

\paragraph{Construction of Expressions using Graph Representations:}By choosing the possible order of operations between primitives, either by considering different paths in directed acyclic graphs or different trees, we can generate expressions. It is evident that not all paths provide meaningful mathematical expressions; e.g. in Figure \ref{fig:explanatory figures} the blue path delivers $x13y$. To ensure that only meaningful expressions are generated, constraints can be considered in the choice of the graph paths or, in the case of trees, in the choice of the children given the parent. It is also clear that the space of possible expressions grows exponentially with the number of nodes; thus the choice of the primitives plays a crucial role on the size of this space.
%to reduce the complexity of this space, most of the times the primitives have to be predefined exploiting domain knowledge.

\begin{figure}[!ht]
    \centering
        \includegraphics[width=0.7\textwidth]{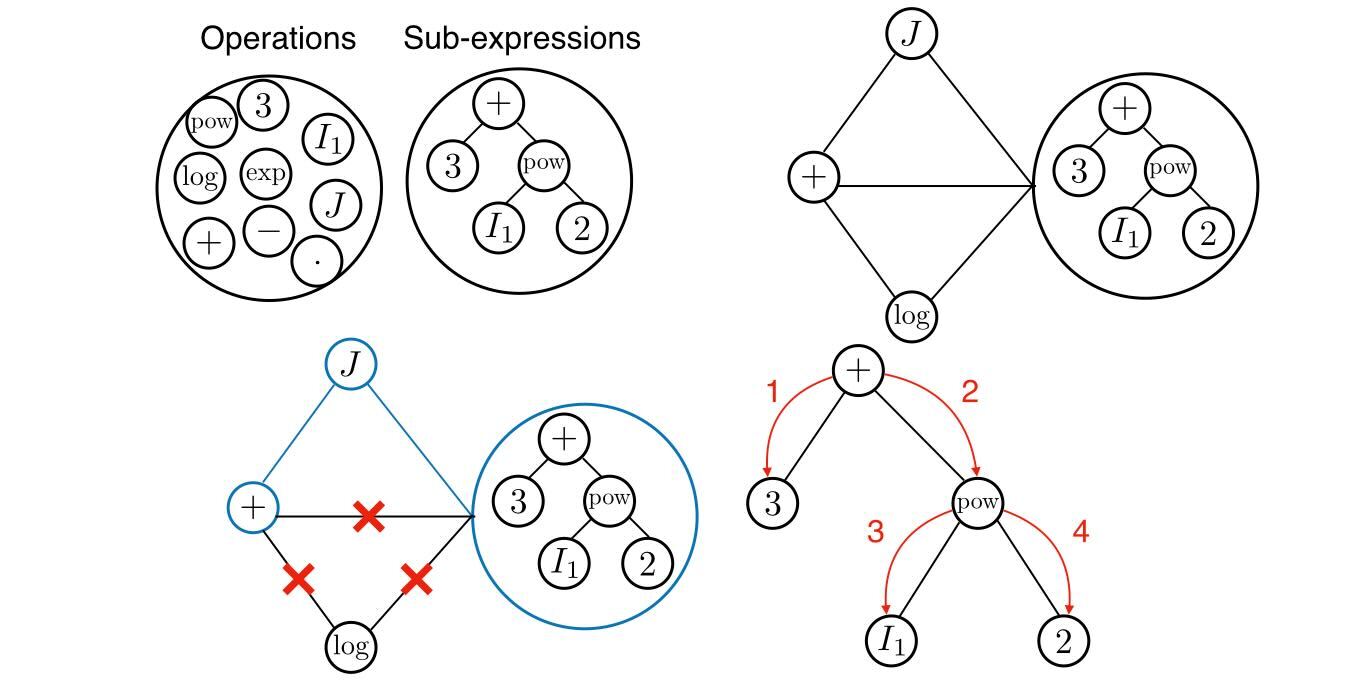}
\caption{Symbolic regression methods can be constructed by choosing i. a set of primitives consisting of operations, and sub-expressions, top left, ii. a representation of the possible functions, i.e. directed acyclic graphs, top right, and iii. sampling valid expressions based on constraints.}
\label{fig:symbolic regression explanation}
\end{figure}

\paragraph{Discovering Expressions from Data (Symbolic Regression)} As discussed in the previous paragraphs, expressions are constructed by choosing orders of operations between nodes and paths that provide meaningful results. Symbolic regression is a systematic way of finding a possible path in a graph or a tree that corresponds to an expression that best fits given data. Different symbolic regressions algorithms differ by their choices of i. primitives, i.e. whether they are only mathematical operations or variables or whether they can also consist of sub-expressions, ii. graph representations, i.e. directed acyclic or tree representations, iii. schemes to enforce that the resulting expressions are meaningful. A further choice of that of the objective function to be minimized to optimize the fit of the given data, see Figure \ref{fig:symbolic regression explanation}. 

\section{Regular Tree Grammars}
\label{section: appendix regular tree grammars}

The purpose of this section is to provide a more detailed description of the properties of RTGs, and more specifically how production rules and trees/sub-trees are defined. Moreover, we present the parsing and generation algorithms with simple examples.

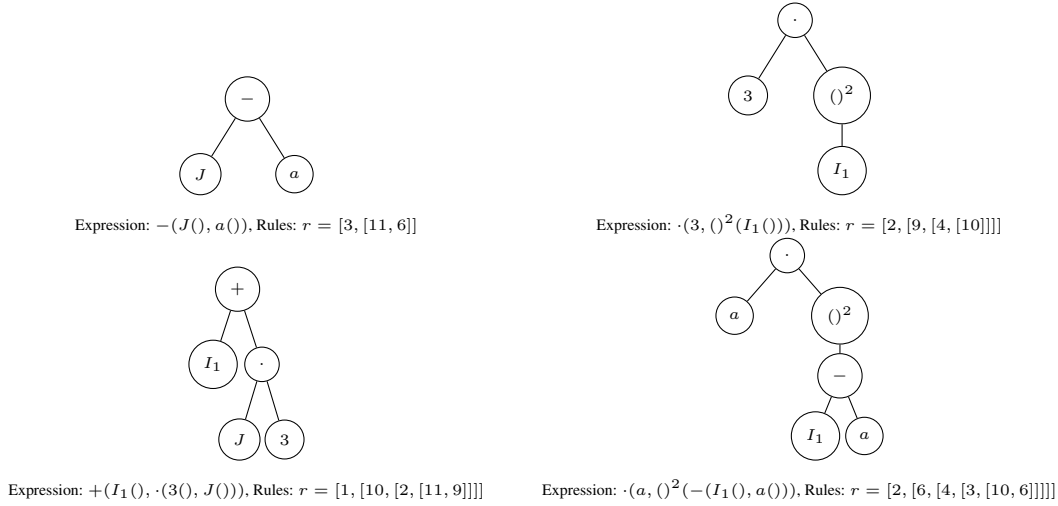
\begin{figure}[!ht]
    \centering
  \begin{subfigure}{0.44\textwidth}
    \centering
    \begin{tikzpicture}[
      level distance=1cm,
      level 1/.style={sibling distance=1.25cm},
      level 2/.style={sibling distance=0.65cm},
      level 3/.style={sibling distance=0.65cm},
      every node/.style={draw, circle, minimum size=0.05em, font=\tiny},
      edge from parent/.style={draw, -, >=stealth},
    ]
         \node {$-$}
          child { node {$J$} }
          child { node {$a$} };
    \end{tikzpicture}
    \caption*{\tiny Expression: $-(J(),a())$, Rules: $r = [3,[11,6]]$}
  \end{subfigure}
  \begin{subfigure}{0.44\textwidth}
    \centering
    \begin{tikzpicture}[
      level distance=1cm,
      level 1/.style={sibling distance=1.25cm},
      level 2/.style={sibling distance=0.65cm},
      level 3/.style={sibling distance=0.65cm},
      every node/.style={draw, circle, minimum size=0.05em, font=\tiny},
      edge from parent/.style={draw, -, >=stealth},
    ]
        \node {$\cdot$}
          child { node {$3$} }
          child { node {$()^2$}
            child { node {$I_1$} }
          };
    \end{tikzpicture}
    \caption*{\tiny Expression: $\cdot ( 3,  ()^2(I_1()))$, Rules: $r = [2,[9, [4,[10]] ]]$}
  \end{subfigure}
  \begin{subfigure}{0.44\textwidth}
    \centering
    \begin{tikzpicture}[
      level distance=1cm,
      level 1/.style={sibling distance=0.65cm},
      level 2/.style={sibling distance=0.6cm},
      level 3/.style={sibling distance=0.65cm},
      every node/.style={draw, circle, minimum size=0.05em, font=\tiny},
      edge from parent/.style={draw, -, >=stealth},
    ]
      \node {$+$}
        child { node {$I_1$} }
        child { node {$\cdot$}
          child { node {$J$} }
          child { node {$3$} }
      };
    \end{tikzpicture}
    \caption*{\tiny Expression: $+ (I_1 (),  \cdot(3(),J()))$, Rules: $r = [1,[10,[2,[11,9]]]]$}
  \end{subfigure}
  \begin{subfigure}{0.44\textwidth}
    \centering
    \begin{tikzpicture}[
      level distance=0.8cm,
      level 1/.style={sibling distance=1.4cm},
      level 2/.style={sibling distance=0.75cm},
      level 3/.style={sibling distance=0.65cm},
      every node/.style={draw, circle, minimum size=0.05em, font=\tiny},
      edge from parent/.style={draw, -, >=stealth},
    ]
    \node {$\cdot$}
        child { node {$a$} }
        child { node {$()^2$}
          child { node {$-$}
            child { node {$I_1$} }
            child { node {$a$} }
          }
        };
    \end{tikzpicture}
    \caption*{\tiny Expression: $\cdot (a,  ()^2 (-(I_1(), a()))$, Rules: $r = [2, [6, [4,[3,[10, 6]]]]]$}
  \end{subfigure}    
  \caption{Examples of sub-trees produced by the grammar. We present the derivations of the sub-trees in Polish notation to emphasize the fact that we are now working with tree grammars.}
    \label{fig:sub-tree examples}
\end{figure}

\paragraph{Properties of Regular Tree Grammars} A RTG is defined as the 4-tuple $\hat{\curlyG} = \{ \Phi, \tilde{\Sigma}, R, S \}$, where $\Phi$ a set of non-terminal symbols, $\tilde{\Sigma}$ a ranked alphabet, $R$ a set of production rules, and $S$ the starting non-terminal symbol. Consider $\tilde{\Sigma} = \{ +:2, a:0, 3:0 \}$ for simplicity and $a + 3$  a derivation of the grammar using a list of production rules. We write $a + 3$ in Polish notation as $+(y,3)$, then we can define a tree as $\hat{w} = s(\hat{t}_1, \hat{t}_2)$, where $s=+$, $\hat{t}_1=a$, and $\hat{t}_2=3$. More generally, a tree with $k$ children, a $k$-ary tree, can be defined as $\hat{w} = s(\hat{t}_1, ..., \hat{t}_k)$, where $s \in \tilde{\Sigma}$ an element of the alphabet, and $\hat{t}_1, ..., \hat{t}_k$ sub-trees defined using rules involving  elements of the alphabet. In this definition $k=0$ denotes leaf nodes, nodes without children, of the tree. The production rules are defined as $\Psi \rightarrow \hat{w}$ where $\Psi \in \Phi$ a non-terminal symbol and $\hat{w}$ a possible sub-tree derived by performing recursive substitutions of non-terminal symbols starting from $\Psi$ using production rules that have $\Psi$ on their left-hand side. More specifically, the production rules for an operation $s$ of arity $k$ of the alphabet is defined $\Psi \rightarrow s(L_1, ..., L_k)$, where $L_1,...,L_k \in \Phi$ non-terminal symbols. In Section \ref{sec:related work} and \ref{sec:formal grammars}, we discussed how the trees produced by grammar are characterized by production rules. In RTGs, the trees are derived in a recursive, hierarchical, manner and we represent this hierarchical structure of rules using nested lists. We provide examples of sub-tree expressions derived starting from symbol $\Psi$ of $\hat{\curlyG}_{NH}$, see Equation \ref{eq:example tree grammar appendix}, together with the nested lists of production rules that derive them in Figure \ref{fig:sub-tree examples}.

We are going to use the expressions in Figure \ref{fig:sub-tree examples} as examples, together with the grammar $\hat{\curlyG}_{NH}$ presented in Equation \ref{eq:example tree grammar appendix}, to explain how the nested list structures work. The list that contains the rules that generate the expression $-(J, a)$ is comprised by two levels. The second level $[11,6]$ corresponds to the rules that generate the left and the right children of $-$, respectively. The first level $[3, ...]$ corresponds to the rule that generates $-$. Therefore, what the nested list $[3,[11,6]]$ represents in words is "the first level of the tree is generated using the rule $(3)$, and on the second level the left child is generated using rule $(11)$ and the right is generated using rule $(6)$. For more complicated cases, the same logic applied. For example in the expression $\cdot (a,  ()^2 (-(I_1(), a()))$, the nested list $[2, [6, [4,[3,[10, 6]]]]]$ can be connected to the tree structure as follows:

\begin{itemize}
    \item The first level of the tree is derived by rule $(2)$ which then created two children, meaning $r= [2, [ , ]]$.
    \item  The left child in the second level is derived by rule $(6)$, $r = [2, [6, ]]$.The right child in the second level is created by first applying rule $(4)$ which created one child,  $r = [2, [6,[4,[ ]]]]$.
    \item  The child in the third level is derived by rule $3$ and creates two children, $r = [2, [6,[4,[ 3, [,] ]]]]$.
    \item The left child of the fourth level is derived by rule $(10)$ and rule $(6)$ is applied to derive the right child, $r = [2, [6,[4,[ 3, [10,6] ]]]]$.
\end{itemize}

\begin{equation}
\label{eq:example tree grammar appendix}
\NORMAL{
    \begin{split}
        S = \{& \ C \  \}, \\
        \Phi  = \{& \ C,  \Psi, L \}, \\
        \tilde{\Sigma} = \{& \ + : 2, - : 2 ,  \cdot :2, ()^2:1,  I_1:0, J:0, \  a:0 \ , 1:0, b:0,  3:0 \  \}, \\
        R  = \{ &  \ C \rightarrow  \ + (\Psi, \Psi), \mathbf{(1)} \\
               &  \ \Psi \rightarrow \cdot ( L, \Psi), \mathbf{(2)} \\
               &  \ \Psi \rightarrow \  - (L , L), \mathbf{(3)} \\
               &  \ \Psi \rightarrow  \ ()^2 (\Psi), \mathbf{(4)} \\
               &  \ \Psi \rightarrow  \ L, \mathbf{(5)} \\
                 &  L \rightarrow 0.5(), \mathbf{(6)} \\ 
                 & L \rightarrow  1(),  \mathbf{(7)}\\ 
                 & L \rightarrow 1.5(), \mathbf{(8)} \\
                 & L \rightarrow 3(), \mathbf{(9)} \\
                 & L \rightarrow I_1(), \mathbf{(10)} \\
                 & L \rightarrow J() \mathbf{(11)} \ \}.
    \end{split}
    }
\end{equation}

Each sub-list in the nested list structure describes one level of the tree hierarchy. Next we discuss the process of extracting this information given a tree and the process of generating a tree given a nested list structure in the general case. For implementing data-driven discovery of constitutive laws, RTGs need to be defined in a \emph{deterministic manner}, meaning that two production rules cannot have the same right hand side. The reason that this property is favorable is that deterministic grammars are \emph{unambiguous}, which means that each list of production rules uniquely generates one tree and vise versa \cite{paassen2022recursive}. In deterministic RTGs the parsing and generation procedures possess a $\mathcal{O}(|\hat{w}|)$ time, where $|\hat{w}|$ the size of the tree, and computation complexity \cite{paassen2022recursive}. 

\paragraph{Tree Parsing}

We discussed how expressions represented as trees can be characterized using grammar rules and provided some examples in Figure \ref{fig:sub-tree examples}. In this section, we are going to discuss the process of extracting a characterization given a tree and a grammar, called parsing. Parsing takes a tree, i.e. $\hat{w} = + (I_1 (),  \cdot(3(),J()))$ as an input and provides a nested list of rules, i.e. $r = [1,[10,[2,[11,9]]]]$ as an output. This is done by traversing the tree structure either starting from the bottom (bottom-up parsing) or starting from the top (top-down parsing), and extracting the rules that generate each level of the hierarchy. Consider a case where $\hat{w} = s(\hat{t}_1, ..., \hat{t}_k)$,  and rules of the form $\Psi \rightarrow s(L_1, ...,L_k)$ that generate any possible tree $\hat{w}$ starting from $\Psi$. Bottom-up parsing finds the rule $r_j$ that provides $\Psi$ given $L_j$ and top-down parsing tries to find the rule $r_j$ that provide $L_j$ given $\Psi$ where $\Psi,L_j \in \Phi$ and $j \in [1, ..., k]$ denotes the indices of the children of a k-ary tree. Regular tree grammars consider a bottom-up parsing approach.

In a RTG, the parsing is done recursively starting from identifying the rule $r_j$ with $L_j$ on its right hand side that generated the sub-tree $\hat{t}_j$ with $L_j$ as a starting point. For each step of the process the algorithm provides the rule $r_j$ with $L_j$ on its left-hand side and the non-terminal $L_j$ for each $j \in [1,...,k]$. Then $R$ is searched for a rule with the non-terminal $\Psi$ on its left-hand side, i.e. $r = \Psi\rightarrow s(L_1, ..., L_k)$, and returns a concatenation of the rules that generate each sub-tree, i.e. $[r, \hat{r}_1, ... ,\hat{r}_k]$. If a rule that contains $\Psi$ on the left hand is not found then the process fails, which means that a tree cannot be produced by the chosen grammar. If the recursive process successfully reaches the starting symbol $S$, then the sentence is in the language $\curlyL(\curlyG)$. We provide the general form of the parsing algorithm in Algorithm \ref{alg:parsing algorithm}.

\begin{algorithm}[!ht]
\caption{The tree parsing algorithm for a RTG $\hat{\curlyG}$ considered in this work.}
\label{alg:parsing algorithm}
\begin{algorithmic}
  \STATE \textbf{Input:} An expression tree of the form $\hat{w} = s ( \hat{t}_1, ... , \hat{t}_k)$.
  \STATE \textbf{Output:} A nested list of rules $[r, \hat{r}_1, ... ,\hat{r}_k]$ that generate $\hat{w}$.
    \STATE \textbf{function} $parsing(\hat{w})$
    \bindent
    \FOR{$j=1,...,k$}
        \STATE $L_j, \hat{r}_j = parsing(\hat{t}_j)$
    \ENDFOR
    \IF{$r = [\hat{r}_1, ..., \hat{r}_k]$ exists with $\hat{r}_1, ..., \hat{r}_k$ rules with $L_1, ..., L_k$ on their right-hand side such that $r: \Psi \rightarrow s(L_1, ..., L_k)$}
    \STATE \textbf{return} $\Psi$, $[r, \hat{r}_1, ..., \hat{r}_k]$
    \ELSE{ \STATE There exists no rule sequence with the non-terminal $\Psi$ on its right-hand side, therefore the expression is not in the language.}
    \ENDIF
    \eindent
\end{algorithmic}
\end{algorithm}

\paragraph{Tree Generation Guided by a List of Rules} 

The inverse process of parsing is generation. The parsing process extracts a nested list of rules from a tree, while the generation constructs a tree based on a list of rules. Therefore, the generation algorithm takes as an input a list, i.e. $r = [1, 10, 2,11,9 ]$ and generates a tree, i.e. $\hat{w} = +(I_1(), \cdot (3(), J()))$, presented in Figure \ref{fig:sub-tree examples}. This process does not require a nested list of rules, but instead a list of rules $r = [\hat{r}_1, ..., \hat{r}_{N_k}]$, where $N_k$ the number of rules, and $S$ the starting non-terminal. This process is performed by traversing the tree from the top to bottom and from left to right.

Consider a case where $\hat{w} = s(\hat{t}_1, ..., \hat{t}_k)$ a tree and $\Psi \rightarrow s(L_1, ..., L_k)$ the form of the grammar rules. The generation substitutes a non-terminal $\Psi$ in a tree according to a rule $\Psi \rightarrow s(L_1, ..., L_k)$, and at the same time stores $L_k, ..., L_1$, where $\Psi, L_j \in \Phi$ and $j \in [1, ..., k]$ denotes the indices of a k-ary tree. For each step of the process, the algorithm recovers the non-terminal $L_j$ on the top of the list, checks if the given rule has $L_j$ on the left-hand side, applies it and then removes $L_j$ from the list. The process stops when the list of non-terminals is empty and all the pre-defined rules have been applied. If the list is not empty then the process fails.

\begin{algorithm}[!ht]
\caption{The tree generation algorithm for a RTG $\hat{\curlyG}$ and a list of rules $[r_1, ..., r_T]$ considered in this work.}
\label{alg:generation algorithm}
\begin{algorithmic}
  \STATE \textbf{Input:}  A list of rules $[\hat{r}_1, ... ,\hat{r}_{N_r}]$ that generate $\hat{w}$.
  \STATE \textbf{Output:} An expression tree of the form $\hat{w} = s ( \hat{t}_1, ... , \hat{t}_k)$.
  \STATE Create an empty list of non-terminals and store the starting symbol $S$.
    \STATE \textbf{function} $generation(\hat{w})$
    \bindent
    \FOR{$j=1,...,N_r$}
        \STATE Recover non-terminal $\Psi$ from the tail of the list.
        \STATE Check if there a exists a non-terminal with $\Psi$ on the left-hand side. 
        \STATE Apply the rule $\Psi \rightarrow s(L_1, ... L_k)$.
        \STATE Replace $\Psi$ in the generation with $s (L_1, ..., L_k)$
        \STATE Store $L_1, ..., L_k$ in the list.
    \ENDFOR
    \IF{The list in non-empty }
    \STATE The process fails.
    \ENDIF
    \eindent
    \STATE return $\hat{w}$
\end{algorithmic}
\end{algorithm}

\paragraph{An Example of Parsing and Generating for a Simple Tree} 
In this paragraph, we present a detailed example of the tree parsing and generation algorithms. We consider a RTG version of $\hat{\curlyG}_{NH}$ and the expression $ \WF = 3 \cdot I_1^2 + (J - 0.5)^2$  to explain the concepts of parsing and generation. We can write $\WF$ in Polish notation as $\WF = + ( \cdot (3(), ()^2 (I_1())), ()^2 (- ( J(), 0.5() )))$ and is be written as a tree:

\begin{equation*}
    \hat{w}= + ( \hat{t}_1, \hat{t}_2),
\end{equation*}
where the sub-trees $\hat{t}_1$ and $\hat{t}_2$ are written as:
\begin{equation*}
    \hat{t}_1 = \cdot (\hat{t}^1_1, \hat{t}^2_1), \quad \hat{t}_2 = ()^2 ( \hat{t}^1_2),
\end{equation*}
the sub-trees $\hat{t}^1_1, \hat{t}^2_1$, and $\hat{t}^1_2 $, as:
\begin{equation*}
    \hat{t}^1_1 =  3(), \quad  \hat{t}^2_1 = ()^2 (\hat{t}^{21}_1 ), \quad \hat{t}^1_2 = - ( \hat{t}^{12}_1 , \hat{t}^{12}_2 ), 
\end{equation*}
and the sub-tree $\hat{t}^{21}_1, \hat{t}^{12}_1$ and $\hat{t}^{12}_2$ as:
\begin{equation*}
\hat{t}^{21}_1 = I_1(), \quad \hat{t}^{12}_1 = J(), \quad \hat{t}^{12}_2 = 0.5(),
\end{equation*}

Now we can apply the parsing algorithm \ref{alg:parsing algorithm} to $\WF$, as follows:

\begin{itemize}
    \item The algorithm begins from parsing $\hat{w}= + (\hat{t}_1, \hat{t}_2)$, which is composed from the sub-trees $\hat{t}_1$, and $\hat{t}_2$. The presence of the two sub-trees triggers recursion and the process considers parsing the sub-trees $\hat{t}_1 = \cdot (\hat{t}^1_1, \hat{t}^2_1)$ and $  \hat{t}_2 = ()^2 ( \hat{t}^1_2)$ first.
    \item We consider the left sub-tree $\hat{t}_1=\cdot (\hat{t}^1_1, \hat{t}^2_1)$ which is composed by $\hat{t}^1_1 =  3()$, and $\hat{t}^2_1 = ()^2 (\hat{t}^{21}_1 )$. For the right sub-tree, the recursion is triggered and the algorithm parses the $\hat{t}^{21}_1 = I_1()$. 
    \item After parsing $\hat{t}^{21}_1$, which is in this case the bottom level, the algorithm continues to the upper levels of the hierarchy extracting the grammar rules that derive each level. 
    \item The algorithm performs the same actions for the right sub-tree $\hat{t}_2 = ()^2 ( \hat{t}^1_2)$.
\end{itemize}

The application of the algorithm \ref{alg:parsing algorithm} can be written in detail as:

\begin{MyIndentedList}
\item for $j=1$ do 
\item $L_1, \hat{r}_1 = parser(\hat{t}_1)$
    \begin{MyIndentedList}
        \item for $i=1$ do 
        \item $L^1_1, \hat{r}^1_1 = parser(\hat{t}^1_1)$
        \item return $L, 9$
        
        \item for $i=2$ do 
        \item $L^2_1, \hat{r}^2_1 = parser(\hat{t}^2_1)$
        \begin{MyIndentedList}
         \item for $l=1$ do 
        \item $L^{21}_1, \hat{r}^{21}_1 = parser(\hat{t}^{21}_1)$
        \begin{MyIndentedList}
        \item for $k=1$ do
        \item $L^{21}_1, \hat{r}^{21}_1 = parser(\hat{t}^{21}_1)$
        \item return $L, 5$
        \end{MyIndentedList}
        \item return $L, [5,[10]]$     
        \end{MyIndentedList}
    return $\Psi$, $[4, [5,[10]]]$
    \end{MyIndentedList}
\item return $\Psi$, $[2,[9,[4,[10]]]]$
\item for $j=2$ do
        \begin{MyIndentedList}
         \item for $i=1$ do 
        \item $L^{1}_2, \hat{r}^{1}_2 = parser(\hat{t}^{1}_2)$
        \begin{MyIndentedList}
        \item for $l=1$ do
        \item $B^{12}_1, \hat{r}^{12}_1 = parser(\hat{t}^{12}_1)$
        \item return $L$, $11$
        \item for $l=2$ do
        \item $B^{12}_2, \hat{r}^{12}_2 = parser(\hat{t}^{12}_2)$
        \item return $L$, $6$
        \end{MyIndentedList}
        \item return $\Psi$, $[3,[12,6]]$
        \end{MyIndentedList}
\item return $\Psi$, $[4, [3,[12,6]]]$
\item return $C$, $[1, [2, [9, [4, [5,[10]]]]], [4, [3,[11,6]]]]$
\end{MyIndentedList}
The final result of the algorithm is the nested list $r=[1, [2, [9, [4, [5,[10]]]]], [4,3, [11,6]]]]$ that characterizes the tree $\hat{w} =  ( \cdot (3(), ()^2 (I_1())), ()^2 (- ( J(), 0.5() )))$. We now consider the inverse process for the previous example to showcase the generation algorithm. In this case, we provide the list of rules $r=[1, 2, 9, 4, 5,10, 3, 11, 6]$ and the starting non-terminal $S$ as an input and get the tree $\hat{w} =  ( \cdot (3(), ()^2 (I_1())), ()^2( - ( J(), 0.5() )))$ as an output following Algorithm \ref{alg:generation algorithm}. The process is provided in detail:

\begin{MyIndentedList}
\item Create an empty list of non-terminals and store the starting symbol $S$
    \begin{MyIndentedList}
    \item for $j=1$ do 
    \item Get $S$ from the list. 
    \item Apply rule $r_1 = S \rightarrow +(\Psi, \Psi)$.
    \item Replace $\Psi$ with $+(\Psi, \Psi)$ in the tree generation.
    \item Now $\hat{w} = +(\Psi, \Psi)$.
    \item Store $\Psi, \Psi$ in the list.
    \item 
    \item for $j=2$ do 
    \item Get $\Psi$ from the tail of the list. 
    \item Apply rule $r_2 = \Psi \rightarrow  \cdot (L, \Psi)$.
    \item Replace $\Psi$ with $\cdot (L, \Psi)$ in the tree generation.
    \item Now $\hat{w} = +(\cdot (L, \Psi), \Psi)$.
    \item Store $L, \Psi$ in the list.
    \item 
    \item for $j=3$ do 
    \item Get $L$ from the tail of the list. 
    \item Apply rule $r_3 = L \rightarrow  3 ()$.
    \item Replace $L$ with $3 ()$ in the tree generation.
    \item Now $\hat{w} = +(\cdot (3(), \Psi), \Psi)$.
    \item Rule $r_3$ did not produce a non-terminal to store.
    \item 
    \item for $j=4$ do 
    \item Get $\Psi$ from the tail of the list. 
    \item Apply rule $r_4 = \Psi \rightarrow  ()^2 (\Psi)$.
    \item Replace $\Psi$ with $()^2 (\Psi)$ in the tree generation.
    \item Now $\hat{w} = +(\cdot (3(), ()^2 (\Psi)), \Psi)$.
    \item Store $\Psi$ in the list.
    \item 
    \item for $j=5$ do 
    \item Get $\Psi$ from the tail of the list. 
    \item Apply rule $r_4 = \Psi \rightarrow  L$
    \item Replace $\Psi$ with $L$ in the tree generation.
    \item Now $\hat{w} = +(\cdot (3(), ()^2 (L)), \Psi)$.
    \item Store $L$ in the list.
    \item 
    \item for $j=6$ do 
    \item Get $L$ from the tail of the list.
    \item Apply rule $r_{10} = L \rightarrow  I_1()$.
    \item Replace $L$ with $I_1()$ in the tree generation.
    \item Now $\hat{w} = +(\cdot (3(), ()^2 (I_1())), \Psi)$.
    \item Rule $r_{10}$ did not produce a non-terminal to store.
    \item 
    \item for $j=7$ do 
    \item Get $\Psi$ from the tail of the list.
    \item Apply rule $r_{4} = \Psi \rightarrow ()^2(\Psi)$.
    \item Replace $\Psi$ with $()^2(\Psi)$ in the tree generation.
    \item Now $\hat{w} = +(\cdot (3(), ()^2 (I_1())), ()^2(\Psi))$.
    \item Store $L, L$ in the list.
    \item 
    \item for $j=8$ do 
    \item Get $\Psi$ from the tail of the list.
    \item Apply rule $r_{3} = \Psi \rightarrow -(L,L)$.
    \item Replace $\Psi$ with $-(L,L)$ in the tree generation.
    \item Now $\hat{w} = +(\cdot (3(), ()^2 (I_1())), ()^2(-(L,L)))$.
    \item Store $L, L$ in the list.
    \item 
    \item for $j=9$ do 
    \item Get $L$ from the tail of the list.
    \item Apply rule $r_{11} = L \rightarrow J()$.
    \item Replace $L$ with $J()$ in the tree generation.
    \item Now $\hat{w} = +(\cdot (3(), ()^2 (I_1())), ()^2(-(J(),L)))$.
    \item Rule $r_{11}$ did not produce a non-terminal to store.
    \item 
    \item for $j=10$ do 
    \item Get $L$ from the tail of the list.
    \item Apply rule $r_{6} = L \rightarrow 0.5()$.
    \item Replace $L$ with $0.5()$ in the tree generation.
    \item Now $\hat{w} = +(\cdot (3(), ()^2 (I_1())), ()^2(-(J(),0.5())))$.
    \item Rule $r_{6}$ did not produce a non-terminal to store.
    \item 
    \end{MyIndentedList}
return $\hat{w} = +(\cdot (3(), ()^2 (I_1())), ()^2(-(J(),0.5())))$.
\end{MyIndentedList}

These two algorithms generalize for the general case of the list of rules $r = [\hat{r}_1,...,\hat{r}_{N_r}]$ and $\hat{w} = s(L_1, ..., L_k)$ that generates $\hat{w}$. 

% \section{Fundamentals: Recursive Tree Variational Auto-encoders}
% \label{section: model details}

\end{document}